\documentclass{aa}  

\usepackage{lscape}
\usepackage{graphicx}
\usepackage{natbib}
\bibpunct{(}{)}{;}{a}{}{,}

\usepackage{txfonts}
\begin{document}
   \title{Chemistry in Infrared Dark Clouds.}

   \author{T. Vasyunina\inst{1}\thanks{Member of the International Max Planck Research School (IMPRS) Heidelberg}, 
   	   H. Linz\inst{1}, Th. Henning\inst{1}, 
           I. Zinchenko\inst{2}, H. Beuther\inst{1}, M. Voronkov\inst{3}}

   \offprints{T.~Vasyunina}

   \institute{Max Planck Institute for Astronomy (MPIA),
              K\"onigstuhl 17, D-69117 Heidelberg, Germany\\
              \email{[vasyunina,linz,henning,beuther]@mpia.de} 
	      \and  
	      Institute of Applied Physics of the Russian Academy
	      of Sciences, Ulyanova 46, 603950 Nizhny Novgorod, Russia\\
	      \email{zin@appl.sci-nnov.ru}	
	      \and  
	      Australia Telescope National Facility, 
	      CSIRO Astronomy and Space  
	      Science, PO Box 76, Epping, NSW 1710, Australia\\
	      \email{maxim.voronkov@csiro.au}	            
}

   \date{Received ; accepted }

 
  \abstract
   {Massive stars play an important role in shaping the structure of galaxies. 
    Infrared dark clouds (IRDCs), with their low temperatures and high densities, have 
    been identified as the potential birthplaces of massive stars. In order
    to understand the formation processes of massive stars the physical
    and chemical conditions in infrared dark clouds have to be characterized.
    }
   { The goal of this paper is to investigate the chemical composition of a
   sample of southern infrared dark clouds. One important aspect of the observations is
   to check, if the molecular abuncances in IRDCs are similar to the
    low-mass pre-stellar cores,
   or whether they show signatures of more evolved evolutionary stages.
   } 
   {We performed observations toward 15 IRDCs in the frequency range between 86 and 93 GHz
    using the 22-m Mopra radio telescope. In total, 13 molecular species comprising 
    N$_2$H$^+$, $^{13}$CS, CH$_3$CN, HC$_3$N, HNC, HCO$^+$, HCN, HNCO, 
    C$_2$H, SiO, H$^{13}$CO$^+$,  H$^{13}$CN and CH$_3$C$_2$H were observed for all targets.
    Hence, we included in general species appropriate for elevated
    densities, where some of them trace the more quiescent gas while
    others are sensitive to more dynamical processes.
   }
   { We detect HNC, HCO$^+$ and HNC emission in all clouds and N$_2$H$^+$ in all IRDCs 
    except one.
    In some clouds we detect SiO emission. Complicated shapes of the HCO$^+$ emission
    line profile are found in all IRDCs. Both signatures indicates the presence of 
    infall and outflow motions and beginning of star formation
    activity, at least in some parts of the IRDCs.
    Where possible, we calculate molecular abundances
    and make a comparison with previously obtained values for low-mass pre-stellar cores and high-mass
    protostellar objects (HMPOs). 
    We show a tendency for IRDCs to have molecular abundances
    similar to low-mass pre-stellar cores rather than to HMPOs abundances on the scale
    of our single-dish observations. 
    }
   {}
   
   \keywords{ISM: clouds, ISM: molecules, Radio lines: ISM, Stars: Formation}

   \authorrunning{T. Vasyunina et al.}
   \maketitle
%

\section{Introduction}

Massive stars play an important role in determining physical, chemical, and morphological
properties of galaxies. The last decade has seen considerable progress in the understanding
of massive star formation \citep{2007ARA&A..45..481Z}. Since their detection by 
\citet{1996A&A...315L.165P} and \citet{1998ApJ...494L.199E},  
cold and dense infrared dark clouds (IRDCs) appear to be the ideal sites to 
investigate the initial conditions for the process of massive star formation.
IRDCs have typical sizes between 1 and 10 pc, masses from several hundreds to
several thousands solar masses and H$_2$ column densities between 2 and 10$\times  10^{23}$~cm$^{-2}$
\citep[e.g.][]{2006ApJ...641..389R,2009A&A...499..149V,2009ApJ...698..324R}.

Apart from continuum observations, molecular line data have been used to 
characterize the properties of IRDCs. The first molecular line data were obtained
by \citet{1998ApJ...508..721C}, who detected H$_2$CO in 10 clouds, 
thus confirming the presence of dense gas. 
Using LVG modeling 
they estimated H$_2$CO abundances of $\sim 10^{-10}$. That is a factor of 50
lower in comparison with low-density clouds and can be explained by
accretion of gas-phase metals onto dust grains in the cold and dense
IRDCs.
Ammonia observations by \citet{2005ApJ...634L..57S} and \citet{2006A&A...450..569P} allowed 
temperature determination of IRDCs in a range from 10 to 20 K. 

While earlier studies \citep{2006ApJ...639..227S} suggested the association of most
IRDCs with the so-called Galactic molecular ring (galactocentric distance of 5
kpc), the \citet{2008ApJ...680..349J} study gave new evidence that the IRDC distribution in
the first and fourth galactic quadrant more closely follows a galactic spiral
arm (the Scutum-Centaurus arm). Since in normal spiral galaxies, OB stars seem
to form primarily in spiral arms, the association of IRDCs with a Milky Way
spiral arm supports the idea that IRDCs are related to high-mass star
formation.

\citet{2008ApJ...678.1049S}
observed N$_2$H$^+$(1-0), HC$_3$N(5-4), CCS(4$_3$-3$_2$), NH$_3$(1,1), (2,2), (3,3)
and CH$_3$OH(7-6) lines toward the massive clumps associated with IRDCs,
to study the chemical conditions in IRDCs.
Analysing the CCS and N$_2$H$^+$ abundance ratio, they conclude that
infrared dark clouds are chemicaly more evolved than low-mass pre-stellar cores.

An estimation of the chemical evolutionary status of IRDCs 
was  performed also by \citet{2009ApJ...705..123G}. 
Using C$^{18}$O, CS and N$_2$H$^+$ abundances, and a chemical evolution code,
they showed that cores where all three lines are
detected appear to be chemically young (10$^{4.5} < $ t $<$ 10$^{5.5}$~years).
Cores where no N$_2$H$^+$ emission is detected are suspected to 
be especially young (t $<$ 10$^2$~years). This suggests
that these regions may not have yet formed massive protostars. 

A molecular survey by \citet{2010ApJ...714.1658S} toward 20 massive clumps, including mid-IR bright
and mid-IR dark sources, showed  
outflow activities and the possible presence of protostars in some clouds.
Their line width analysis allowed them to reconstruct the distribution of the molecular 
species at the different evolutionary stages of massive clumps.

While many characteristics of infrared dark clouds were determined during the last ten years, 
some of their properties are still not well known.
Among the open questions is their chemical state.
What is the chemistry in IRDCs? Is it really
different from the chemistry in low-mass dark clouds?

To enlarge the sample of well characterized IRDCs, we started a program
to study gas and dust properties of southern infrared dark clouds 
\citep[][Paper I hereafter]{2009A&A...499..149V}.  A set
of 15 clouds has been selected in the pre--Spitzer era by visual examination of the extinction 
contrast of the MSX 8.3~$\mu$m images.
In the meantime, the Spitzer satellite has since succeeded MSX and provided a far
higher spatial resolution and sensitivity.
GLIMPSE mid-infrared images of our regions were retrieved from the Spitzer Archive.
Continuum 1.2~mm maps were obtained with the SIMBA bolometer array at the SEST telescope. 
A 1.2~mm and 8~$\mu$m study of these southern IRDCs showed that
these objects are not just 
distant Taurus-like clouds, but a distinct type of clouds with a clear trend to   
higher H$_2$ column densities. 
It was found that the true peak column densities, extracted from millimeter data, 
exceed  3 $\times  10^{23}$~cm$^{-2}$ (or 1~g~cm$^{-2}$), 
which has been proposed as a threshold for
 high-mass star-forming clouds \citep{2008Natur.451.1082K}.

This paper is our next step toward the understanding of the nature of the
IRDCs. Here we present our investigations of the chemical composion of  
southern clouds.
We perform molecular line observations in the 3 mm band 
with the Mopra single-dish radio telescope.
Combining molecular line data and H$_2$ column densities from
the previous study we estimate molecular
abundances and compare them with results for low-mass pre-stellar
cores.
Analysis of molecular lines provides not only information about
chemistry, but also about physical processes in molecular clouds.
For instance,
the presence of SiO emission and HCO$^+$ extended wings are evidences for outflow
activity in a cloud. Specific line shapes can indicate infall motion. 
Detection or non-detection of some species help to determine the evolutionary
status of our targets.

The paper is organized in the following way. In Sect. 2 we describe our 
target selection, the selected molecular lines, observational and technical 
details. In Sect. 3 we present the results of the qualitative analysis of obtained
molecular line spectra, line parameters and abundance estimates.
We discuss the obtained results in Sect. 4 and conclude in Sect. 5.

\section{Target and line selection, observations and data reduction}

\subsection{Target selection}

A sample of 15 southern infrared dark clouds was 
selected by visual examination of the
MSX 8.3~$\mu$m images for the presence of high-contrast dark clouds. 
1.2~mm continuum data were obtained for these objects with the 
SIMBA/SEST telescope.
Together with millimeter data we used 8\,$\mu$m IRAC data from the Spitzer  
Galactic Legacy Infrared Mid-Plane Survey Extraordinaire \citep[GLIMPSE,][]{2003PASP..115..953B} 
to investigate the physical properties of the extinction and emission material
(see Paper I).
Based on our millimeter and mid-IR data and taking into account single-dish
beam sizes at millimeter wavelengths,
we select several points for observations in every cloud.
The criterion for selection was the presence of either a 1.2~mm emission peak,
a 8~$\mu$m emission peak, or a 8~$\mu$m extinction peak. 
Such a criterion let us cover IRDCs at the different evolutionary stages.
Millimeter emission and mid-IR extinction indicate the presence of 
cold gas, typical for more quiescent regions.
A typical source of 8~$\mu$m emission
usually contains a small infrared cluster and, therefore, is likely  
to be at a more advanced evolutionary stage than most dark regions in  
this IRAC band.
The targets are listed in Table~\ref{table:main}.
3-color Spitzer/GLIMPSE images of IRDCs together with the telescope and beam positions
are presented in the on-line material Fig.~\ref{3color1}-\ref{3color3}.

\subsection{Line selection}

In order to probe the dense and cold
gas in IRDCs,
we need appropriate tracers. The 3-mm band offers a large selection of molecular
transitions. In particular, rotational transitions with low quantum
numbers are
accessible. In the following, we introduce the 3-mm lines we have chosen
for our study.

{\bf N$_2$H$^+$: }
This species is
known to be a selective tracer of quiescent gas \citep[e.g.,][]{2002ApJ...572..238C}
and is particularly suitable for studying the structure
and kinematics of cold star-forming cores. The hyperfine structure allows to
reliably measure the optical depth. 
N$_2$H$^+$ is known to be a ``late depleter'', thus it is not strongly affected by freeze
out on grain surfaces (which prevents the use of the more common CO or CS transition). 
This makes it a robust tool for scrutinizing the 
highly structured interiors of massive star--forming regions where warm massive
protostars might co--exist with younger cold and massive cloud cores.
N$_2$H$^+$ has been - detected in both low- and high-
mass pre-stellar cores and infrared dark clouds 
\citep[e.g.,][]{2001ApJS..136..703L,2003A&A...405..639P,2005A&A...442..949F,2006ApJS..166..567R}. 

{\bf $^{13}$CS:}
This molecule is a very good tracer of dense gas \citep[e.g.,][]{1996A&AS..115...81B}
due to its high dipole moment.
It was detected in  low-mass pre-stellar cores \citep[e.g.,][]{2002ApJ...569..815T}
and high-mass star forming regions \citep[e.g.,][]{2008MNRAS.386..117J}.
In contrast to N$_2$H$^+$, in cold dark clouds the CS emission vanishes toward
the core center due to depletion, but can be used to trace layers surrounding the central cores.

{\bf CH$_3$CN:}
This molecule is considered to be a tracer of warm and dense regions 
\citep[e.g.,][]{2000A&A...354.1036K,2005ApJS..157..279A, 2005IAUS..231...67V}.s
Chemical models involving only gas phase reactions as well as models which
take into account grain-surface chemistry, show that CH$_3$CN is only detectable 
in an environment with elevated temperatures 
(see \citet{2006MNRAS.367..553P} and references therein).
It was detected in more evolved massive star-formation regions, like Sgr B2 
\citep{2008MNRAS.386..117J} 
or G305.2+0.2 \citep{2006MNRAS.365..321W}, thus confirming the theoretical 
predictions.
We include CH$_3$CN  in our "cold-gas-survey", because among infrared 
dark points we have quite a significant number of regions, where
star-formation processes might have already started. 

{\bf HC$_3$N:}
This molecule belongs to an important group of interstellar molecules
- the cyanopolyynes, HC$_{2n+1}$N. and is a valuable tracer of physical 
conditions in molecular clouds. Since its first detection in space 
(Turner, 1971) HC$_3$N has been found in every type of molecular cloud
from giant molecular clouds associated with H II regions to circumstellar
envelopes.
It was shown that HC$_3$N transitions have low optical depth and
indicate the presence of denser gas than other high density tracers
\citep{1996ApJ...460..343B}.

{\bf HNC:}
This molecule is a commonly used tracer of dense gas in molecular clouds.
The abundance ratio HCN/HNC strongly depends
on the temperature and in the case of the Orion molecular cloud
decreases from 80 near the warm core to 5 on the colder edges
\citep{1986ApJ...310..383G,1992A&A...256..595S}.
The recent theoretical work by \citet{2010MNRAS.404..518S} confirmed that this ratio
should be around one for cold molecular clouds.
This line is a triplet, but the spread of the hyperfine
components is only 0.21 MHz or 0.4 km/s. 

{\bf  HCO$^+$ and H$^{13}$CO$^+$:}
HCO$^+$ is known to be a good tracer of the dense
gas especially of embedded molecular outflows 
\citep[e.g.,][]{2001A&A...376..271C,2001ApJ...549..425H}.
HCO$^+$ is an abundant
molecule, with abundances especially enhanced around regions of
higher fractional ionization. It is also enhanced by the presence
of outflows where shock-generated radiation fields are present
\citep{2000MNRAS.313..461R,2004MNRAS.351.1054R}.

The emission from the H$^{13}$CO$^+$
isotopologue is mostly optically thin and traces similar
gas densities as HCO$^+$. A comparison between the
generally optically thick HCO$^+$ line and the optically thin H$^{13}$CO$^+$
line yields information on the bulk motion of the gas in the region.

{\bf HCN  and H$^{13}$CN:}
This molecule has been suggested as ubiquitous high 
density gas tracer. Moreover, the HCN molecule is known to be a good tracer
of infall motions in low-mass star-forming regions. However, for the high-mass
cores this can be different. 
Here HCN may become an unreliable infall tracer because of a higher level of 
turbulence \citep{2008ASPC..387...38R} and outflow signatures \citep{2007A&A...470..269Z}.

{\bf HNCO: }
\citet{2000A&A...361.1079Z} showed that HNCO integrated intensities correlate well
with those of thermal SiO emission in massive clouds. This can indicate a spatial
coexistence of the two species and may hint to a common production mechanism,
presumably based on shock chemistry.

{\bf  C$_2$H:} 
Chemical models predict that C$_2$H is
only well centered on the sources when they are very young. At later stages it gets destroyed
in the central cores, but is replenished in the outer shells \citep{2008ApJ...675L..33B}.

{\bf SiO:} 
It can trace shocked gas 
potentially associated with energetic young outflows. Hence, this line can reveal
star formation activity even for cores where no Spitzer/MIPS sources are apparent
\citep[e.g.,][]{2007MNRAS.381L..30L}. For a few of our IRDCs 
we have indications from Spitzer/GLIMPSE imaging that shocked gas exists, 
often, but not exclusively, at the edges of the IRDCs. Tracers are the
``green fuzzies'' found in the 4.5$\ \mu$m GLIMPSE channel 
\citep{2008AJ....136.2391C,2009ApJS..181..360C} and generally 
attributed to pure rotational IR lines of H$_2$
\citep{2010AJ....140..196D}.

{\bf  CH$_3$C$_2$H:}
A good tracer of dense gas appropriate for early stages of
star forming regions \citep{1994ApJ...431..674B}.
This molecule can be used as a good thermometer
in a dense environment.

The covered molecular lines and transitions are
summarized in Table \ref{table:lines}.

\subsection{Mopra observations and data reduction}

The observations were made with the 22-m Mopra radio telescope, operated
by the Australia Telescope National Facility (ATNF), in the position
switching mode. 
In total we spent 7 minutes on source and 7 minutes on the OFF position.
Our targets are dense molecular condensations within larger
molecular clouds
with often widespread molecular emission. Therefore, we refrained from using
one standard
OFF position throw. Instead, OFF positions were chosen individually for every
target region and were approximately 8-10$'$ away from the source.

The Mopra spectrometer (MOPS) offers 
zoom mode configurations with  
the possibility to observe up to 16 sub-bands of 138 MHz each
within a total frequency range of 8.3 GHz.
This set up delivers a velocity
resolution of $\sim$ 0.11 km/s. 

The observations were carried out on 9-11 May 2008 with the 3mm band 
receiver. We put the central frequency for the 8.3 GHz block 
to 89270 MHz and thus covered the 
range from 85 to 93 GHz. In this range we distributed 13 zoom windows,
which covered the lines listed in Table~\ref{table:lines}.

System temperature measurements were performed every 30 minutes and
a pointing scan every hour. 
Typical system temperatures (measured with the common 
chopper-wheel technique) during the observations were 170-210 K.
At Mopra observatory, SiO masers are used to correct the telescope
pointing, giving a pointing accuracy better than 10$''$.
The main beam of the telescope varies between  36$''$ at 86 GHz and 
33$''$ at 115 GHz and
the main beam efficiency varies between 0.49 at 86 GHz and
0.44 at 100 GHz \citep{2005PASA...22...62L}  

Mopra data are originally stored in RPFITS format. Using the
ATNF Spectral line Analysis package (ASAP), we transformed these 
raw data into ascii files which were then fed  into GILDAS
for further analysis. 
The typical rms level in the obtained spectra is about 0.12$--$0.16~K.
We give 1$\sigma$ errors in Tables \ref{table:N2H}-\ref{table:H13CO}.

\begin{table*}

\caption{List of observed IRDCs.} 
\label{table:main}      
\begin{tabular}{r c c c c c c}        
\hline\hline             
\noalign{\smallskip}
Name & R.A. & Decl. & distance$^a$ & T$^c$ & N(H$_2$)$^d$  & category$^e$\\
& (J2000.0) & (J2000.0) &(kpc) & (K) & *10$^{22}$ cm$^{-2}$  & \\   

\noalign{\smallskip}   
\hline                        
\noalign{\smallskip}

IRDC308.13-1 &    13 37 01.582 &  -62 44 34.01 &4.8    & 13.5  &   1.0 &  Q   \\ 
IRDC308.13-2 &    13 37 00.418 &  -62 43 41.01 &4.0    & 13.5  &   1.0 &  M   \\ 
IRDC308.13-3 &    13 37 02.163 &  -62 43 39.01 &4.0    & 13.5  &   1.0 &  M    \\ 
IRDC309.13-1 &    13 45 17.521 &  -62 22 02.84 &3.9    & 16.3  &   0.6 &  M    \\
IRDC309.13-2 &    13 45 22.610 &  -62 23 27.48 &3.9    & 14.7  &   0.5 &  M    \\ 
IRDC309.13-3 &    13 45 16.775 &  -62 25 37.25 &3.9    & 35.4  &   0.4 &  A   \\ 
IRDC309.37-1 &    13 48 38.532 &  -62 46 17.55 &3.4    & 31.4  &   0.9 &  A   \\ 
IRDC309.37-2 &    13 47 56.116 &  -62 48 33.46 &3.4    & 15.7  &   0.8 &  A    \\ 
IRDC309.37-3 &    13 48 39.383 &  -62 47 22.39 &3.4    & 15.7  &   0.7 &  Q   \\ 
IRDC310.39-1 &    13 56 01.359 &  -62 14 18.29 &4.2    & 27.4  &   1.2 &  A    \\ 
IRDC310.39-2 &    13 56 00.759 &  -62 13 59.80 &4.2    & 27.4  &   1.2 &  A   \\ 
IRDC312.36-1 &    14 11 27.752 &  -61 29 27.18 &4.5    & 13.5  &   1.1 &  A    \\
IRDC312.36-2 &    14 11 56.773 &  -61 29 25.78 &4.0    & 14.4  &   0.3 &  Q   \\ 
IRDC313.72-1 &    14 22 53.158 &  -61 14 41.00 &3.3    & 19.9  &   0.4 &  A    \\ 
IRDC313.72-2 &    14 22 57.151 &  -61 14 10.84 &3.3    & 19.9  &   0.4 &  A   \\ 
IRDC313.72-3 &    14 23 02.720 &  -61 13 39.64 &3.3    & 19.9  &   0.3 &  Q	\\ 
IRDC313.72-4 &    14 23 04.533 &  -61 14 46.00 &3.3    & 19.9  &   0.4 &  Q	\\ 
IRDC316.72-1 &    14 44 19.000 &  -59 44 29.00 &2.7    & 26.1  &   1.2 &  M   \\ 
IRDC316.76-1 &    14 44 56.000 &  -59 48 08.00 &2.7    & 22.6  &   4.1 &  A    \\ 
IRDC316.72-2 &    14 44 15.400 &  -59 43 20.00 &2.7    & 24.3  &   1.3 &  Q    \\ 
IRDC316.76-2 &    14 45 00.500 &  -59 48 44.00 &2.8    & 23.2  &   4.8 &  A   \\ 
IRDC317.71-1 &    14 51 06.905 &  -59 16 11.03 &3.0    & 15.6  &   1.2 &  Q    \\ 
IRDC317.71-2 &    14 51 10.975 &  -59 17 01.73 &3.0    & 16.6  &   3.5 &  A    \\ 
IRDC317.71-3 &    14 51 19.667 &  -59 17 43.77 &3.2    & 15.6  &   0.6 &  Q	\\ 
IRDC320.27-1 &    15 07 56.251 &  -57 54 32.11 &2.1    & 15.3  &   0.8 &  Q    \\ 
IRDC320.27-2 &    15 07 31.616 &  -57 53 27.24 &2.2    & 16.1  &   0.5 &  Q    \\
IRDC320.27-3 &    15 07 35.077 &  -57 54 13.98 &2.1    & 16.1  &   0.4 &  Q    \\ 
IRDC321.73-1 &    15 18 26.387 &  -57 22 00.14 &2.2    & 22.0  &   1.0 &  M	\\ 
IRDC321.73-2 &    15 18 01.693 &  -57 22 02.00 &2.2    & 11.7  &   1.7 &  M	\\ 
IRDC321.73-3 &    15 18 01.065 &  -57 21 24.48 &2.1    & 11.7  &   1.7 &  A	\\ 
IRDC013.90-1 &    18 17 33.378 &  -17 06 36.70 &2.5    & 12.9  &   2.6 &  M   \\
IRDC013.90-2 &    18 17 19.350 &  -17 09 23.69 &2.4    & 13.4  &   1.1 &  Q   \\ 
IRDC316.45-1 &    14 44 51.515 &  -60 30 55.00 &3.1    & 15.4  &   0.7 &  M	\\ 
IRDC316.45-2 &    14 44 47.095 &  -60 31 30.89 &3.1    & 14.2  &   0.5 &  M	\\ 
IRDC318.15-1 &    14 55 57.704 &  -59 29 04.12 &3.0    & 17.6  &   0.5 &  M   \\ 
IRDC318.15-2 &    14 55 58.334 &  -59 28 30.52 &2.9    & 17.6  &   0.4 &  Q   \\ 
IRDC309.94-1 &    13 50 54.970 &  -61 44 21.00 &5.3$^b$& 48.8  &   5.2 &  A   \\ 

\noalign{\smallskip}   
\hline                                   
\end{tabular}

Notes: Columns are name, right ascension, declination, distance, kinetic temperature,
	H$_2$ column density and type according to mid-IR classification.

$^a$     At first, to estimate the kinematic distances to our IRDCs, we used HCO$^+$ line velocities
	(see Vasyunina et al. 2009). But more detailed investigation showed that 
	HCO$^+$ lines, as a rule, have complex line shapes and the $v_{\rm LSR}$ positions
	can be shifted up to 2 km/s in comparison with its optically thin isotopologue
	H$^{13}$CO$^+$. We cannot use H$^{13}$CO$^+$ for distance determination,
	since it is much weaker and we detected it not in all cases. 
	Thus, for distance determination we decided to 
	use N$_2$H$^+$, which is optically thin and distinguishable for all regions except IRDC309.37-2.
        Despite the significant difference in velocities (up to 2 km/s) between HCO$^+$ 
	and N$_2$H$^+$, the kinematic distances did not change drastically compared with
	the values in Paper I.
	
$^b$    from \citet{2001PASJ...53.1037S}
	
$^c$    Temperatures were derived based on (1,1) and (2,2) ammonia transitions observed with the 
	the 64-m Parkes radio telescope (Linz et al. in prep.). 	

$^d$    To estimate H$_2$ column densities in every point we used 1.2 mm data
        from SIMBA/SEST adopting the Mopra telescope beam size.  	

$^e$    "A" indicates "active" cores, "M" - "middle", "Q" - "quiescent".	
	
\end{table*}

\begin{table*}
\caption{Observed molecular species.} 
\label{table:lines} 
\centering    
\begin{tabular}{r c l l l l l} 
\hline\hline
\noalign{\smallskip} 
Molecule  & Transition & Rest frequency &  A   & g$_u$    & E$_u$ &  Comments\\
          &            & (GHz)          & (*10$^{-5}$s$^{-1}$) &    &(K)&   \\
\noalign{\smallskip}
\hline\hline
\noalign{\smallskip} 
                    
CH$_3$C$_2$H   & 5$_3$-4$_3$           &   85.442 600 & 0.129778  &  44     & 77.37 &Tracer of dense gas,\\				  
               & 5$_2$-4$_2$           &   85.450 765 & 0.170373  &  22     & 41.22 & good thermometer   \\
               & 5$_1$-4$_1$           &   85.455 665 & 0.194760  &  22     & 19.53 &			  \\
               & 5$_0$-4$_0$           &   85.457 299 & 0.202908  &  22     & 12.30   & 		    \\

\noalign{\smallskip}

H$^{13}$CN     & 1$_1$-0$_1$	     &   86.338 735 &            &         &	 &Tracer of dense gas, \\
               & 1$_2$-0$_1$	     &   86.340 167 & 2.8        &  9      & 4.14&infall motions      \\
               & 1$_0$-0$_1$	     &   86.342 256 &            &         &	 &		       \\

\noalign{\smallskip}
	       
H$^{13}$CO$^+$ & 1-0                 &   86.754 330 & 2.8        &  3      & 4.16& Tracer of dense gas \\

\noalign{\smallskip}

SiO            & 2-1                 &   86.847 010 & 2.0        &  5      & 6.25 & Trace shoked gas	\\ 

\noalign{\smallskip}

C$_2$H         & 1-0 3/2-1/2 F = 2-1 &  87.316 925  & 0.152757   &  5      & 4.19   &Tracer of early	\\ 
               &                     &              &            &         &	    &stages of  	\\ 
               &                     &              &            &         &	    &star formation	\\  

\noalign{\smallskip}
	       
HNCO           & 4$_{0,4}$-3$_{0,3 }$&  87.925 238  & 0.878011   &  9      & 10.55 & Indicate the presence of\\ 
 	       &	             &              &            &         &	   & denser gas, than other high  \\ 
 	       &	             &              &            &         &	   & density tracers  \\

\noalign{\smallskip}

HCN            & 1$_1$-0$_1$ 	     &  88.630 4157 &            &         &	   &Tracer of dense gas,\\ 
               & 1$_2$-0$_1$ 	     &  88.631 8473 & 2.4        &  9      & 4.25  &infall motions   \\ 
               & 1$_0$-0$_1$ 	     &  88.633 9360 &            &         &	   &			      \\ 

\noalign{\smallskip}
	       
HCO$^+$        & 1-0                 &  89.188 526  & 3.0        &  3      & 4.28  & Tracer of dense gas, \\ 
               &                     &              &            &         &	   & outflows \\ 

\noalign{\smallskip}

HNC            & 1$_0$-0$_1$ 	      &  90.663 450  &            &         &	  & Tracer of dense gas \\ 
               & 1$_2$-0$_1$ 	      &  90.663 574  & 2.7        &  3      & 4.35&			       \\ 
               & 1$_1$-0$_1$ 	      &  90.663 656  &            &         &	  &			       \\ 

\noalign{\smallskip}
	       
HC$_3$N        & 10-9                &  90.978 989  & 5.81300    &  21     & 24.02  & Indicate the presence of\\ 
 	       &	             &              &            &         &	    & denser gas, than other high  \\ 
 	       &	             &              &            &         &	    & density tracers  \\

\noalign{\smallskip}
                      
CH$_3$CN       & 5$_4$-4$_4$           &  91.959 206 & 2.27824    &  22     & 127.60 &Tracer of warm and  \\ 
               & 5$_3$-4$_3$           &  91.971 465 & 4.05228    &  44     & 77.58  &dense regions \\ 
               & 5$_2$-4$_2$           &  91.980 089 & 5.31863    &  22     & 41.84  &  \\ 
               & 5$_1$-4$_1$           &  91.985 316 & 6.07995    &  22     & 20.39  &  \\ 
               & 5$_0$-4$_0$           &  91.987 089 & 6.33432    &  22     & 13.25  & \\ 

\noalign{\smallskip}
	       
$^{13}$CS      & 2-1                 &  92.494 303   & 1.41254    &  10     & 6.66   & Tracer of dense gas \\

\noalign{\smallskip}
 
N$_2$H$^+$     & 1$_{11}$-0$_{01}$ &  93.171 621 &              &         &	       & Tracer of quiescent gas,\\ 
               & 1$_{11}$-0$_{22}$ &  93.171 917 &              &         &	       &	"late depleter"\\ 
               & 1$_{11}$-0$_{10}$ &  93.172 053 &              &         &	       &	\\ 
               & 1$_{21}$-0$_{21}$ &  93.173 480 &              &         &	       &	\\ 
               & 1$_{21}$-0$_{32}$ &  93.173 777 & 3.8534       &  27     & 4.47       &	 \\ 
               & 1$_{21}$-0$_{11}$ &  93.173 967 &              &         &	       &       \\ 
               & 1$_{01}$-0$_{12}$ &  93.176 265 &              &         &	       &	\\ 

\noalign{\smallskip}
\hline                                  
\end{tabular}

Notes: Columns are species, transition, rest frequency, Einstein A coefficients, degeneracy, energy
of the upper lewel, comments.

\end{table*}

\begin{figure}
\centering
  \includegraphics[width=7cm]{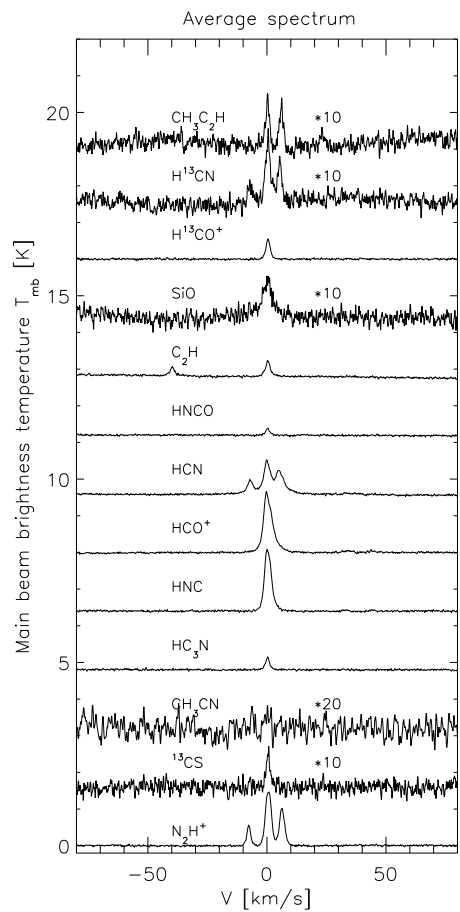}
  \caption{Average spectrum. Obtained by averaging all spectra for every species 
  	   with equal weight after shifting
  	   all of them to the same reference velocity. Several weak detections are amplified
	   by a factor of 10, and CH$_3$CN is amplified by a factor of 20 for plotting.
	   }
  \label{Fig:average}
\end{figure}

\begin{figure*}
\centering
\includegraphics[width=5cm]{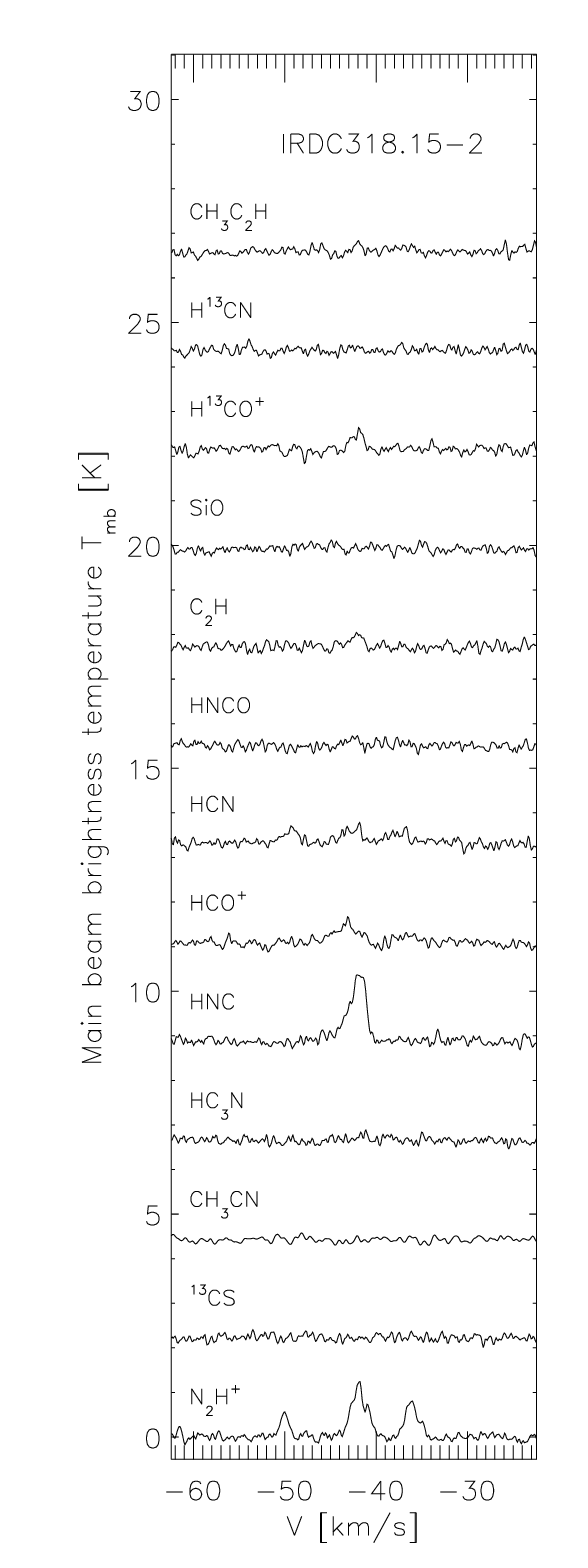}
\includegraphics[width=5cm]{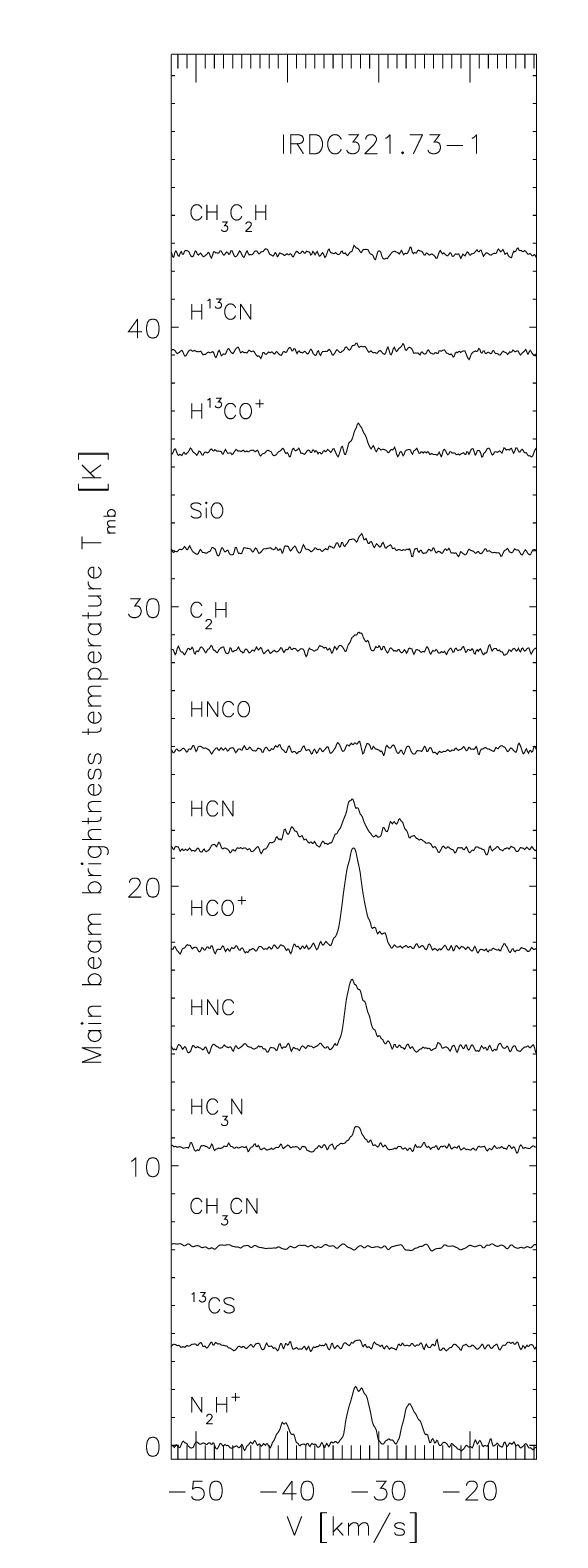}
\includegraphics[width=5cm]{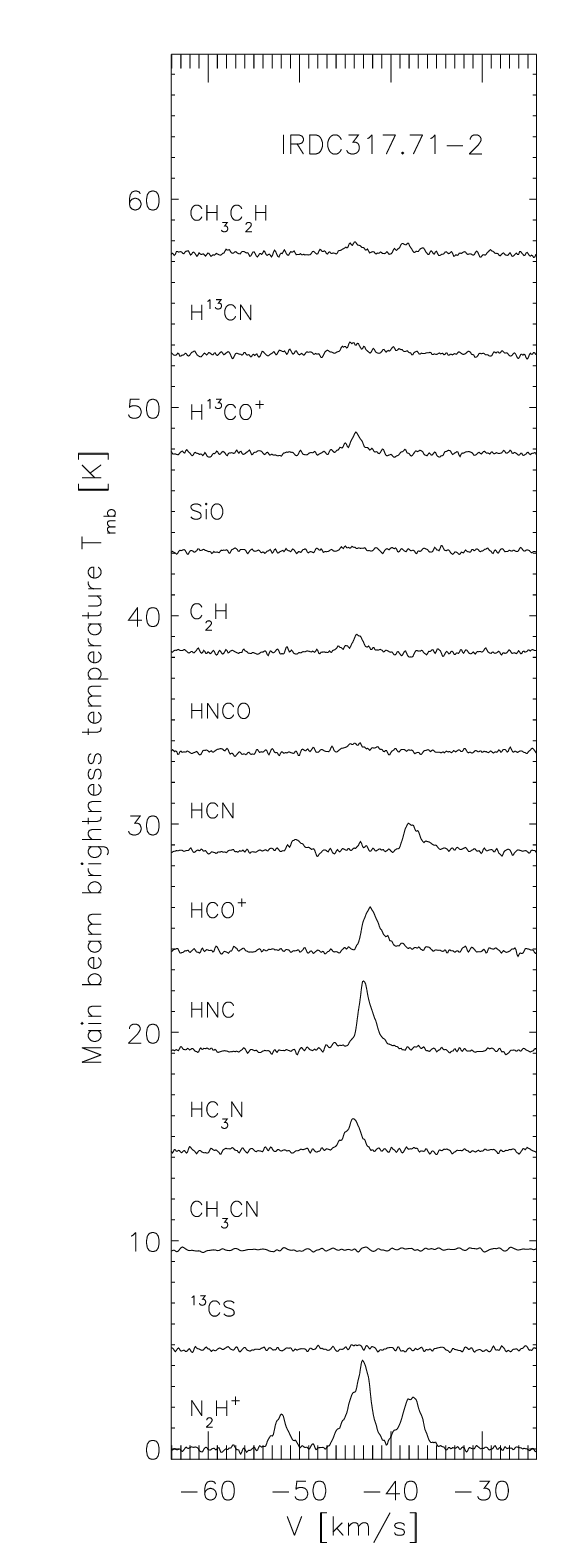}
  \caption{Line spectra for different mid-IR categories of IRDCs. (left) "Quiescent" cloud, 
  	   (middle) "Middle" cloud, (right) "Active" cloud.}
  \label{Fig:spectra}
\end{figure*}

\section{Results}

\subsection{Qualitative assessment}

Figure~\ref{Fig:average} presents the average spectrum of an infrared dark cloud.
To produce it we shifted the  spectra of all our IRDCs to the same reference velocity and averaged
spectra for each species with equal weight. The "average spectra" technique enables us to
reduce the noise level and recognize weak lines, which are not detectable in the single spectra.
Spectra for every 
single point are available in the on-line material (see Fig. ~\ref{appfig1}-\ref{appfig18}).
For all IRDCs we have quite strong and clear detections of the HNC, HCO$^+$ and HCN molecules.
In all cases we see a non-gaussian shape of HCO$^+$ with "shoulders" or double-peaked profiles.
We will discuss the asymmetries of HCO$^+$ in comparison with the optically thin
H$^{13}$CO$^+$ line in Sect.~\ref{Sect:infall} in more detail.

N$_2$H$^+$ is detected everywhere, except for IRDC309.37-2. As it was
expected, we cannot distinguish all 7 hyperfine components because of the large 
line width. 
While one hyperfine component (1$_{01}$-0$_{12}$) constitutes a distinct line peak, the
other six transitions merge into two satellites.

HC$_3$N, HNCO and C$_2$H show rather weak emission and were detected 
in 18, 13 and 24 positions, respectively.

According to the previous studies 
\citep[e.g.][]{2006ApJS..166..567R,2008ApJ...680..349J},
CS is a common molecule in IRDCs.
Still, we detected only very weak $^{13}$CS(2-1) emission in three clouds.
The low detection rate of $^{13}$CS(2-1) in our case 
can be explained by 50 times less abundant  $^{13}$CS in comparison with CS.
Also depletion of CS (and $^{13}$CS) can play a significant role here \citep{2009A&A...503..859B}.
Another molecule, CH$_3$C$_2$H, was detected only in IRDC316.76-1 and IRDC317.71-2
and shows very weak emission in IRDC316.72-1, IRDC316.72-2 and IRDC316.76-2.

Even in the most evolved regions IRDC316.76-1 and IRDC316.76-2, 
where strong mid-IR and millimeter emission are present, we did not detect 
CH$_3$CN as a typical hot core tracer.
Since CH$_3$CN emission arises from a warm and compact region, one explanation 
of this non-detection can be the low spatial resolution of the telescope and relatively
low signal to noise ratio in our spectra.
Using the "average spectra" technique  we were able 
to reduce the noise level and to find very weak CH$_3$CN emission (see Fig.~\ref{Fig:average}).
This detection indicates that CH$_3$CN is present at least in some parts of our clouds,
but there is an abundance deficit for this molecule in IRDCs.
This result is in agreement with the previous work by \citet{2007ApJ...668..348B}.

According to \citet{2008AJ....136.2391C} and \citet{2009ApJS..181..360C} it is more likely to 
see SiO, as an outflow tracer, in the clouds, where there is emission both at 4.5 and 24 $\mu$m.
However, we detected SiO lines with extended line wings only in three 
regions with mid-IR  emission: IRDC313.72-1, IRDC313.72-2 and IRDC316.76-1. 
Other potentially interesting regions with emission in mid-IR 
either show very weak SiO emission (IRDC317.71-2, IRDC309.94-1), 
or no SiO emission at all (IRDC309.37-1, IRDC310.39-2). 
The strongest SiO line was detected in the IRDC321.73-1. 
Within the Mopra beam this region contains a very weak source at 24~$\mu$m
and so weak emission at GLIMPSE 4.5~$\mu$m that it was not identified as 
extended green objects (EGOs) in \citet{2008AJ....136.2391C}.
We classify this object as "middle" (see below).
Detection of SiO emission in such source
indicates that absence or extreme weakness of mid-IR emission does not mean that star formation
processes are not taking place in a cloud 
\citep[cf.][]{2007MNRAS.381L..30L} and shows the necessity of spectral line
observations to identify outflows and star formation activity in molecular clouds.


Based on mid-infrared SPITZER data,
\citet{2009ApJS..181..360C} subdivided infrared dark clouds cores
in to "active" and "quiescent". Cores were classified as "active", if they
showed emission both at 8 and 24 $\mu$m, and as "quiescent" if they
contained neither IR emission signatures.
We added a "middle" stage to this classification whenever detect
emission at 24 micron only and not at the shorter wavelength (see Table \ref{table:main}).
In the Fig.~\ref{Fig:spectra} we presented the spectra for typical clouds from every category.
\citet{2009ApJS..181..360C} have shown that "active" cores have smaller sizes, higher densities and
more pronounced water and methanol maser activity than the "quiescent" cores.
However, from the qualitative analysis of the Mopra spectra, 
we do not see clear molecular signatures of any of these three categories of sources.

\begin{figure}
\centering
  \includegraphics[width=6cm,angle=90]{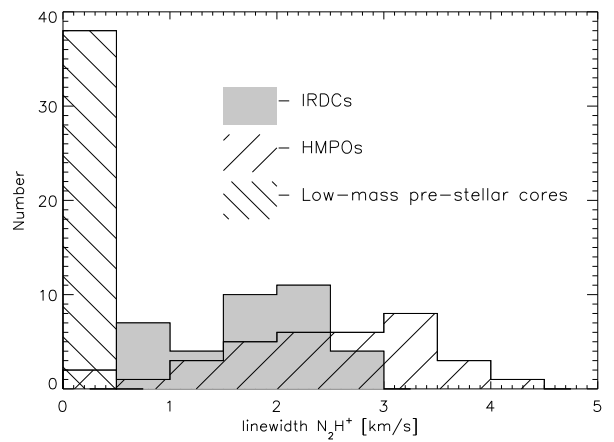}
  \caption{Distribution of N$_2$H$^+$ line width for our IRDCs sample, 
  	   more evolved regions from \citep{2003A&A...405..639P}
	   and low-mass pre-stellar cores from \citet{2001ApJS..136..703L}.
	   }
  \label{Fig:W_N2H}
\end{figure}

\begin{figure}
\centering
  \includegraphics[width=9cm]{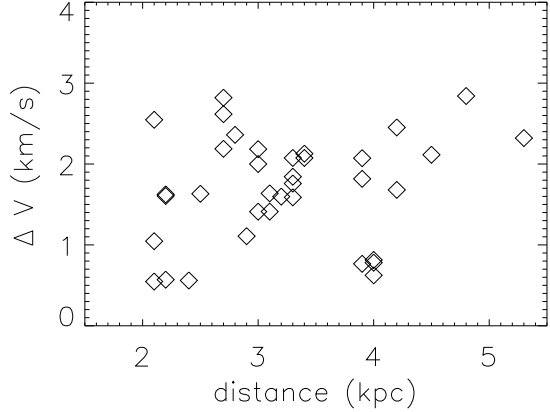}
  \caption{Comparison between N$_2$H$^+$ line width and kinematic distance to the objects for our IRDCs sample. 
	   }
\label{Fig:width_dist}
\end{figure}

\subsection{Line parameters}\label{Sect:lines-parameters}

Using the GILDAS/CLASS software, we fit the whole N$_2$H$^+$ structure, resulting in 
LSR velocities, full linewidths at half maximum,
and optical depths (Tables~\ref{table:N2H}).
For the lines where there is no hyperfine splitting ($^{13}$CS, HC$_3$N, HCO$^+$, HNCO,
SiO, H$^{13}$CO$^+$),
or it cannot be resolved (HNC), we estimate integrated and peak line intensities, 
LSR velocities and full linewidths at half maximum  from gaussian fits (Tables~\ref{table:13CS}-\ref{table:H13CO}).
For all lines we estimated integrated areas, measured by summing
the channels between suitable velocity limits under the line (Table~\ref{table:area}).
To compare our results with the line parameters of other IRDCs, low-mass starless cores and more evolved 
high-mass clouds we chose N$_2$H$^+$. 
This line has a relatively simple line shape - without extended shoulders and self-absorption features, 
and it is detected in 97\% of the clouds.
Moreover, N$_2$H$^+$ is widely detected in low- and high-mass pre-stellar and 
protostellar cores.

We find that the line widths for our IRDCs vary in the range from 0.6 to 2.8~km/s and  are in agreement with the results
obtained for other IRDCs by \citet{2006ApJS..166..567R}, where
the line width was 0.5-4.0~km/s with the mean value 2.2~km/s.
In low-mass pre-stellar cores \citep{2001ApJS..136..703L} strong N$_2$H$^+$ emission 
was detected in 75\% of the objects. Lines in low-mass regions are narrow with line widths in the range of
$\sim$0.2-0.4~km/s, and a mean value of 0.3~km/s, what corresponds to the thermal linewidth.
High-mass protostellar objects show N$_2$H$^+$  emission  in all selected targets
\citep{2003A&A...405..639P,2005A&A...442..949F}.
In comparison to low-mass cores, these targets have broader lines 0.5-3.5 km/s.
Figure~~\ref{Fig:W_N2H} presents the distribution of the N$_2$H$^+$ line width for 
all three types of objects. 

Broader IRDC lines can indicate more turbulent processes within clouds. However, 36$''$ beam,
corresponds to 0.17 pc at 1~kpc distance and 0.9 pc at 5~kpc. Thus,
larger and larger volumes are covered by the same beam for regions further away and hence might
offer a simple explanation for larger line widths.
To check if the line widths become larger for more distant clouds, on the Fig.~\ref{Fig:width_dist}
we plot the line width against the kinematic distance. For this comparison we use our N$_2$H$^+$ data  
from the Table~\ref{table:N2H}.
For our objects the correlation coefficient between the two parameters is only 0.2.
Thus, large line widths of high-mass clouds cannot be explained exclusively by the large distances. In the same vein,
clumping, which can easily be hidden within a 36$''$ beam, can only partly explain 
such elevated line widths \citep{2006A&A...450..569P}.
The linewidth - size relationship also cannot explain high line width values in IRDCs. 
\citet{2001ApJ...551..852H} report that a clear line width - size relation is just
observed in larger clouds complexes of $>$~ 10 pc in size. The IRDC clumps
targeted by us are much smaller with typical sizes of around 1 pc.
Furthermore,
previous studies have already indicated that the line widths in high-mass
star-forming clumps are much higher than would be predicted by the usual
relations between line width and size determined from regions of lower mass
\citep[e.g.][]{2003ApJS..149..375S}.

Comparing line parameters for "quiescent", "middle" and "active" regions,
we can see that there is a trend to have larger line widths and higher integrated intensities 
in more evolved objects. The mean line widths are 1.4, 1.7 and 2.2~km/s  
and the integrated intensities are 4.7, 7.1 and 20.6~K~km/s
for "quiescent", "middle" and "active" sources respectively.
These results are in good agreement with \citet{2009ApJS..181..360C} 
who concluded that "active" sources are more evolved and present further
evolutionary stages in comparison with  "quiescent" sources.

We also utilized N$_2$H$^+$ to estimate kinematic distances
to our objects (see Table \ref{table:main}). 
In case of IRDC309.37-2 the N$_2$H$^+$ line was not detected and we used HCO$^+$ to 
estimate the distance.
Since we observed several points within one cloud, distance determination helps us to 
confirm that clumps with associated mid-IR emission in IRDC309.37, IRDC310.39 and IRDC371.71 are connected
with the other "dark" parts of the respective clouds and are not just a
projection effect.

\begin{figure}
\centering
  \includegraphics[width=8cm]{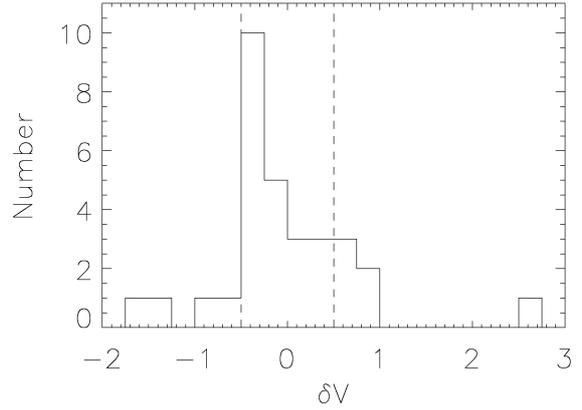}
  \caption{Number distribution of $\delta V =  (V_{HCO^+}-V_{H^{13}CO^+}) / \Delta V_{H^{13}CO^+} $ 
           for our sample (see Sect.~\ref{Sect:infall}). Dashed lines indicate the ~$\pm$5$\sigma_{\delta \rm V}$ level.
	   }
  \label{Fig:deltaV}
\end{figure}

\subsection{HCO$^+$ and H$^{13}$CO$^+$ line profiles}\label{Sect:infall}

As was already mentioned, the HCO$^+$ lines show a non-gaussian profile 
in almost all clouds. In some cases even its isotopoloque H$^{13}$CO$^+$
has a complicated line shape. Moreover, we see a significant shift, up to 2 km/s, between 
the peak positions of the optically thick HCO$^+$ line and optically thin H$^{13}$CO$^+$. 
A quantitative estimate of this asymmetry could provide us with  information about 
dynamical processes in the clouds.

The most common way to extract  line asymmetries is based on the comparison of optically thin 
and optically thick  line positions:
$\delta V =  (V_{thick}-V_{thin}) / \Delta V_{thin} $ \citep{1997ApJ...489..719M}.
This estimate was widely used both for low- and high mass star formation regions.
\citep{2005A&A...442..949F,2003A&A...405..639P, 2006MNRAS.367..553P,2001ApJS..136..703L}.
Using this criterion, we can subdivide our IRDCs in to "blue shifted" clouds with $\delta V < 0 $
and "red shifted" clouds with $\delta V > 0 $. 
The blue excess could be due to infall motions and red excess expanding motions or outflow.
Among the sources, where both HCO$^+$ and H$^{13}$CO$^+$
were detected, we identified 19 "blue shifted" and 12 "red shifted" clouds 
(Fig.~\ref{Fig:deltaV}, Table~\ref{table:H13CO}).
In order to exclude the sources, where differences between the V$_{lsr}$ values for
different species are dominated by measurement errors, we have to determine the threshold value.
For the low-mass starless cores \citet{1997ApJ...489..719M} determined this threshold as
5$\sigma$, where $\sigma$ is the typical $\delta$V error.
In our case  H$^{13}$CO$^+$, which we use as optically thin component, for some sources is quite weak,
which leads to the 5$\sigma$ value equal 0.5.
Most of our clouds have $| \delta V | < 0.5 $,
only 4 objects show a sufficient blue shift, and 6 a red shift.
From these 10 candidates we also have to exclude sources with very weak and/or
noisy spectra.
After this selection we determine one infall candidate: IRDC313.72-4,
and three sources show evidence of expanding motions: IRDC317.71-1, IRDC317.71-2 and IRDC317.71-3.
However, this method allows us only to identify infall or outflow candidates,
but does not provide any quantitative values (e.g. infall speed).

\begin{figure*}
\centering
  \includegraphics[width=5cm,height=6cm,angle=90]{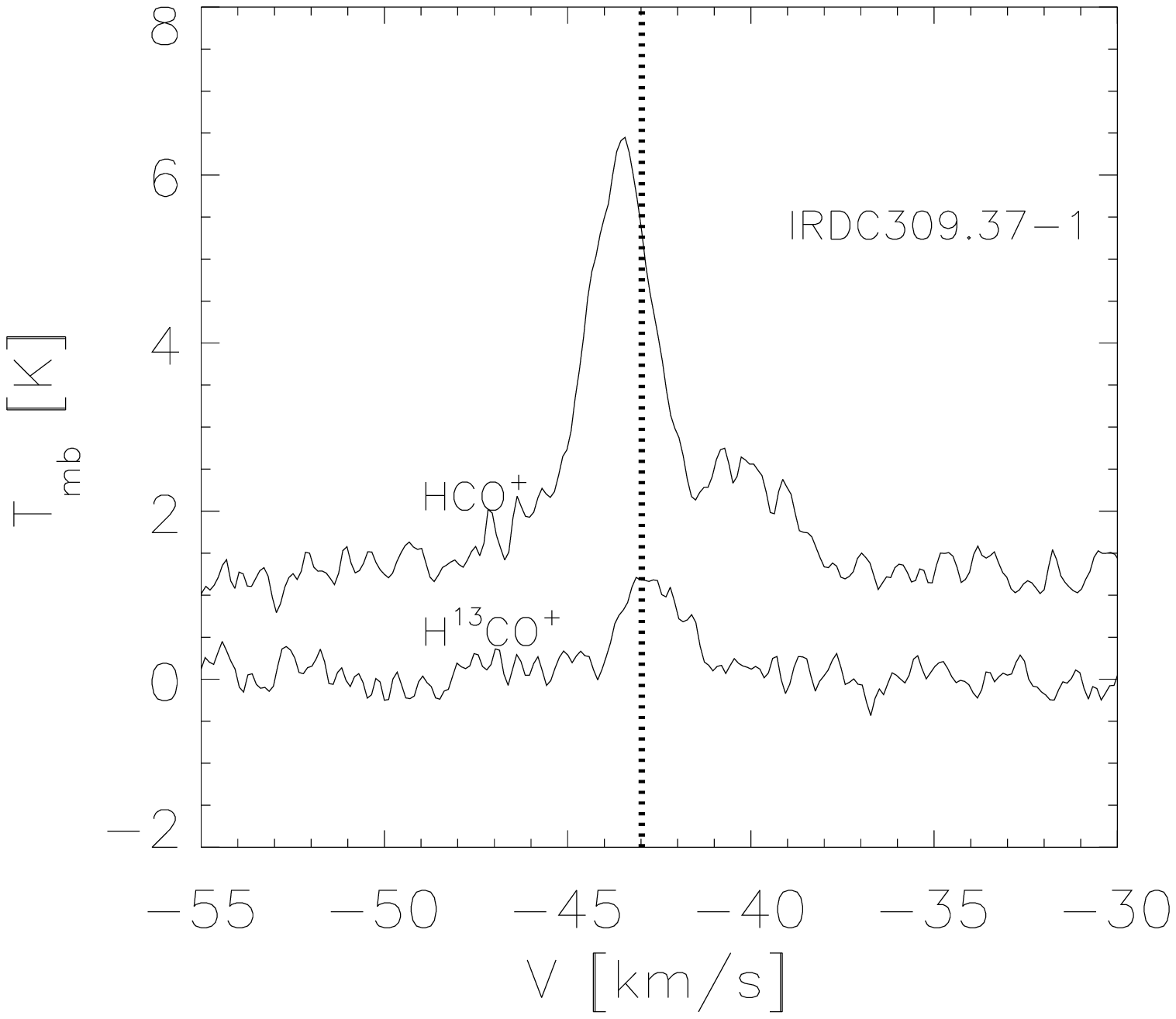}
  \includegraphics[width=5cm,height=6cm,angle=90]{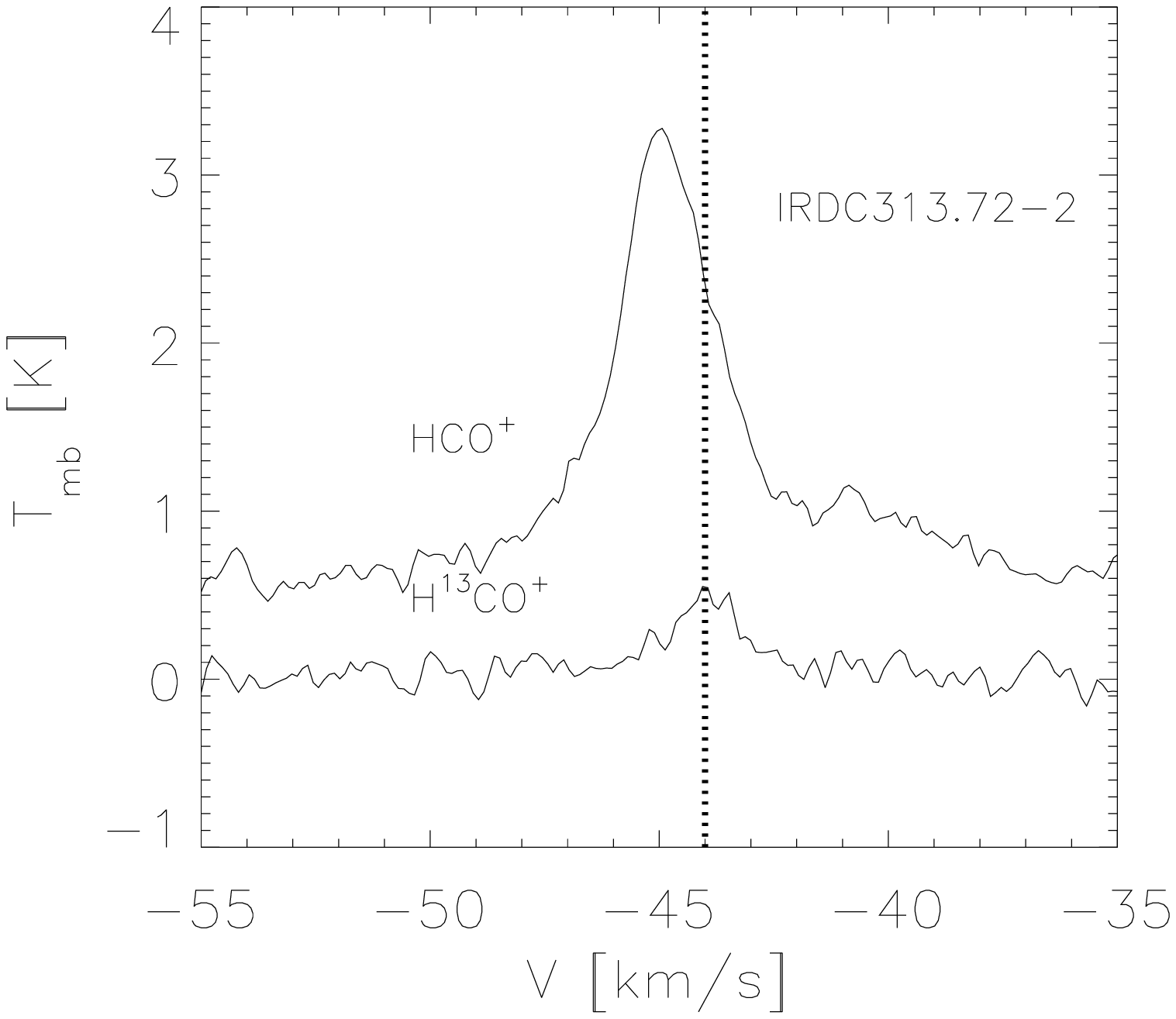}
  \includegraphics[width=5cm,height=6cm,angle=90]{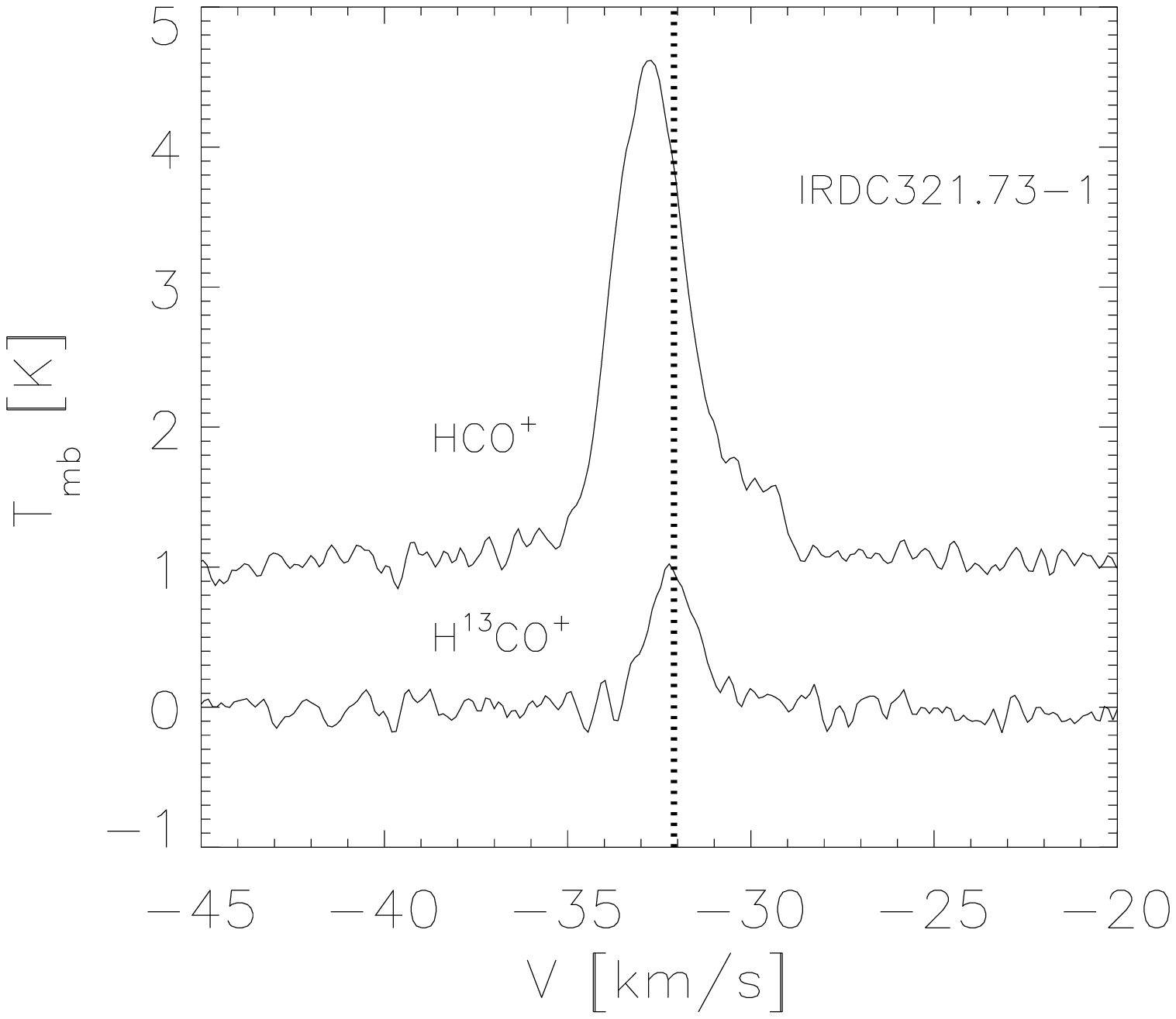}
  \caption{Samples of the HCO$^+$ and H$^{13}$CO$^+$ line profiles. The vertical lines indicate the 
           V$_{lsr}$ positions measured from H$^{13}$CO$^+$ line profiles.
	   }
  \label{Fig:infall}
\end{figure*}

Another method to estimate line asymmetries quantitatively was described
in \citet{1996ApJ...465L.133M}. They used an analytic two layer radiative transfer model 
to obtain theoretical spectra, compare them with the wide range of observed double-peak profiles and 
extract a formula for infall speed (see Eq. 9 from \citet{1996ApJ...465L.133M}).
This method allows us to estimate the infall velocity in a cloud
by measuring blue and red peak parameters of the optically thick line.
Among all our targets we selected 3 clouds with clear red shoulders in HCO$^+$ 
and non-complicated shape of H$^{13}$CO$^+$ (see Fig.~\ref{Fig:infall}). 
The estimated infall velocities are around 2 km/s. 
This value is much higher than the sound 
speed in such an environment - 0.3 km/s, or a typical infall speed for low-mass
pre-stellar cores 0.05-0.09 km/s \citep{2001ApJS..136..703L}.

\begin{figure*}
\centering
  \includegraphics[width=9cm,height=7cm]{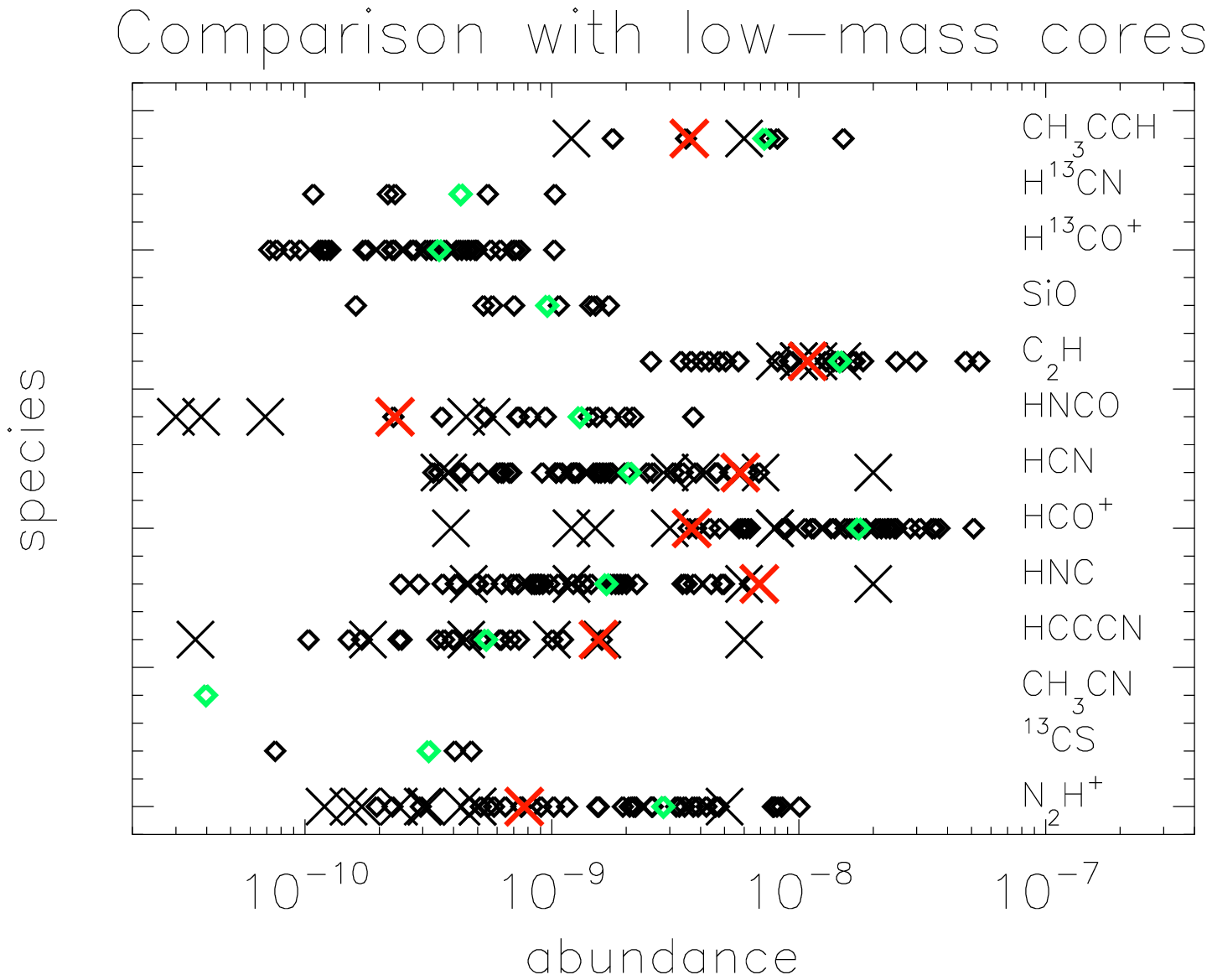}
  \includegraphics[width=9cm,height=7cm]{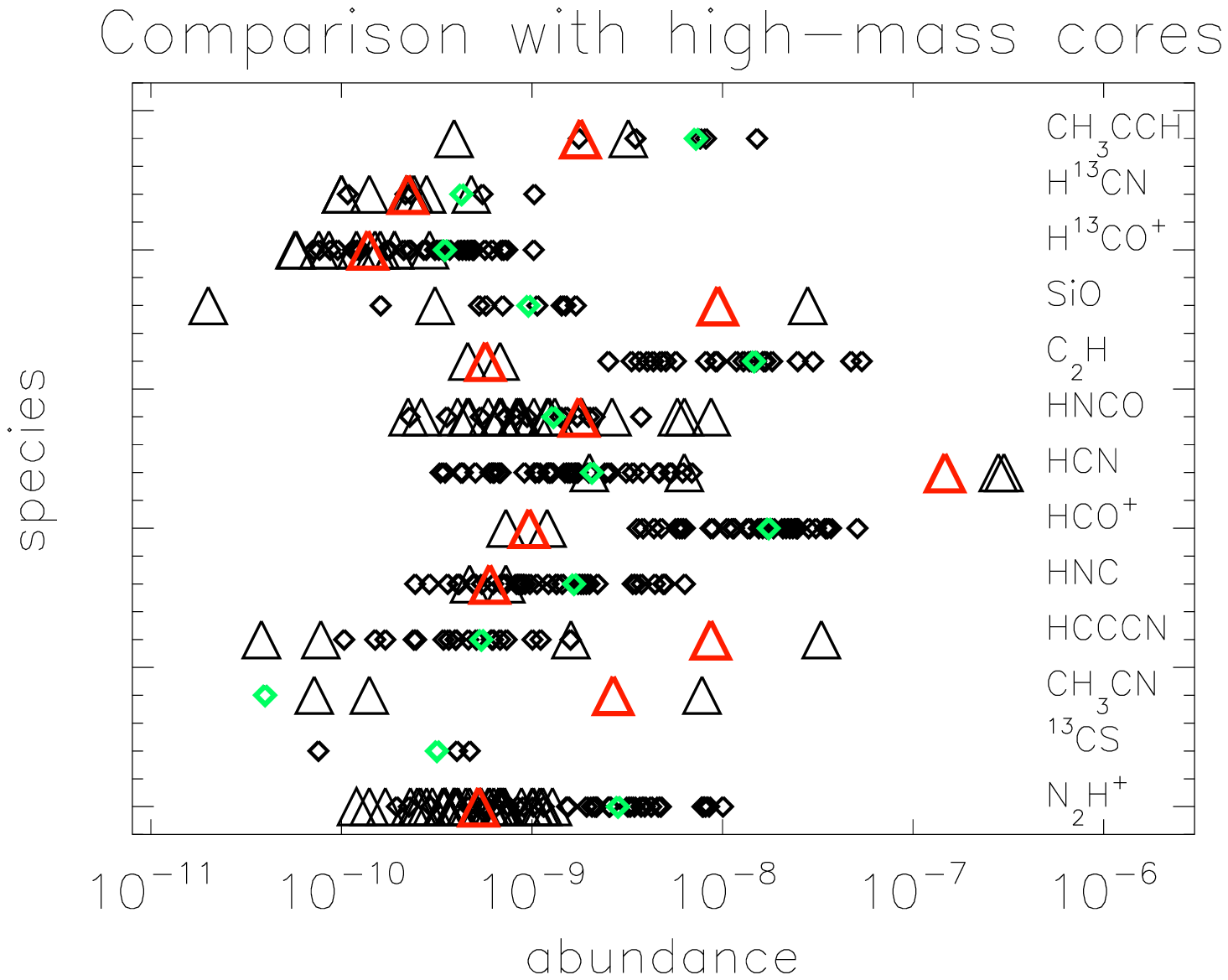}
  \caption{(left) Comparison of the molecular abundances of the IRDCs with low-mass pre-stellar cores
                  from \citet{2009A&A...505.1199P}, \citet{2006A&A...455..577T}, 
                  \citet{2004A&A...416..603J} and \citet{1992IAUS..150..171O}.
  		  Black diamonds indicate our IRDC abundances, green diamond indicates mean values for IRDCs.
		  Black crosses indicate low-mass starless cores, red cross indicates mean value for low-mass starless cores.
           (right) Comparison of the molecular abundances of the IRDCs with HMPOs
	           from \citet{1987ApJ...315..621B}, \citet{1997A&AS..124..205H},
                   \citet{2003A&A...405..639P} and \citet{2000A&A...361.1079Z,2009MNRAS.395.2234Z}.
	   	   Black diamonds indicate our IRDC abundances, green diamond indicate mean values 
		   of the corresponding species for IRDCs.
		   Black triangles indicate HMPOs, red triangles indicate mean values 
		   of the corresponding species for HMPOs.
	   	   On both panels the upper limit abundance value for CH$_3$CN for IRDCs was estimated from the 
	   	   average spectrum.
	   }
  \label{Fig:abundances}
\end{figure*}

\subsection{Derived quantities: column densities and abundances}\label{Sect:column}
 
To estimate column densities we assume LTE conditions and optically thin emission
and use the following equation:

%

\begin{equation}
\begin{array}{lll}
	N_{\rm tot} &=&\frac {8 \pi }{\lambda^3 A}  \frac {g_l}{g_u} \frac {1}{J_\nu (T_{ex})-J_\nu(T_{bg})}  \frac {1}{1-exp(-h \nu /kT_{ex})} \times\\
	&\times &  \frac {Q_{rot}}{g_l exp(-E_l /kT_{ex}) }  \int T_{\rm mb} d \upsilon ,\\
\end{array}
\label{Equ:column}
\end{equation}

\noindent
where $\lambda$ is the rest wavelength of the transition, A is the 
Einstein coefficient,
g$_u$ is the upper state degeneracy, J$_\nu (T_{ex})$ and J$_\nu (T_{bg})$ are
the equivalent Rayleigh-Jeans excitation and background temperatures, Q$_{rot}$ is the partition
function, and E$_l$ is the energy of the lower level \citep{2009AJ....138.1101L}. 
In our calculations for g$_u$, A and E$_l$ we used values from 
The Cologne Database for Molecular Spectroscopy (CDMS)
\citep{2001A&A...370L..49M,2005JMoSt.742..215M}
(Table~\ref{table:lines}).
For the excitation temperatures $T_{ex}$ in all cases,
except HCO$^+$ and H$^{13}$CO$^+$ (see below), we assume ammonia kinetic temperatures
(Linz et al. in prep.) (Table~\ref{table:main}). 
We calculate  the partition function Q$_{rot}$ for every source by interpolating 
data from the CDMS for the particular source temperature $T_{ex}$. 

Integrated intensities $\int T_{\rm mb} d \upsilon$ for every species, were measured by summing the 
channels between suitable velocity limits under the corresponding molecular
line (Table~\ref{table:area}).

The optically thin assumption is suitable for most of our lines: 
$^{13}$CS, HC$_3$N, HNCO, C$_2$H, SiO, H$^{13}$CO$^+$,  H$^{13}$CN, CH$_3$C$_2$H.
In case of optically thick emission (e.g. HCN, HCO$^+$, HNC) Eq.~\ref{Equ:column} gives only
a lower limit. However, 
for N$_2$H$^+$ and HCO$^+$ we can make a better column density estimation. 
In the case of N$_2$H$^+$ we use the advantage of the hyperfine structure and
estimate optical depths from the hyperfine components fitting.
Then column densities for N$_2$H$^+$ can be calculated by multiplying Eq.~\ref{Equ:column} by 
a factor of $\tau/(1-e^{-\tau})$.
For HCO$^+$ the presence of an optically thin isotopoloque allows us to estimate
more realistic values for HCO$^+$. 
Using
 an algorithm described in \citet{2006MNRAS.367..553P} and assuming that HCO$^+$ is optically thick, 
H$^{13}$CO$^+$ is optically thin and their relative abundance ratio $X=[HCO^+]/[H^{13}CO^+]$
equals 50, we calculated HCO$^+$ column densities for the sources, where both 
HCO$^+$ and H$^{13}$CO$^+$ lines were detected. 
We do not apply the same algorithm for the pair HCN-H$^{13}$CN, since H$^{13}$CN
is detected only in four objects.
In spite of non-detection of CH$_3$CN in any single spectra, the "average spectra"
technique let us recognize the very weak CH$_3$CN (5$_0$-4$_0$) and CH$_3$CN (5$_1$-4$_1$) lines
(see Fig~\ref{Fig:average}). 
We use this average spectra to estimate the integrated intensity $\int T_{\rm mb} d \upsilon$ and derive
the upper limit column density for this species.

To transfer column densities to abundances we use the N(species)/N(H$_2$) ratio,
where N(H$_2$) values were estimated from the 1.2 mm SIMBA/SEST data  (Paper I)
adopting the Mopra telescope beam size (see Table ~\ref{table:main}).
In the case of CH$_3$CN 
we use the average value for the molecular hydrogen column density of 1.2$\times  10^{22}$~cm$^{-2}$
and derived upper limit for CH$_3$CN  abundances of 4.0.$\times  10^{-11}$.
Abundances for other species are presented in Table \ref{table:abundances}. 

For our column density and abundance estimates, systematic errors play the dominant role.
One aspect is the assumption that all
lines are optically thin. In case of moderately optically thick lines,
the true column densities might be higher by a factor of 2\,--\,3 for
our objects.
We also utilize kinetic temperatures derived from ammonia observations
as excitation temperatures. This can give another factor of 2-5
to the estimated values.
Systematic errors are also present in the molecular hydrogen column
densities, derived from the millimeter continuum measurements, 
where we made assumptions about the dust model and equal
gas and dust temperatures.
All these assumptions give rise to a systematic error of an order of
magnitude in molecular abundances which we obtained.

\subsection{Are the objects really cold?}\label{Sect:temperatures}

To estimate kinetic temperatures in our clouds we observed the ammonia (1,1) and (2,2) inversion 
transitions with the Parkes radio telescope.
These transitions are known as a good thermometer for cold, dense gas. 
We derived kinetic temperatures for most IRDCs between 10 and 30~K
(Linz et al. in prep.).
Ammonia observations show the presence of cold gas in IRDCs.
However, the Parkes telescope beam is 72$''$  or approximately 
1 pc at the typical IRDCs distances.
With the Mopra telescope (telescope beam 36$''$) we can trace a smaller
volume and use other molecular lines to identify warmer regions.

One way is to consider the HCN/HNC abundance ratio.
This ratio strongly depends
on the temperature and in the case of the Orion molecular cloud
decreases from 80 near the warm core to 5 on the colder edges
\citep{1986ApJ...310..383G,1992A&A...256..595S,2010MNRAS.404..518S}.
For our IRDC sample the estimated HCN/HNC
abundance ratio values are not higher than 3. This confirms,
that we deal with quite cold clouds.

Beside ammonia, symmetric top molecules like CH$_3$C$_2$H can be 
used as probes of kinetic temperature in molecular clouds
\citep{1994ApJ...431..674B}.
We detect CH$_3$C$_2$H in 5 targets. In two objects the
5$_2$-4$_2$, 5$_1$-4$_1$ and 5$_0$-4$_0$ components are 
clearly seen. This allows us to estimate rotational temperatures in
these regions, using the algorithm given in \citet{1994ApJ...431..674B}.
Derived temperatures equal 17~K for IRDC317.71-2 and 38~K for IRDC316.76-1.
The temperature for IRDC317.71-2 is similar to the value we have from ammonia
observations, whereas the temperature for IRDC316.76-1 is 15~K higher.
This indicates that in some clouds a warmer component can be present
in addition to cold gas.
In the case where there is also a warm gas component and the real gas temperature in some
targets is higher, we underestimated molecular abundances. The abundances for 
IRDC316.76-1 with T=38~K become a factor of 1.2-1.7 higher. However,
taking into account all our assumptions and the still
large telescope beam, this difference is still within the error limit.

\subsection{Special objects}\label{Sect:objects}

In this subsection we present infrared dark clouds,
where line spectra show some unique features.

{\bf IRDC317.71}
One of the remarkable features in this object is the  
ratios of HCN hyperfine components, which are different 
from other IRDCs (see Fig~\ref{Fig:spectra}). 
In contrast to N$_2$H$^+$, the HCN hyperfine component intensity ratio is not a constant value.
\citet{1998AJ....115.1111A} show for low-mass dark clouds, that
HCN hyperfine components can have different intensity ratios in different sources.
For most of our IRDCs we detected that the ratios 
F=2-1 to F=0-1  and F=2-1 to F=1-1 components are different but larger than 1.
For IRDC317.71-1 these ratios are 0.8  and 0.4 and for IRDC317.71-2 are 0.5 and 0.2 respectively.

Another interesting feature is the large shift between H$^{13}$CO$^+$
and HCO$^+$ emission peaks.
H$^{13}$CO$^+$ emission peak is shifted to the
blue side in comparison with HCO$^+$ and corresponds to the extinction in the
optically thick component. 
Also IRDC317.71-2 is one of the five locations where CH$_3$C$_2$H
was detected.
As we can see from the previous section, the CH$_3$C$_2$H excitation 
analysis shows the presence of a warmer gas component  
thus indicating a more evolved star forming region.

Most probably the HCN spectrum can be explained by foreground absorption
in low-excitation molecular gas. Similar hyperfine anomalies are
frequently seen in cold dark clouds where they are explained in this way
(absorption in the envelope). The same effect is probably seen in HCO$^+$.
The HCO$^+$ line asymmetry apparently indicates the velocity shift of
the foreground absorption. If attributed to the envelope, this shift
indicates an expansion. In addition, HCO$^+$ perhaps traces a
contribution from an outflow.

IRDC317.71-2 corresponds to the bright emission source in all mid-IR Spitzer/GLIMPSE
bands.
This, together with all spectral features, which we discuss above, 
indicate ongoing star-formation processes
in IRDC317.71-2. Cross correlation of the mid-IR data and near-IR 2MASS K-band images show
that the bright mid-IR source hides a young stellar cluster.

{\bf IRDC321.73}
In IRDC321.73-1 (Fig~\ref{Fig:spectra}), in spite of no emission in GLIMPSE bands and
only a very weak source at 24~$\mu$m, we detected clear SiO emission.
This indicates the presence of a shock and, probably, outflow activity
in this region.
The presence of the red HCO$^+$ shoulder may be an evidence of infall 
motion. The other two points in this cloud (IRDC321.73-2 and IRDC321.73-3) show a
more complicated than average shape of HCO$^+$. The 
rest of the lines, even optically thin like N$_2$H$^+$, also have non-gaussian shapes.
This can be explained by the presence of several kinematic components.

\begin{landscape}
\begin{table}

\caption{Abundances of the IRDCs. a(b) denotes a*10$^b$} 
\label{table:abundances}      
\begin{tabular}{r c c c c c c c c c c c c c}        
\hline\hline             
\noalign{\smallskip}
Name & N$_2$H$^+$ &$^{13}$CS &HC$_3$N& HNC& HCO$^+$& HCN &HNCO& C$_2$H &SiO &H$^{13}$CO$^+$ &
H$^{13}$CN &CH$_3$C$_2$H & category$^a$\\   

\noalign{\smallskip}   
\hline                        
\noalign{\smallskip}
                   
IRDC308.13-1 & 5.5(-10)  &	   &		  & 8.6(-10) &  6.3(-09)& 1.0(-09) &		&	      & 	  & 1.2(-10) &  	&	   &  Q     \\ 
IRDC308.13-2 & 2.9(-10)  &	   &		  & 4.9(-10) &  	& 6.0(-10) &		&  4.3(-09)   & 	  &	     &  	&	   &  M    \\ 
IRDC308.13-3 & 2.2(-10)  &	   &		  & 5.4(-10) &  	& 6.8(-10) &		&	      & 	  &	     &  	&	   &  M      \\ 
IRDC309.13-1 & 1.1(-09)  &	   &		  & 9.7(-10) &  3.7(-08)& 1.5(-09) &		&  5.0(-09)   & 	  & 7.4(-10) &  	&	   &  M      \\
IRDC309.13-2 & 7.5(-10)  &	   &		  & 9.1(-10) &  	& 1.5(-09) &		&	      & 	  &	     &  	&	   &  M 	\\ 
IRDC309.13-3 & 9.0(-10)  &	   &		  & 6.2(-10) &  2.8(-08)& 2.0(-09) &		&	      & 	  & 5.6(-10) &  	&	   &  A        \\ 
IRDC309.37-1 & 8.5(-09)  &	   &	7.3(-10)  & 4.9(-09) &  2.4(-08)& 6.1(-09) &  3.7(-09)  &  4.7(-08)   & 	  & 4.9(-10) &  	&	   &  A   \\ 
IRDC309.37-2 &           &	   &		  & 2.8(-10) &  	& 6.3(-10) &		&	      & 	  &	     &  	&	   &  A \\ 
IRDC309.37-3 & 1.0(-09)  &	   &		  & 1.4(-09) &  1.7(-08)& 1.9(-09) &		&	      & 	  & 3.4(-10) &  	&	   &  Q   \\ 
IRDC310.39-1 & 3.7(-09)  &	   &	6.2(-10)  & 2.0(-09) &  1.3(-08)& 3.3(-09) &  3.5(-10)  &  1.6(-08)   & 1.6(-10)  & 2.7(-10) &  	&	   &  A      \\ 
IRDC310.39-2 & 2.0(-09)  &	   &	3.9(-10)  & 1.3(-09) &  8.9(-09)& 2.4(-09) &		&  1.3(-08)   & 	  & 1.7(-10) &  	&	   &  A      \\ 
IRDC312.36-1 & 7.5(-10)  &	   &	1.0(-10)  & 8.2(-10) &  1.1(-08)& 1.2(-09) &  7.2(-10)  &  4.7(-09)   & 	  & 2.2(-10) &  	&	   &  A 	\\
IRDC312.36-2 & 5.1(-10)  &	   &		  & 1.0(-09) &  5.1(-08)& 1.2(-09) &		&	      & 	  & 1.0(-09) &  	&	   &  Q    \\ 
IRDC313.72-1 & 3.1(-09)  &	   &	1.5(-09)  & 3.7(-09) &  2.4(-08)& 4.6(-09) &  2.1(-09)  &  2.4(-08)   & 1.5(-09)  & 4.8(-10) &  	&	   &  A  \\ 
IRDC313.72-2 & 3.2(-09)  & 4.7(-10)&	1.1(-09)  & 3.5(-09) &  2.3(-08)& 4.6(-09) &  1.7(-09)  &  1.8(-08)   & 1.6(-09)  & 4.6(-10) &  	&	   &  A  \\ 
IRDC313.72-3 & 3.4(-09)  &	   &		  & 4.8(-09) &  2.1(-08)& 5.3(-09) &  1.9(-09)  &  1.5(-08)   & 1.4(-09)  & 4.3(-10) &  	&	   &  Q  \\ 
IRDC313.72-4 & 2.5(-09)  &	   &		  & 3.3(-09) &  1.8(-08)& 3.8(-09) &		&	      & 5.7(-10)  & 3.7(-10) &  	&	   &  Q    \\ 
IRDC316.72-1 & 7.9(-09)  &	   &	3.9(-10)  & 4.4(-09) &  3.0(-08)& 5.2(-09) &  1.4(-09)  &  2.9(-08)   & 	  & 6.1(-10) &5.4(-10)  &8.1(-09)  &  M 	\\ 
IRDC316.76-1 & 1.0(-08)  &4.0(-10) &	1.0(-09)  & 6.3(-09) &  1.1(-08)& 6.8(-09) &  1.5(-09)  &  5.3(-08)   & 7.0(-10)  & 2.2(-10) &1.0(-09)  &1.5(-08)  &  A       \\ 
IRDC316.72-2 & 4.7(-09)  &	   &	2.3(-10)  & 2.2(-09) &  2.2(-08)& 2.5(-09) &		&  1.6(-08)   & 	  & 4.5(-10) &  	&7.6(-09)  &  Q    \\ 
IRDC316.76-2 & 2.1(-09)  &7.5(-11) &  2.4(-10)    & 1.2(-09) &  5.7(-09)& 1.7(-09) &		&  1.6(-08)   & 	  & 1.1(-10) &2.1(-10)  &1.7(-09)  &  A        \\ 
IRDC317.71-1 & 2.8(-09)  &	   &	1.7(-10)  & 8.4(-10) &  1.0(-08)& 5.0(-10) &  8.1(-10)  &  5.6(-09)   & 	  & 2.1(-10) &  	&	   &  Q  \\ 
IRDC317.71-2 & 1.5(-09)  &	   &	3.6(-10)  & 5.0(-10) &  4.7(-09)& 3.3(-10) &  5.3(-10)  &  3.6(-09)   & 	  & 9.5(-11) &1.0(-10)  &3.4(-09)  &  A 	 \\ 
IRDC317.71-3 & 1.5(-09)  &	   &		  & 7.2(-10) &  8.6(-09)& 6.2(-10) &		&	      & 5.2(-10)  & 1.7(-10) &  	&	   &  Q 	    \\
IRDC320.27-1 & 2.1(-09)  &	   &		  & 6.9(-10) &  1.5(-08)& 1.0(-09) &		&	      & 	  & 3.0(-10) &  	&	   &  Q 	    \\ 
IRDC320.27-2 & 4.6(-10)  &	   &		  & 4.1(-10) &  	& 6.7(-10) &		&	      & 	  &	     &  	&	   &  Q 	    \\
IRDC320.27-3 & 1.9(-10)  &	   &		  & 3.6(-10) &  	& 4.3(-10) &		&	      & 	  &	     &  	&	   &  Q 	    \\ 
IRDC321.73-1 & 3.9(-09)  &	   &	5.1(-10)  & 1.8(-09) &  1.3(-08)& 3.1(-09) &		&  1.4(-08)   & 1.0(-09)  & 2.7(-10) &  	&	   &  M 	  \\
IRDC321.73-2 & 1.9(-09)  &	   &	6.1(-10)  & 7.2(-10) &  4.3(-09)& 9.1(-10) &		&  4.0(-09)   & 	  & 8.7(-11) &  	&	   &  M        \\
IRDC321.73-3 & 5.8(-10)  &	   &	4.6(-10)  & 8.9(-10) &  5.9(-09)& 1.0(-09) &		&  3.3(-09)   & 	  & 1.1(-10) &  	&	   &  A        \\
IRDC013.90-1 & 6.4(-10)  &	   &	1.5(-10)  & 4.0(-10) &  3.8(-09)& 4.2(-10) &		&  2.5(-09)   & 	  & 7.6(-11) &  	&	   &  M      \\
IRDC013.90-2 & 8.3(-10)  &	   &		  & 2.4(-10) &  6.1(-09)& 3.4(-10) &		&	      & 	  & 1.2(-10) &  	&	   &  Q      \\ 
IRDC316.45-1 & 8.1(-09)  &	   &	6.8(-10)  & 1.3(-09) &  1.6(-08)& 1.2(-09) &  2.2(-10)  &  8.1(-09)   & 	  & 3.2(-10) &  	&	   &  M 	\\ 
IRDC316.45-2 & 7.8(-09)  &	   &		  & 1.9(-09) &  3.5(-08)& 1.7(-09) &  9.4(-10)  &  1.1(-08)   & 	  & 7.0(-10) &  	&	   &  M      \\ 
IRDC318.15-1 & 4.2(-09)  &	   &		  & 1.7(-09) &  2.0(-08)& 1.6(-09) &		&  1.2(-08)   &-	  & 4.1(-10) &  	&	   &  M  \\ 
IRDC318.15-2 & 4.5(-09)  &	   &		  & 1.8(-09) &  3.4(-08)& 1.4(-09) &		&  9.3(-09)   & 	  & 6.9(-10) &  	&	   &  Q  \\ 
IRDC309.94-1 & 2.5(-09)  &	   &	3.4(-10)  & 8.9(-10) &  3.5(-09)& 1.2(-09) &  7.3(-10)  &  9.1(-09)   & 	  & 7.1(-11) &2.3(-10)  &	   &  A \\ 

\noalign{\smallskip}   
\hline                        
\noalign{\smallskip}

\multicolumn{13}{c}{Average abundances} \\
\noalign{\smallskip}
IRDCs        &   2.8(-09)  &	3.1(-10)   &	 5.4(-10)&  1.6(-09)	&  1.7(-08)&  2.0(-09)   &    1.2(-09)  &      1.4(-08) &    9.5(-10)	 &  3.4(-10)	&   4.2(-10)  &     7.2(-09)\\
low-mass     &   7.7(-10)  &		   &	 1.5(-09)&  6.9(-09)	&  3.7(-09)&  5.8(-09)   &    2.3(-10)  &      1.1(-08) &	    	 &         	&             &     3.6(-09)\\
high-mass    &   5.2(-10)  &		   &	 8.7(-09)&  6.0(-10)	&  9.6(-10)&  1.5(-07)   &    1.8(-09)  &      5.7(-10) &    9.4(-09)	 &  1.4(-10)	&   2.2(-10)  &     1.8(-09)\\

\noalign{\smallskip}   



\hline                                 
\end{tabular}

$^a$    "A" indicates "active" cores, "M" - "middle", "Q" - "quiescent".

\end{table} 
\end{landscape}

\section{Discussion}

\citet{1998ApJ...508..721C} and \citet{2006A&A...450..569P} mention a difference in the 
ammonia and formaldehyde abundances and hence a possible chemical difference between
low-mass pre-stellar cores and IRDCs. 
To study the chemical conditions in IRDCs, \citet{2008ApJ...678.1049S}
observed N$_2$H$^+$(1-0), HC$_3$N(5-4), CCS(4$_3$-3$_2$), NH$_3$(1,1), (2,2), (3,3)
and CH$_3$OH(7-6) lines toward the massive clumps associated with IRDCs.
They estimated the CCS and N$_2$H$^+$ abundance ratio and used low-mass
prestellar core abundances from \citet{1998ApJ...506..743B} for the comparison. 
The result of this analysis showed that IRDCs have lower N(CCS)/N(N$_2$H$^+$) 
ratio values and hence, they can be chemically more evolved than low-mass pre-stellar cores.
However, \citet{2008ApJ...678.1049S} used particular low-mass objects for the comparison 
and did not take into account the pre-stellar cores L1512 and L63, 
where N(CCS)/N(N$_2$H$^+$) ratio is low and is
in agreement with values obtained for their IRDC sample.

Here we want to determine 
if the molecular abundances in IRDCs are similar to the
low-mass pre-stellar cores,
or whether they show signatures of more evolved evolutionary stages.
To answer this question we compare our  IRDC abundances 
with the data available in the literature for low-mass starless clouds
and more evolved high-mass objects.
We analyse the dispersion of the particular abundance values and the difference
between mean abundances for all available species.
As an additional criteria, we compare values of the HNC/HCN abundance ratio for different 
types of objects.

\subsection{Abundance comparison with low-mass starless cores}

To make a comparison with low-mass cold clouds, we used molecular abundances
from \citet{2009A&A...505.1199P}, \citet{2006A&A...455..577T}, 
\citet{2004A&A...416..603J} and \citet{1992IAUS..150..171O}
 (Fig.~\ref{Fig:abundances} left panel).  
Comparing the dispersion of the abundance values we find
that for some species, like HC$_3$N, there 
is a large spread in abundance for the low-mass sample, but higher and 
lower values occur with regard to the IRDC values.
In case of HNCO we compute abundances for low-mass starless cores
using new results by \citet{2009ApJ...690L..27M} and adopting the H$_2$ column densities 
from Table 3 of \citet{2010A&A...516A.105M}.
That gives us quite low abundance values for  B1, L1527-b, and L1544.
However, cores like L183 and TMC-1 reach HNCO abundances which are overlapping with IRDC values
and  just a factor of 2 - 3 lower than the IRDC average abundance.
The same situation, in principle, is found for HCO$^+$, where 
an overlap between the low-mass core and IRDC HCO$^+$ abundances exist for the 
objects L183 and TMC-1 (both attaining $8.0 \times  10^{-9}$ according to \citet{1992IAUS..150..171O}). 
There is a  tendency for higher abundances in IRDCs compared to low-
mass cores  for N$_2$H$^+$. However, also here we found with L1544 a 
prominent example of a low-mass prestellar core with a rather high N$_2$H$^+$ 
abundance of $5.0 \times  10^{-9}$ \citep{2004A&A...416..603J}.

The difference between mean abundances of IRDCs and low-mass
objects is a factor of 2-3 for all species and around 5 in the extreme case of HNCO
(see Table~\ref{table:abundances}). Taking into account
a large spread in abundances for every molecule within our IRDCs sample
and systematic errors (see Sect.~\ref{Sect:column}),
such differences in average abundances between the low-mass
pre-stellar cores and IRDCs cannot be used to claim a clear
and unequivocal chemical distinction between the two groups of objects.

\subsection{Abundance comparison with high-mass protostellar objects}

The next class of objects, which is interesting for the comparison with our results, is 
the class of more evolved high-mass protostellar objects (HMPOs).
HMPOs are bright at mid- and far-IR wavelengths and characterized by  
higher gas temperatures, compared to IRDCs.
It is assumed that these objects present one of the early stages of high-mass star formation, where 
the central protostar has a mass $>$ 8 $\rm M_\odot$ and is still accreting.
\citep[e.g.][]{2002ApJ...566..931S,2007prpl.conf..165B}.
For the comparison we use data from \citet{1987ApJ...315..621B}, \citet{1997A&AS..124..205H},
\citet{2003A&A...405..639P} and \citet{2000A&A...361.1079Z,2009MNRAS.395.2234Z}.
HMPOs and IRDCs have different physical conditions (e.g. temperatures) and, hence, 
are expected to show different chemistry. 

From our comparison we can see, that high-mass  protostellar objects
show higher  mean HC$_3$N and HCN abundances, since some HMPOs have extremely high abundances of
these species.
C$_2$H, on the contrary, has the mean abundance 
a factor of 20 lower in the case of HMPOs.
We did not detect CH$_3$CN emission in any single IRDCs spectra. 
However,
using the "average spectra" technique we reduce the noise level and find very weak CH$_3$CN emission 
and estimate its upper abundance limit.  
This upper limit is lower in comparison with the HMPOs abundances from single-dish
studies by \citep{1997A&AS..124..205H} and \citep{1987ApJ...315..621B} (see Fig. ~\ref{Fig:abundances} right panel).
Previous theoretical \citep{2004A&A...414..409N} and observational \citep{2007ApJ...668..348B} 
studies of high-mass star forming regions showed 
that a low amount of CH$_3$CN  is typical for the earliest evolutionary stages.

The average abundances of HC$_3$N, HCO$^+$, HCN and C$_2$H for 
IRDCs and HMPOs differ by a factor of 16\,--\,75.
These differences are significant,
even when considering all of the included assumptions and
uncertainties.

\subsection{HCN/HNC abundance ratio}

Another criterion that we use to determine the chemical status of IRDCs 
is the HCN/HNC abundance ratio.
For the low-mass prestellar cores from \citet{2006A&A...455..577T}, 
\citet{2004A&A...416..603J} and \citet{1992IAUS..150..171O} this ratio is $\le$ 1.
Values of the HCN/HNC abundance ratio for HMPOs are higher and reach values up to 13
(see \citet{1997A&AS..124..205H} and \citet{1987ApJ...315..621B})
and are around 80 in the extreme regions like Orion 
\citep{1986ApJ...310..383G,1992A&A...256..595S}.
We find the HCN/HNC abundance ratios for our IRDCs $\sim 1$.
This value is in agreement with the
theoretical predictions for the cold clouds \citep{2010MNRAS.404..518S}
and values for the low-mass pre-stellar cores.\\

Considering the comparison between IRDCs, low-mass star-less cores and HMPOs
and assuming that the HCN/HNC 
abundance ratio strongly depends on temperature and is enhanced in active 
star-forming cores \citep{1986ApJ...310..383G,1992A&A...256..595S}, our results 
support the idea that IRDCs present rather low mass star-less core chemistry than 
HMPOs chemistry. 
However, the number of low-mass pre-stellar cores and  HMPOs, which we
used for the analysis, may be not sufficient in a strict
statistical sense.
Therefore, to make more solid statement about the evolutionary status of 
IRDCs and be more confident, we need to extend the line samples, add sulfur bearing species,
and perform further studies, including chemical modeling.

Beside the comparison with low- and high-mass cloud abundances, 
we also compare molecular abundances of "quiescent", "middle" and "active" 
regions within our IRDC sample. 
In accordance with recent results by \citet{2010arXiv1008.0871B},
we do not detect any significant difference among these three categories.

\section{Conclusions}

In this paper, we present  3~mm molecular line observations with 
the 22-m Mopra radio telescope. In total 13 molecular lines were
observed for all IRDCs. The results of our study can be summarized as follows:

\begin{list}{}{}

\item[1.] Using H$_2$ column densities from the previous investigation, we estimate
molecular abundances of all species. 
We show that there is a tendency for the IRDCs to have molecular abundances 
similar to the low-mass pre-stellar core rather than to the
HMPOs abundances.
However, the derived abundances come with uncertainties of around one
order of magnitude. Furthermore, also the comparison abundances for
low--mass cores and HMPOs can be affected by considerable
uncertainties, especially, when abundances have been computed in the
literature by combining heterogeneous data sets (regarding beam sizes
etc.). To make more solid statements about the evolutionary status of
IRDCs therefore calls for subsequent systematical studies.

\item[2.] According to the classification of \citet{2009ApJS..181..360C}, we subdivided our 
clouds to "quiescent" and "active" and added "middle" class to them. We have found a trend
for more evolved regions to have higher line widths and integrated intensities.
However, we do not detect clear evidence of different chemistry in these
three groups.

\item[3.] Comparison of the line width and integrated intensities of the IRDCs and 
low-mass dark clouds show several times higher values for IRDCs.
Broader and more intense lines mean that in IRDCs we have more  turbulent conditions
   in comparison with low-mass clouds.

\item[4.]  We detect the SiO emission in some clouds and complicated shapes of the HCO$^+$ emission
line profile in all IRDCs, which indicates the presence of infall and outflow motions and the
beginning of star formation
activity, at least in some parts of the IRDCs.

\item[5.]  The analysis of the two available CH$_3$C$_2$H excitation diagrams and
detection of the very weak  CH$_3$CN (5$_0$-4$_0$) and CH$_3$CN (5$_1$-4$_1$) lines on the "average
spectra" indicate the presence of a warm gas component in some IRDCs. However, these
warm regions are compact and cannot be resolved with single-dish observations.

\end{list}




\begin{acknowledgements}

The 22-m Mopra antenna is part of the  
Australia Telescope, which is funded by the Commonwealth of
Australia for operations as a National Facility managed by CSIRO.
The University of New South Wales Digital Filter Bank used for the
observations with the Mopra Telescope was provided with support from the
Australian Research Council.
This research has made use of the NASA/ IPAC Infrared Science Archive, which is operated by 
the Jet Propulsion Laboratory, California Institute of Technology, under contract with the 
National Aeronautics and Space Administration.
 NASA's Astrophysics Data System was used to assess
the literature given in the references.

We thank the anonymous referee and Malcolm Walmsley  
for valuable comments and suggestions that help to improve this
work.
We also wish to thank Eric Herbst, Arjan Bik and Sarah Ragan for useful discussions and Sergej Koposov
for the assistance in software installation.

IZ was partially supported by RFBR.
\end{acknowledgements}


\begin{thebibliography}{72}
\expandafter\ifx\csname natexlab\endcsname\relax\def\natexlab#1{#1}\fi

\bibitem[{{Afonso} {et~al.}(1998){Afonso}, {Yun}, \&
  {Clemens}}]{1998AJ....115.1111A}
{Afonso}, J.~M., {Yun}, J.~L., \& {Clemens}, D.~P. 1998, \aj, 115, 1111

\bibitem[{{Araya} {et~al.}(2005){Araya}, {Hofner}, {Kurtz}, {Bronfman}, \&
  {DeDeo}}]{2005ApJS..157..279A}
{Araya}, E., {Hofner}, P., {Kurtz}, S., {Bronfman}, L., \& {DeDeo}, S. 2005,
  \apjs, 157, 279

\bibitem[{{Battersby} {et~al.}(2010){Battersby}, {Bally}, {Jackson},
  {Ginsburg}, {Shirley}, {Schlingman}, \& {Glenn}}]{2010arXiv1008.0871B}
{Battersby}, C., {Bally}, J., {Jackson}, J.~M., {et~al.} 2010, ArXiv e-prints

\bibitem[{{Benjamin} {et~al.}(2003){Benjamin}, {Churchwell}, {Babler}, {Bania},
  {Clemens}, {Cohen}, {Dickey}, {Indebetouw}, {Jackson}, {Kobulnicky},
  {Lazarian}, {Marston}, {Mathis}, {Meade}, {Seager}, {Stolovy}, {Watson},
  {Whitney}, {Wolff}, \& {Wolfire}}]{2003PASP..115..953B}
{Benjamin}, R.~A., {Churchwell}, E., {Babler}, B.~L., {et~al.} 2003, \pasp,
  115, 953

\bibitem[{{Benson} {et~al.}(1998){Benson}, {Caselli}, \&
  {Myers}}]{1998ApJ...506..743B}
{Benson}, P.~J., {Caselli}, P., \& {Myers}, P.~C. 1998, \apj, 506, 743

\bibitem[{{Bergin} {et~al.}(1994){Bergin}, {Goldsmith}, {Snell}, \&
  {Ungerechts}}]{1994ApJ...431..674B}
{Bergin}, E.~A., {Goldsmith}, P.~F., {Snell}, R.~L., \& {Ungerechts}, H. 1994,
  \apj, 431, 674

\bibitem[{{Bergin} {et~al.}(1996){Bergin}, {Snell}, \&
  {Goldsmith}}]{1996ApJ...460..343B}
{Bergin}, E.~A., {Snell}, R.~L., \& {Goldsmith}, P.~F. 1996, \apj, 460, 343

\bibitem[{{Beuther} {et~al.}(2007){Beuther}, {Churchwell}, {McKee}, \&
  {Tan}}]{2007prpl.conf..165B}
{Beuther}, H., {Churchwell}, E.~B., {McKee}, C.~F., \& {Tan}, J.~C. 2007,
  Protostars and Planets V, 165

\bibitem[{{Beuther} \& {Henning}(2009)}]{2009A&A...503..859B}
{Beuther}, H. \& {Henning}, T. 2009, \aap, 503, 859

\bibitem[{{Beuther} {et~al.}(2008){Beuther}, {Semenov}, {Henning}, \&
  {Linz}}]{2008ApJ...675L..33B}
{Beuther}, H., {Semenov}, D., {Henning}, T., \& {Linz}, H. 2008, \apjl, 675,
  L33

\bibitem[{{Beuther} \& {Sridharan}(2007)}]{2007ApJ...668..348B}
{Beuther}, H. \& {Sridharan}, T.~K. 2007, \apj, 668, 348

\bibitem[{{Blake} {et~al.}(1987){Blake}, {Sutton}, {Masson}, \&
  {Phillips}}]{1987ApJ...315..621B}
{Blake}, G.~A., {Sutton}, E.~C., {Masson}, C.~R., \& {Phillips}, T.~G. 1987,
  \apj, 315, 621

\bibitem[{{Bronfman} {et~al.}(1996){Bronfman}, {Nyman}, \&
  {May}}]{1996A&AS..115...81B}
{Bronfman}, L., {Nyman}, L., \& {May}, J. 1996, \aaps, 115, 81

\bibitem[{{Carey} {et~al.}(1998){Carey}, {Clark}, {Egan}, {Price}, {Shipman},
  \& {Kuchar}}]{1998ApJ...508..721C}
{Carey}, S.~J., {Clark}, F.~O., {Egan}, M.~P., {et~al.} 1998, \apj, 508, 721

\bibitem[{{Caselli} {et~al.}(2002){Caselli}, {Benson}, {Myers}, \&
  {Tafalla}}]{2002ApJ...572..238C}
{Caselli}, P., {Benson}, P.~J., {Myers}, P.~C., \& {Tafalla}, M. 2002, \apj,
  572, 238

\bibitem[{{Chambers} {et~al.}(2009){Chambers}, {Jackson}, {Rathborne}, \&
  {Simon}}]{2009ApJS..181..360C}
{Chambers}, E.~T., {Jackson}, J.~M., {Rathborne}, J.~M., \& {Simon}, R. 2009,
  \apjs, 181, 360

\bibitem[{{Codella} {et~al.}(2001){Codella}, {Bachiller}, {Nisini}, {Saraceno},
  \& {Testi}}]{2001A&A...376..271C}
{Codella}, C., {Bachiller}, R., {Nisini}, B., {Saraceno}, P., \& {Testi}, L.
  2001, \aap, 376, 271

\bibitem[{{Cyganowski} {et~al.}(2008){Cyganowski}, {Whitney}, {Holden},
  {Braden}, {Brogan}, {Churchwell}, {Indebetouw}, {Watson}, {Babler},
  {Benjamin}, {Gomez}, {Meade}, {Povich}, {Robitaille}, \&
  {Watson}}]{2008AJ....136.2391C}
{Cyganowski}, C.~J., {Whitney}, B.~A., {Holden}, E., {et~al.} 2008, \aj, 136,
  2391

\bibitem[{{De Buizer} \& {Vacca}(2010)}]{2010AJ....140..196D}
{De Buizer}, J.~M. \& {Vacca}, W.~D. 2010, \aj, 140, 196

\bibitem[{{Egan} {et~al.}(1998){Egan}, {Shipman}, {Price}, {Carey}, {Clark}, \&
  {Cohen}}]{1998ApJ...494L.199E}
{Egan}, M.~P., {Shipman}, R.~F., {Price}, S.~D., {et~al.} 1998, \apjl, 494,
  L199+

\bibitem[{{Fuller} {et~al.}(2005){Fuller}, {Williams}, \&
  {Sridharan}}]{2005A&A...442..949F}
{Fuller}, G.~A., {Williams}, S.~J., \& {Sridharan}, T.~K. 2005, \aap, 442, 949

\bibitem[{{Gibson} {et~al.}(2009){Gibson}, {Plume}, {Bergin}, {Ragan}, \&
  {Evans}}]{2009ApJ...705..123G}
{Gibson}, D., {Plume}, R., {Bergin}, E., {Ragan}, S., \& {Evans}, N. 2009,
  \apj, 705, 123

\bibitem[{{Goldsmith} {et~al.}(1986){Goldsmith}, {Irvine}, {Hjalmarson}, \&
  {Ellder}}]{1986ApJ...310..383G}
{Goldsmith}, P.~F., {Irvine}, W.~M., {Hjalmarson}, A., \& {Ellder}, J. 1986,
  \apj, 310, 383

\bibitem[{{Helmich} \& {van Dishoeck}(1997)}]{1997A&AS..124..205H}
{Helmich}, F.~P. \& {van Dishoeck}, E.~F. 1997, \aaps, 124, 205

\bibitem[{{Heyer} {et~al.}(2001){Heyer}, {Carpenter}, \&
  {Snell}}]{2001ApJ...551..852H}
{Heyer}, M.~H., {Carpenter}, J.~M., \& {Snell}, R.~L. 2001, \apj, 551, 852

\bibitem[{{Hofner} {et~al.}(2001){Hofner}, {Wiesemeyer}, \&
  {Henning}}]{2001ApJ...549..425H}
{Hofner}, P., {Wiesemeyer}, H., \& {Henning}, T. 2001, \apj, 549, 425

\bibitem[{{Jackson} {et~al.}(2008){Jackson}, {Finn}, {Rathborne}, {Chambers},
  \& {Simon}}]{2008ApJ...680..349J}
{Jackson}, J.~M., {Finn}, S.~C., {Rathborne}, J.~M., {Chambers}, E.~T., \&
  {Simon}, R. 2008, \apj, 680, 349

\bibitem[{{Jones} {et~al.}(2008){Jones}, {Burton}, {Cunningham}, {Menten},
  {Schilke}, {Belloche}, {Leurini}, {Ott}, \& {Walsh}}]{2008MNRAS.386..117J}
{Jones}, P.~A., {Burton}, M.~G., {Cunningham}, M.~R., {et~al.} 2008, \mnras,
  386, 117

\bibitem[{{J{\o}rgensen} {et~al.}(2004){J{\o}rgensen}, {Sch{\"o}ier}, \& {van
  Dishoeck}}]{2004A&A...416..603J}
{J{\o}rgensen}, J.~K., {Sch{\"o}ier}, F.~L., \& {van Dishoeck}, E.~F. 2004,
  \aap, 416, 603

\bibitem[{{Kalenskii} {et~al.}(2000){Kalenskii}, {Promislov}, {Alakoz},
  {Winnberg}, \& {Johansson}}]{2000A&A...354.1036K}
{Kalenskii}, S.~V., {Promislov}, V.~G., {Alakoz}, A., {Winnberg}, A.~V., \&
  {Johansson}, L.~E.~B. 2000, \aap, 354, 1036

\bibitem[{{Krumholz} \& {McKee}(2008)}]{2008Natur.451.1082K}
{Krumholz}, M.~R. \& {McKee}, C.~F. 2008, \nat, 451, 1082

\bibitem[{{Ladd} {et~al.}(2005){Ladd}, {Purcell}, {Wong}, \&
  {Robertson}}]{2005PASA...22...62L}
{Ladd}, N., {Purcell}, C., {Wong}, T., \& {Robertson}, S. 2005, Publications of
  the Astronomical Society of Australia, 22, 62

\bibitem[{{Lee} {et~al.}(2001){Lee}, {Myers}, \&
  {Tafalla}}]{2001ApJS..136..703L}
{Lee}, C.~W., {Myers}, P.~C., \& {Tafalla}, M. 2001, \apjs, 136, 703

\bibitem[{{Lee} {et~al.}(2009){Lee}, {Stanimirovi{\'c}}, {Ott}, {van Loon},
  {Bolatto}, {Jones}, {Cunningham}, {Devine}, \&
  {Oliveira}}]{2009AJ....138.1101L}
{Lee}, M., {Stanimirovi{\'c}}, S., {Ott}, J., {et~al.} 2009, \aj, 138, 1101

\bibitem[{{Lo} {et~al.}(2007){Lo}, {Cunningham}, {Bains}, {Burton}, \&
  {Garay}}]{2007MNRAS.381L..30L}
{Lo}, N., {Cunningham}, M., {Bains}, I., {Burton}, M.~G., \& {Garay}, G. 2007,
  \mnras, 381, L30

\bibitem[{{Marcelino} {et~al.}(2010){Marcelino}, {Br{\"u}nken}, {Cernicharo},
  {Quan}, {Roueff}, {Herbst}, \& {Thaddeus}}]{2010A&A...516A.105M}
{Marcelino}, N., {Br{\"u}nken}, S., {Cernicharo}, J., {et~al.} 2010, \aap, 516,
  A105+

\bibitem[{{Marcelino} {et~al.}(2009){Marcelino}, {Cernicharo}, {Tercero}, \&
  {Roueff}}]{2009ApJ...690L..27M}
{Marcelino}, N., {Cernicharo}, J., {Tercero}, B., \& {Roueff}, E. 2009, \apjl,
  690, L27

\bibitem[{{Mardones} {et~al.}(1997){Mardones}, {Myers}, {Tafalla}, {Wilner},
  {Bachiller}, \& {Garay}}]{1997ApJ...489..719M}
{Mardones}, D., {Myers}, P.~C., {Tafalla}, M., {et~al.} 1997, \apj, 489, 719

\bibitem[{{M{\"u}ller} {et~al.}(2005){M{\"u}ller}, {Schl{\"o}der}, {Stutzki},
  \& {Winnewisser}}]{2005JMoSt.742..215M}
{M{\"u}ller}, H.~S.~P., {Schl{\"o}der}, F., {Stutzki}, J., \& {Winnewisser}, G.
  2005, Journal of Molecular Structure, 742, 215

\bibitem[{{M{\"u}ller} {et~al.}(2001){M{\"u}ller}, {Thorwirth}, {Roth}, \&
  {Winnewisser}}]{2001A&A...370L..49M}
{M{\"u}ller}, H.~S.~P., {Thorwirth}, S., {Roth}, D.~A., \& {Winnewisser}, G.
  2001, \aap, 370, L49

\bibitem[{{Myers} {et~al.}(1996){Myers}, {Mardones}, {Tafalla}, {Williams}, \&
  {Wilner}}]{1996ApJ...465L.133M}
{Myers}, P.~C., {Mardones}, D., {Tafalla}, M., {Williams}, J.~P., \& {Wilner},
  D.~J. 1996, \apjl, 465, L133+

\bibitem[{{Nomura} \& {Millar}(2004)}]{2004A&A...414..409N}
{Nomura}, H. \& {Millar}, T.~J. 2004, \aap, 414, 409

\bibitem[{{Ohishi} {et~al.}(1992){Ohishi}, {Irvine}, \&
  {Kaifu}}]{1992IAUS..150..171O}
{Ohishi}, M., {Irvine}, W.~M., \& {Kaifu}, N. 1992, in IAU Symposium, Vol. 150,
  Astrochemistry of Cosmic Phenomena, ed. {P.~D.~Singh}, 171--+

\bibitem[{{Padovani} {et~al.}(2009){Padovani}, {Walmsley}, {Tafalla}, {Galli},
  \& {M{\"u}ller}}]{2009A&A...505.1199P}
{Padovani}, M., {Walmsley}, C.~M., {Tafalla}, M., {Galli}, D., \& {M{\"u}ller},
  H.~S.~P. 2009, \aap, 505, 1199

\bibitem[{{Perault} {et~al.}(1996){Perault}, {Omont}, {Simon}, {Seguin},
  {Ojha}, {Blommaert}, {Felli}, {Gilmore}, {Guglielmo}, {Habing}, {Price},
  {Robin}, {de Batz}, {Cesarsky}, {Elbaz}, {Epchtein}, {Fouque}, {Guest},
  {Levine}, {Pollock}, {Prusti}, {Siebenmorgen}, {Testi}, \&
  {Tiphene}}]{1996A&A...315L.165P}
{Perault}, M., {Omont}, A., {Simon}, G., {et~al.} 1996, \aap, 315, L165

\bibitem[{{Pillai} {et~al.}(2006){Pillai}, {Wyrowski}, {Carey}, \&
  {Menten}}]{2006A&A...450..569P}
{Pillai}, T., {Wyrowski}, F., {Carey}, S.~J., \& {Menten}, K.~M. 2006, \aap,
  450, 569

\bibitem[{{Pirogov} {et~al.}(2003){Pirogov}, {Zinchenko}, {Caselli},
  {Johansson}, \& {Myers}}]{2003A&A...405..639P}
{Pirogov}, L., {Zinchenko}, I., {Caselli}, P., {Johansson}, L.~E.~B., \&
  {Myers}, P.~C. 2003, \aap, 405, 639

\bibitem[{{Purcell} {et~al.}(2006){Purcell}, {Balasubramanyam}, {Burton},
  {Walsh}, {Minier}, {Hunt-Cunningham}, {Kedziora-Chudczer}, {Longmore},
  {Hill}, {Bains}, {Barnes}, {Busfield}, {Calisse}, {Crighton}, {Curran},
  {Davis}, {Dempsey}, {Derragopian}, {Fulton}, {Hidas}, {Hoare}, {Lee}, {Ladd},
  {Lumsden}, {Moore}, {Murphy}, {Oudmaijer}, {Pracy}, {Rathborne}, {Robertson},
  {Schultz}, {Shobbrook}, {Sparks}, {Storey}, \&
  {Travouillion}}]{2006MNRAS.367..553P}
{Purcell}, C.~R., {Balasubramanyam}, R., {Burton}, M.~G., {et~al.} 2006,
  \mnras, 367, 553

\bibitem[{{Ragan} {et~al.}(2009){Ragan}, {Bergin}, \&
  {Gutermuth}}]{2009ApJ...698..324R}
{Ragan}, S.~E., {Bergin}, E.~A., \& {Gutermuth}, R.~A. 2009, \apj, 698, 324

\bibitem[{{Ragan} {et~al.}(2006){Ragan}, {Bergin}, {Plume}, {Gibson}, {Wilner},
  {O'Brien}, \& {Hails}}]{2006ApJS..166..567R}
{Ragan}, S.~E., {Bergin}, E.~A., {Plume}, R., {et~al.} 2006, \apjs, 166, 567

\bibitem[{{Rathborne} {et~al.}(2006){Rathborne}, {Jackson}, \&
  {Simon}}]{2006ApJ...641..389R}
{Rathborne}, J.~M., {Jackson}, J.~M., \& {Simon}, R. 2006, \apj, 641, 389

\bibitem[{{Rawlings} {et~al.}(2004){Rawlings}, {Redman}, {Keto}, \&
  {Williams}}]{2004MNRAS.351.1054R}
{Rawlings}, J.~M.~C., {Redman}, M.~P., {Keto}, E., \& {Williams}, D.~A. 2004,
  \mnras, 351, 1054

\bibitem[{{Rawlings} {et~al.}(2000){Rawlings}, {Taylor}, \&
  {Williams}}]{2000MNRAS.313..461R}
{Rawlings}, J.~M.~C., {Taylor}, S.~D., \& {Williams}, D.~A. 2000, \mnras, 313,
  461

\bibitem[{{Redman} {et~al.}(2008){Redman}, {Khanzadyan}, {Loughnane}, \&
  {Carolan}}]{2008ASPC..387...38R}
{Redman}, M.~P., {Khanzadyan}, T., {Loughnane}, R.~M., \& {Carolan}, P.~B.
  2008, in Astronomical Society of the Pacific Conference Series, Vol. 387,
  Massive Star Formation: Observations Confront Theory, ed. {H.~Beuther,
  H.~Linz, \& T.~Henning}, 38--+

\bibitem[{{Saito} {et~al.}(2001){Saito}, {Mizuno}, {Moriguchi}, {Matsunaga},
  {Onishi}, {Mizuno}, \& {Fukui}}]{2001PASJ...53.1037S}
{Saito}, H., {Mizuno}, N., {Moriguchi}, Y., {et~al.} 2001, \pasj, 53, 1037

\bibitem[{{Sakai} {et~al.}(2010){Sakai}, {Sakai}, {Hirota}, \&
  {Yamamoto}}]{2010ApJ...714.1658S}
{Sakai}, T., {Sakai}, N., {Hirota}, T., \& {Yamamoto}, S. 2010, \apj, 714, 1658

\bibitem[{{Sakai} {et~al.}(2008){Sakai}, {Sakai}, {Kamegai}, {Hirota},
  {Yamaguchi}, {Shiba}, \& {Yamamoto}}]{2008ApJ...678.1049S}
{Sakai}, T., {Sakai}, N., {Kamegai}, K., {et~al.} 2008, \apj, 678, 1049

\bibitem[{{Sarrasin} {et~al.}(2010){Sarrasin}, {Abdallah}, {Wernli}, {Faure},
  {Cernicharo}, \& {Lique}}]{2010MNRAS.404..518S}
{Sarrasin}, E., {Abdallah}, D.~B., {Wernli}, M., {et~al.} 2010, \mnras, 404,
  518

\bibitem[{{Schilke} {et~al.}(1992){Schilke}, {Walmsley}, {Pineau Des Forets},
  {Roueff}, {Flower}, \& {Guilloteau}}]{1992A&A...256..595S}
{Schilke}, P., {Walmsley}, C.~M., {Pineau Des Forets}, G., {et~al.} 1992, \aap,
  256, 595

\bibitem[{{Shirley} {et~al.}(2003){Shirley}, {Evans}, {Young}, {Knez}, \&
  {Jaffe}}]{2003ApJS..149..375S}
{Shirley}, Y.~L., {Evans}, II, N.~J., {Young}, K.~E., {Knez}, C., \& {Jaffe},
  D.~T. 2003, \apjs, 149, 375

\bibitem[{{Simon} {et~al.}(2006){Simon}, {Jackson}, {Rathborne}, \&
  {Chambers}}]{2006ApJ...639..227S}
{Simon}, R., {Jackson}, J.~M., {Rathborne}, J.~M., \& {Chambers}, E.~T. 2006,
  \apj, 639, 227

\bibitem[{{Sridharan} {et~al.}(2005){Sridharan}, {Beuther}, {Saito},
  {Wyrowski}, \& {Schilke}}]{2005ApJ...634L..57S}
{Sridharan}, T.~K., {Beuther}, H., {Saito}, M., {Wyrowski}, F., \& {Schilke},
  P. 2005, \apjl, 634, L57

\bibitem[{{Sridharan} {et~al.}(2002){Sridharan}, {Beuther}, {Schilke},
  {Menten}, \& {Wyrowski}}]{2002ApJ...566..931S}
{Sridharan}, T.~K., {Beuther}, H., {Schilke}, P., {Menten}, K.~M., \&
  {Wyrowski}, F. 2002, \apj, 566, 931

\bibitem[{{Tafalla} {et~al.}(2002){Tafalla}, {Myers}, {Caselli}, {Walmsley}, \&
  {Comito}}]{2002ApJ...569..815T}
{Tafalla}, M., {Myers}, P.~C., {Caselli}, P., {Walmsley}, C.~M., \& {Comito},
  C. 2002, \apj, 569, 815

\bibitem[{{Tafalla} {et~al.}(2006){Tafalla}, {Santiago-Garc{\'{\i}}a}, {Myers},
  {Caselli}, {Walmsley}, \& {Crapsi}}]{2006A&A...455..577T}
{Tafalla}, M., {Santiago-Garc{\'{\i}}a}, J., {Myers}, P.~C., {et~al.} 2006,
  \aap, 455, 577

\bibitem[{{Vasyunina} {et~al.}(2009){Vasyunina}, {Linz}, {Henning}, {Stecklum},
  {Klose}, \& {Nyman}}]{2009A&A...499..149V}
{Vasyunina}, T., {Linz}, H., {Henning}, T., {et~al.} 2009, \aap, 499, 149
  (Paper I)

\bibitem[{{Viti}(2005)}]{2005IAUS..231...67V}
{Viti}, S. 2005, in IAU Symposium, Vol. 231, Astrochemistry: Recent Successes
  and Current Challenges, ed. {D.~C.~Lis, G.~A.~Blake, \& E.~Herbst}, 67--76

\bibitem[{{Walsh} \& {Burton}(2006)}]{2006MNRAS.365..321W}
{Walsh}, A.~J. \& {Burton}, M.~G. 2006, \mnras, 365, 321

\bibitem[{{Zhang} {et~al.}(2007){Zhang}, {Sridharan}, {Hunter}, {Chen},
  {Beuther}, \& {Wyrowski}}]{2007A&A...470..269Z}
{Zhang}, Q., {Sridharan}, T.~K., {Hunter}, T.~R., {et~al.} 2007, \aap, 470, 269

\bibitem[{{Zinchenko} {et~al.}(2009){Zinchenko}, {Caselli}, \&
  {Pirogov}}]{2009MNRAS.395.2234Z}
{Zinchenko}, I., {Caselli}, P., \& {Pirogov}, L. 2009, \mnras, 395, 2234

\bibitem[{{Zinchenko} {et~al.}(2000){Zinchenko}, {Henkel}, \&
  {Mao}}]{2000A&A...361.1079Z}
{Zinchenko}, I., {Henkel}, C., \& {Mao}, R.~Q. 2000, \aap, 361, 1079

\bibitem[{{Zinnecker} \& {Yorke}(2007)}]{2007ARA&A..45..481Z}
{Zinnecker}, H. \& {Yorke}, H.~W. 2007, \araa, 45, 481

\end{thebibliography}

\Online


\begin{appendix}
\section{3-color Spitzer/Glimpse images.}

   \begin{figure*}
   \centering
     \includegraphics[width=9cm]{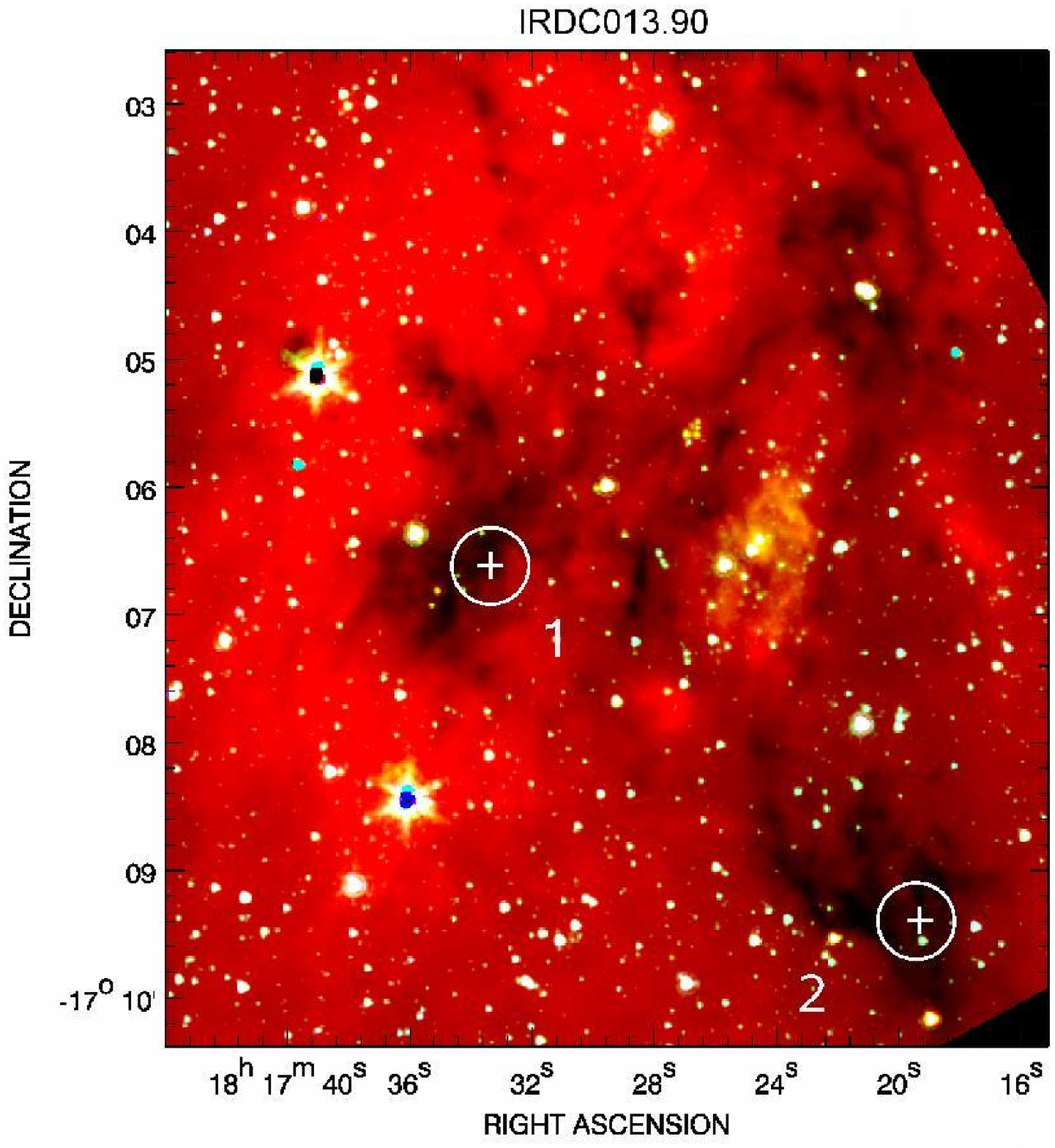}  
     \includegraphics[width=9cm]{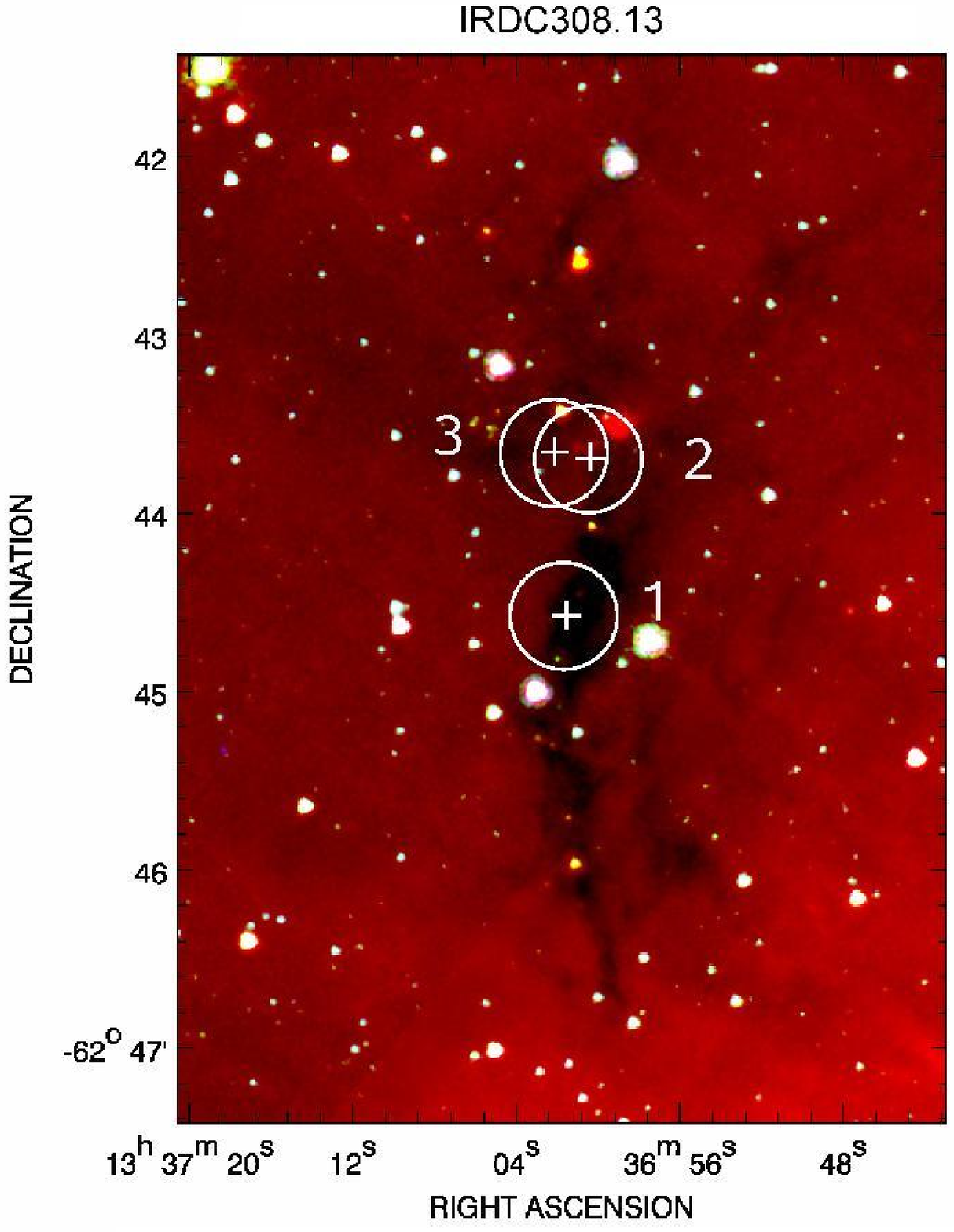}
     \includegraphics[width=9cm]{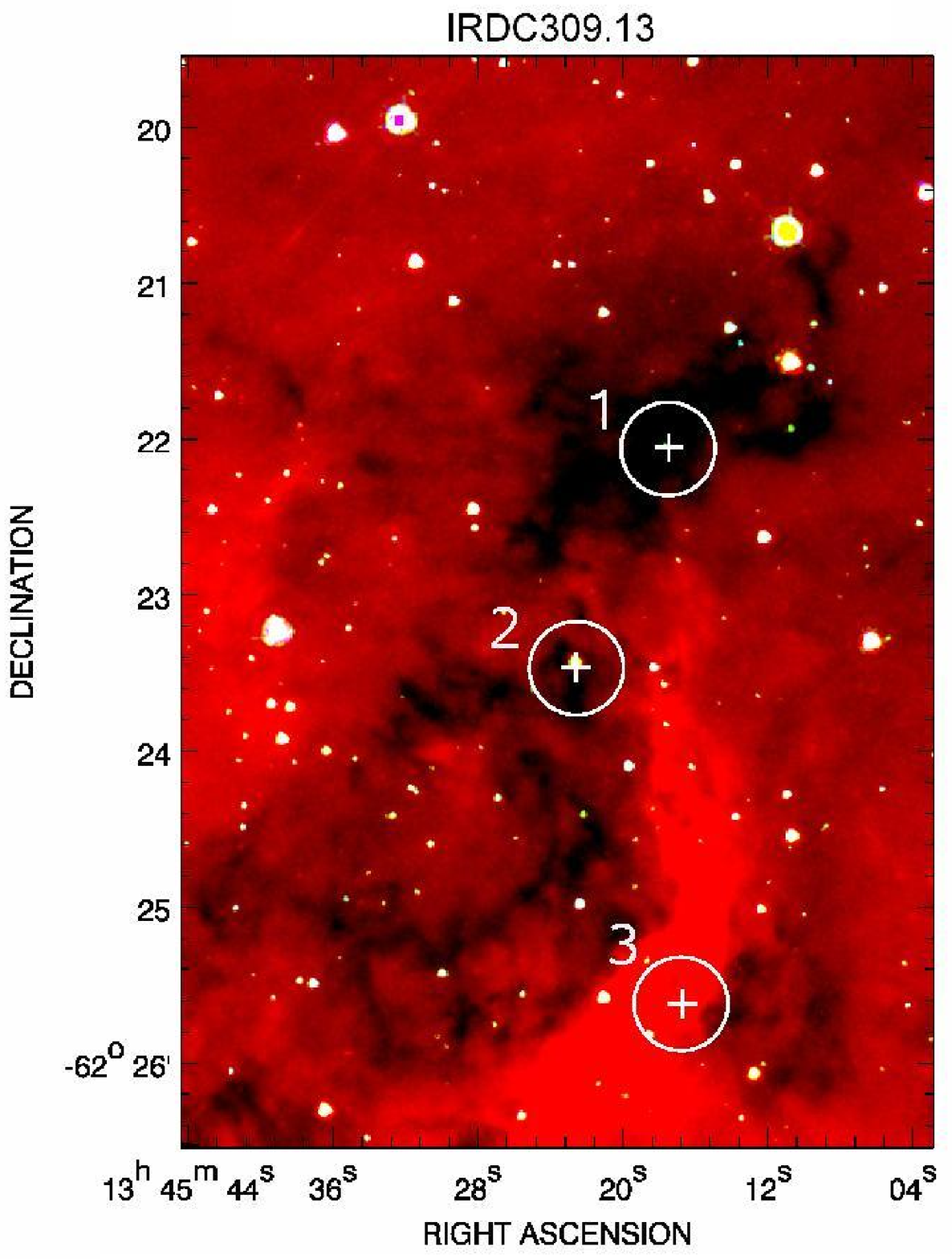}
     \includegraphics[width=9cm]{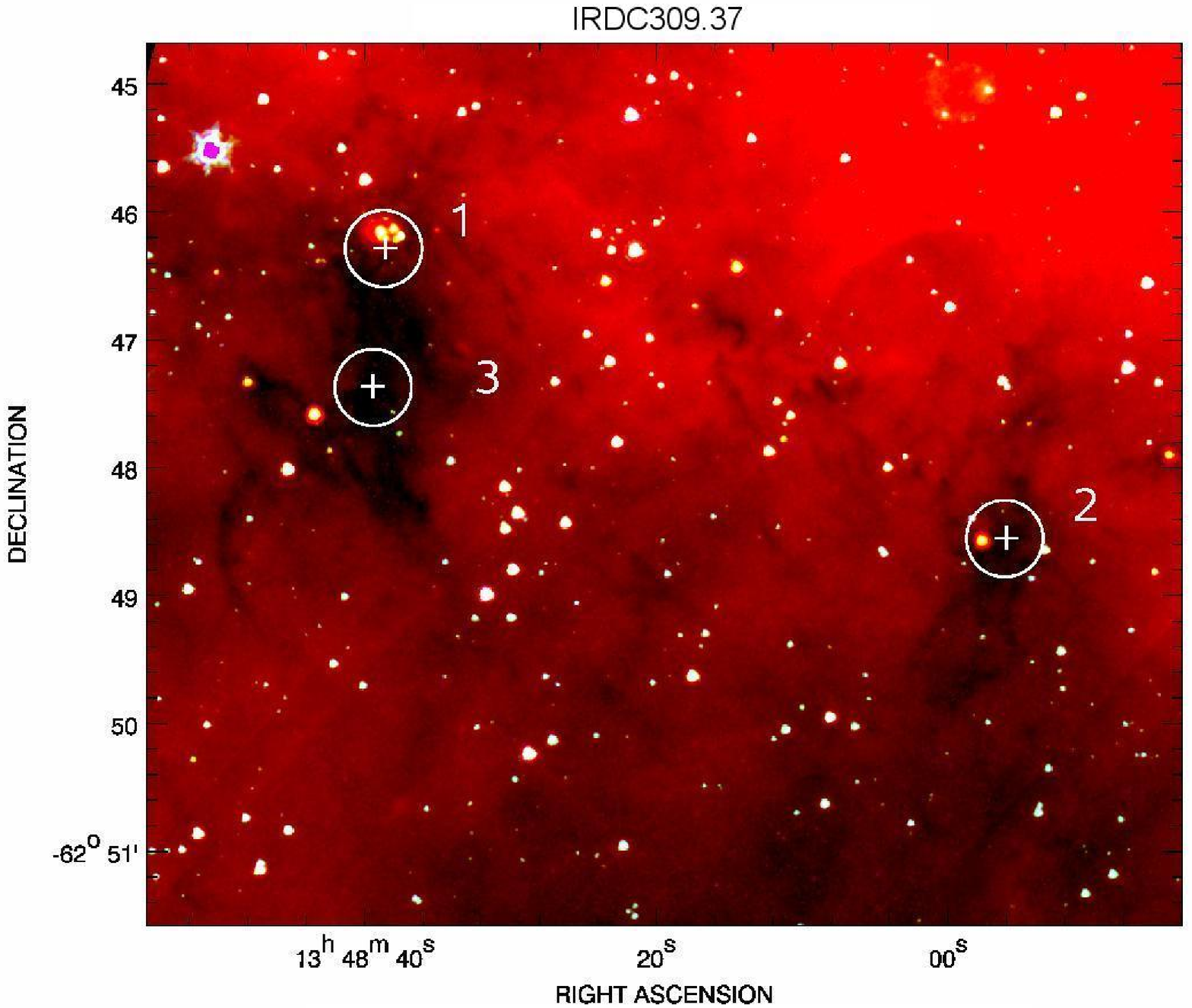}
     \includegraphics[width=9cm]{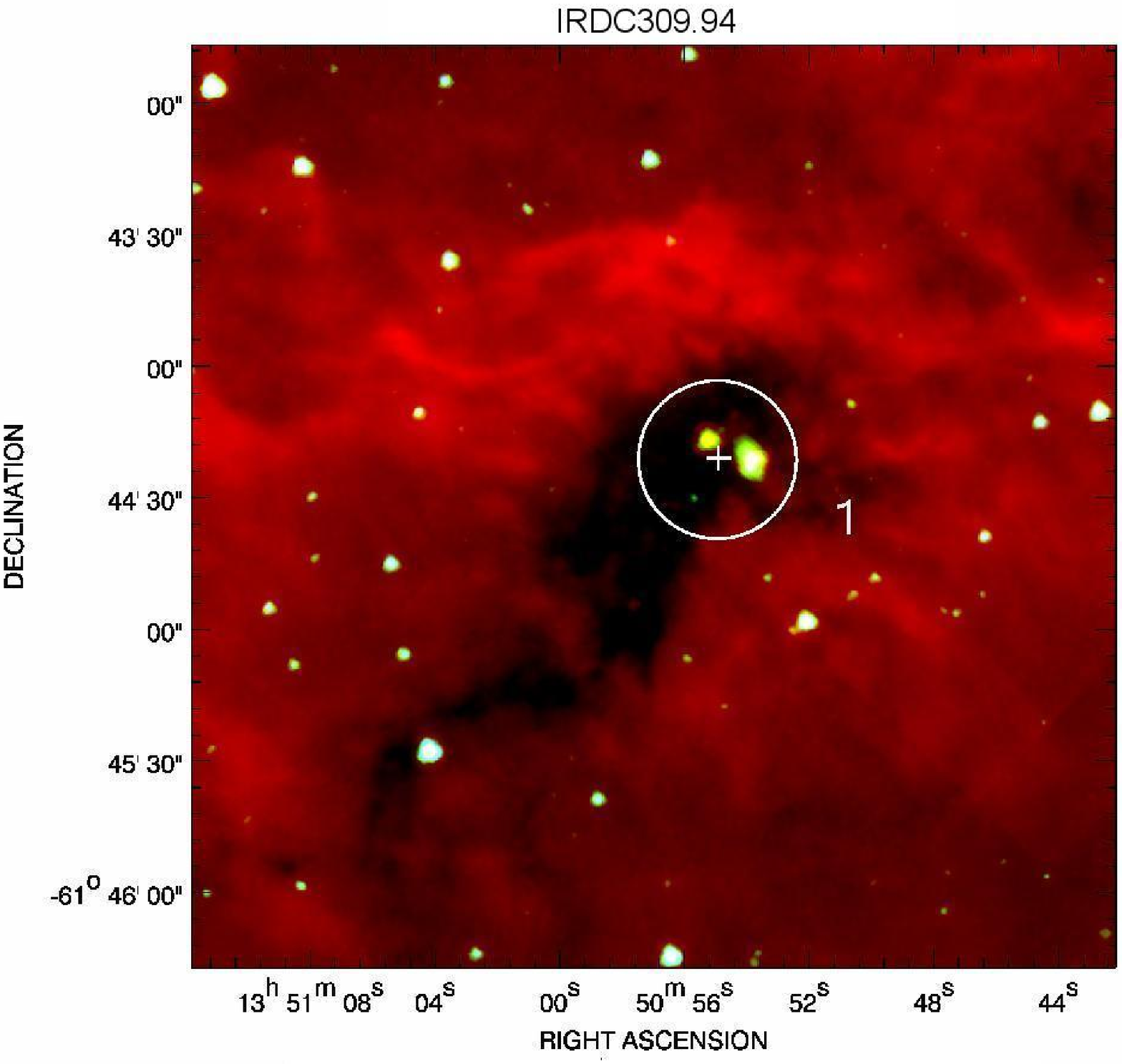}
     \includegraphics[width=9cm]{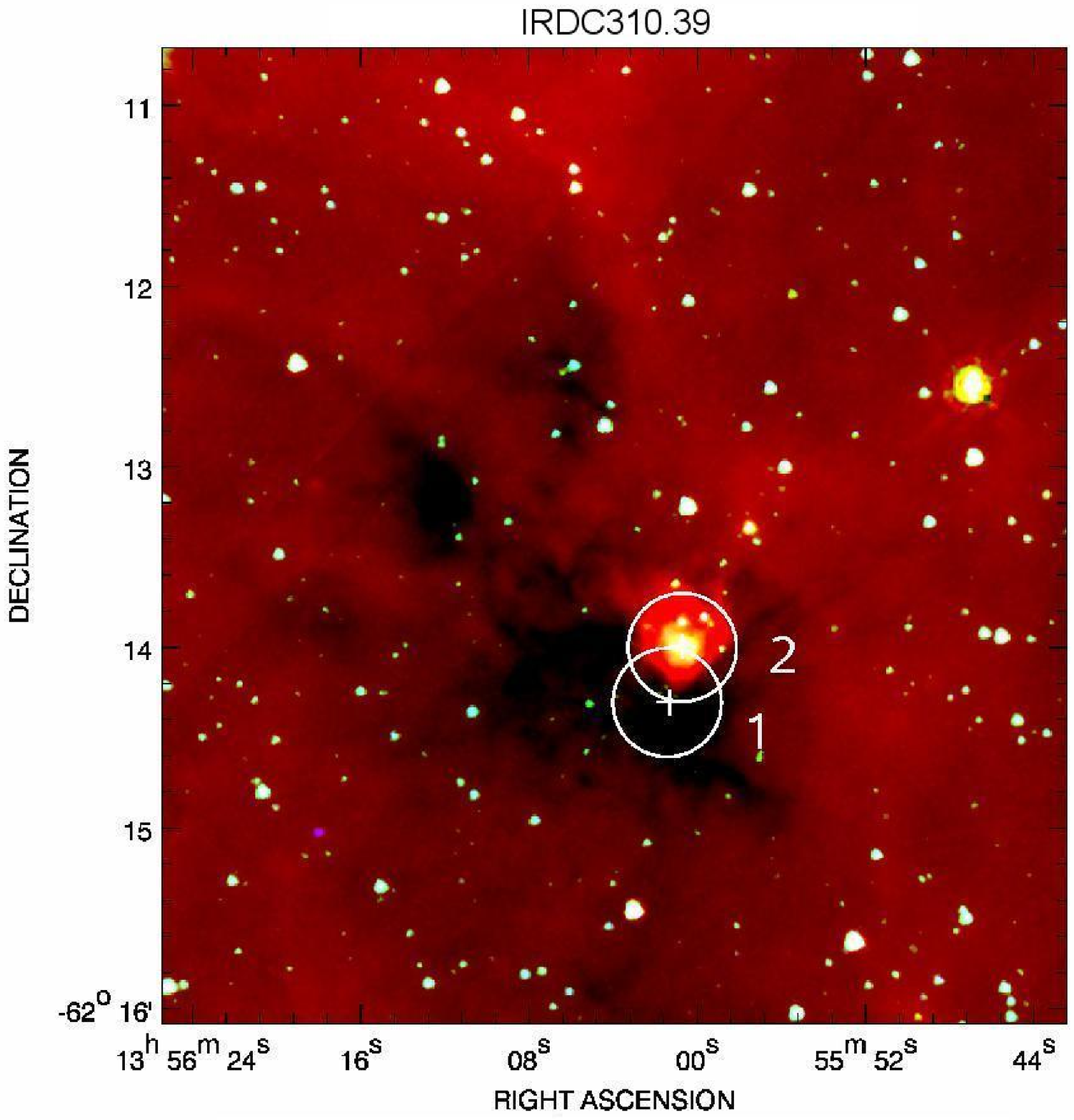}
     \caption{3-color Spitzer/GLIMPSE image of the Infrared Dark Cloud, where 3.6 $\mu$m is blue, 
              4.5$\mu$m is green and 8$\mu$m is red. Circles mark observed positions and 
	      show the beam size.}
     \label{3color1}
   \end{figure*} 
\clearpage
   \begin{figure*}
   \centering
     \includegraphics[width=9cm]{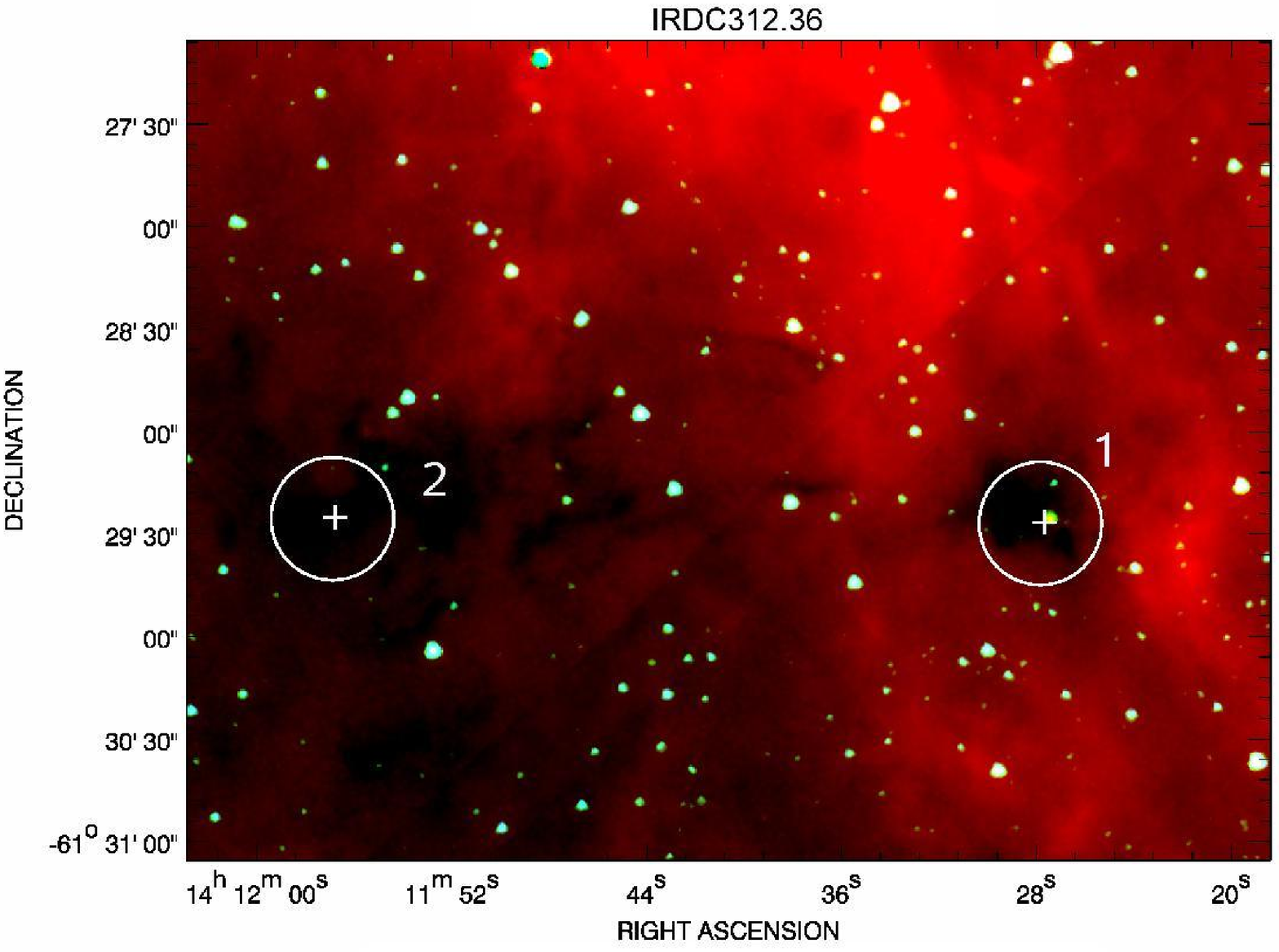}
     \includegraphics[width=9cm]{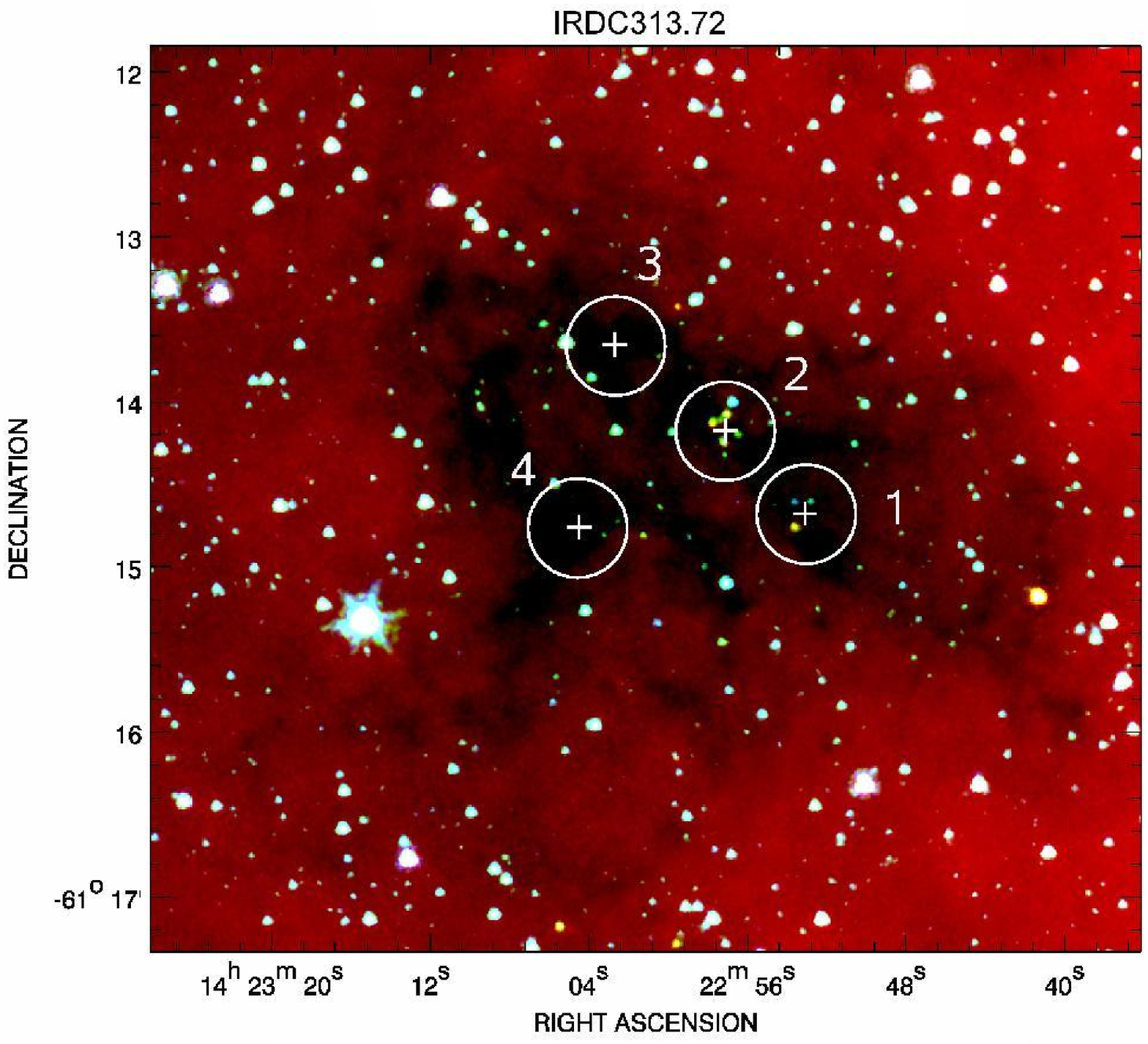}
     \includegraphics[width=9cm]{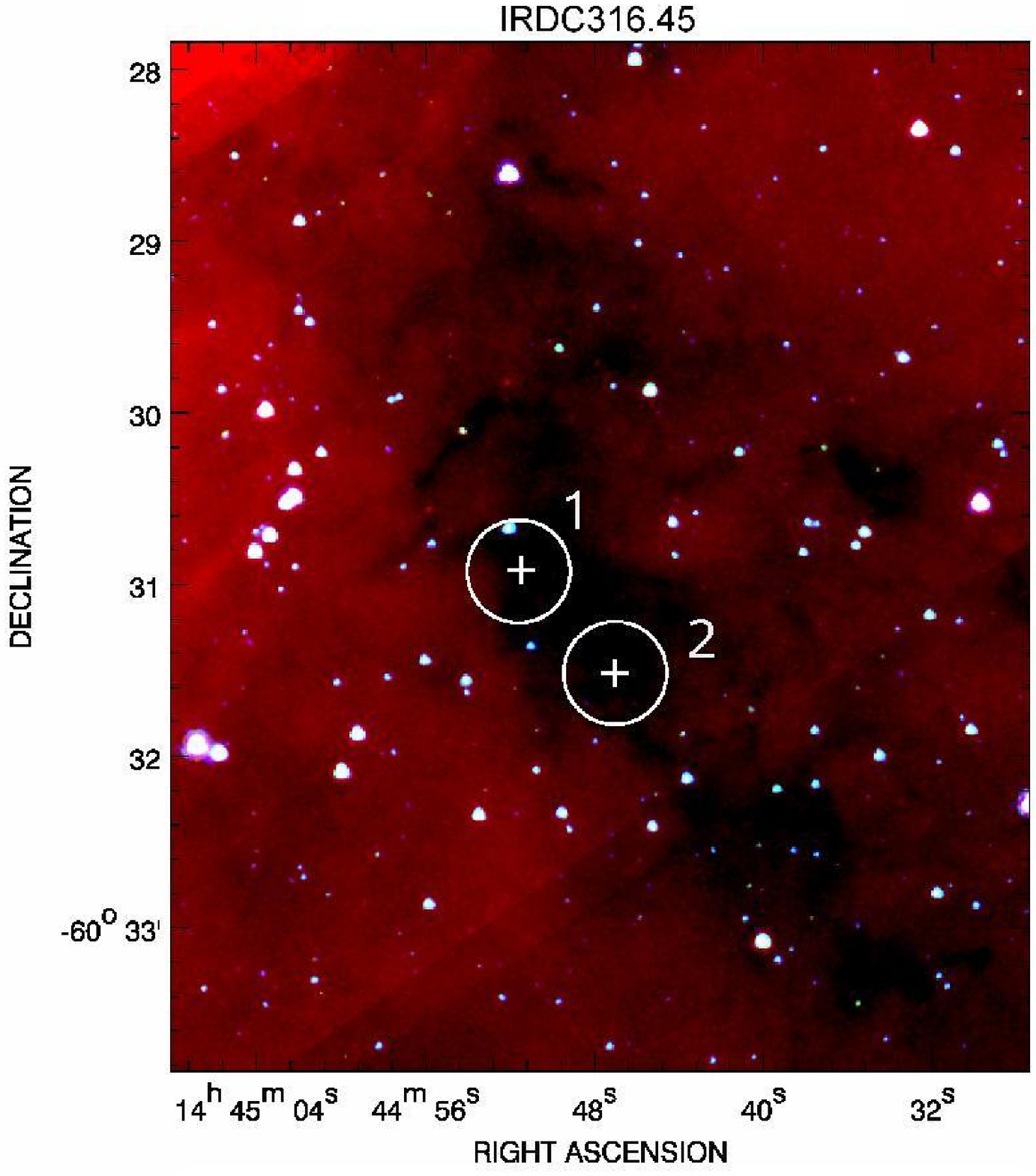}
     \includegraphics[width=9cm]{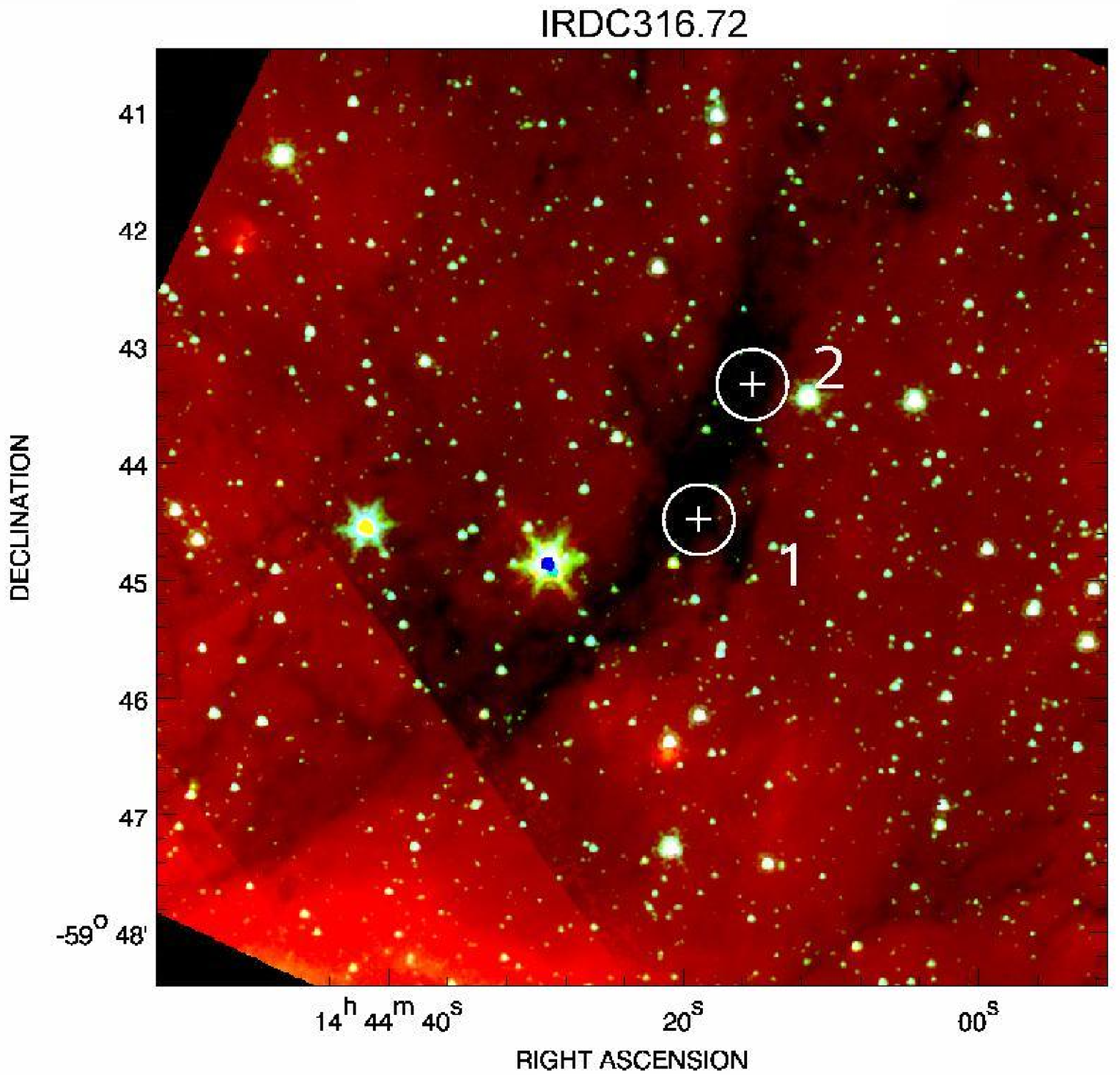}
     \includegraphics[width=9cm]{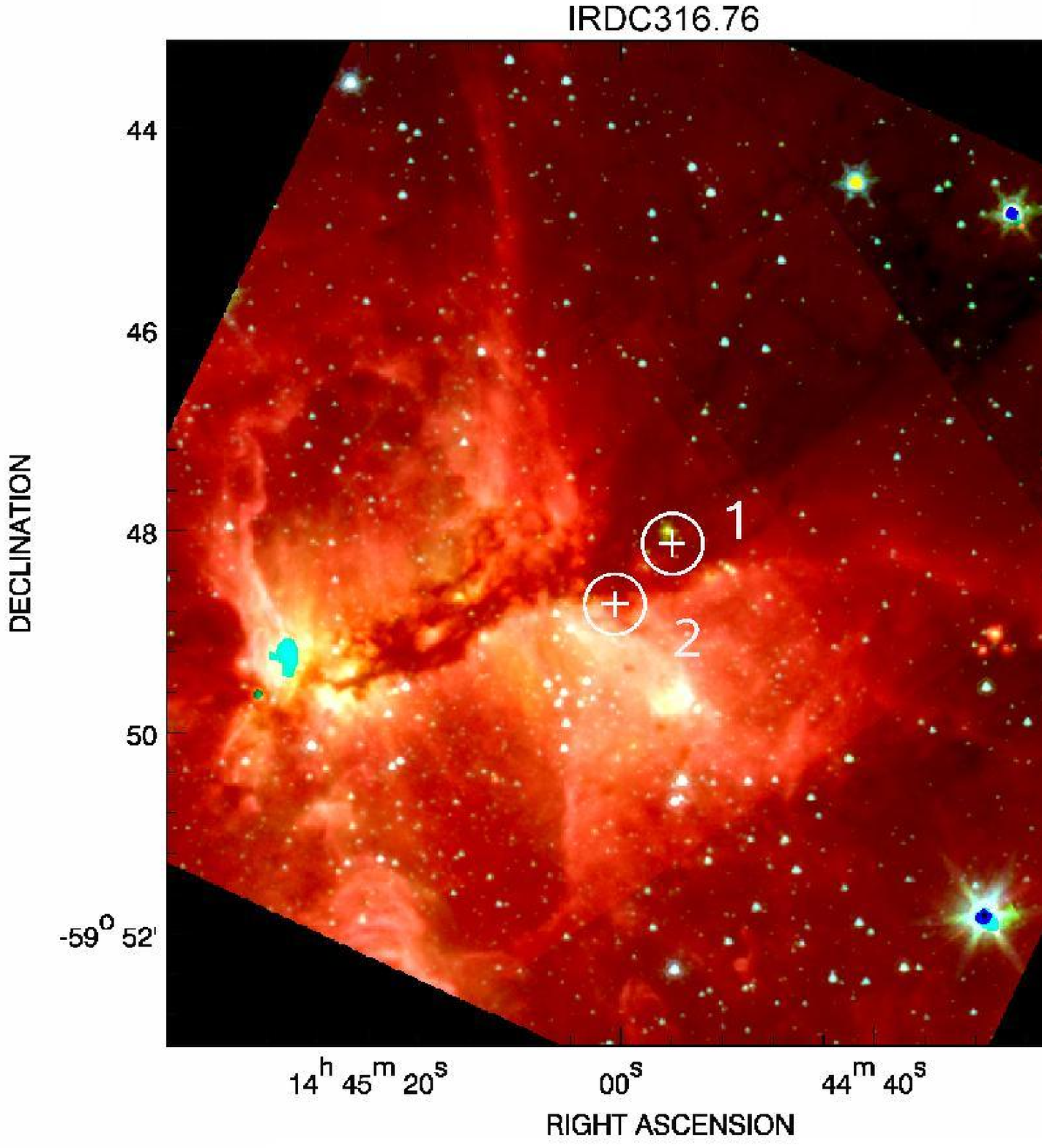}
     \includegraphics[width=9cm]{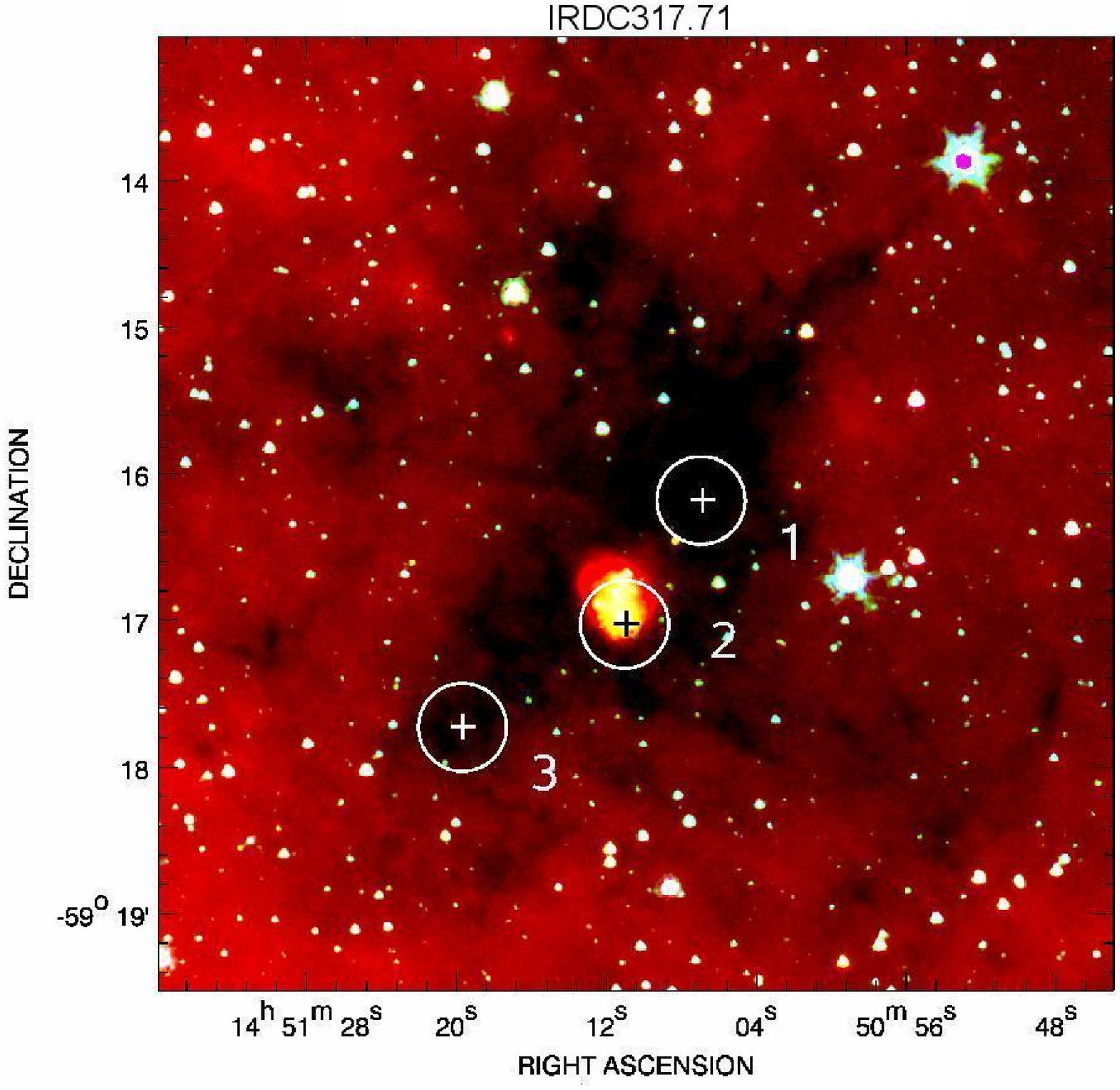}	     
     \caption{3-color Spitzer/GLIMPSE image of the Infrared Dark Cloud, where 3.6 $\mu$m is blue, 
              4.5$\mu$m is green and 8$\mu$m is red. Circles mark observed positions and 
	      show the beam size.}
     \label{3color2}
   \end{figure*} 
\clearpage
   \begin{figure*}
   \centering
     \includegraphics[width=9cm]{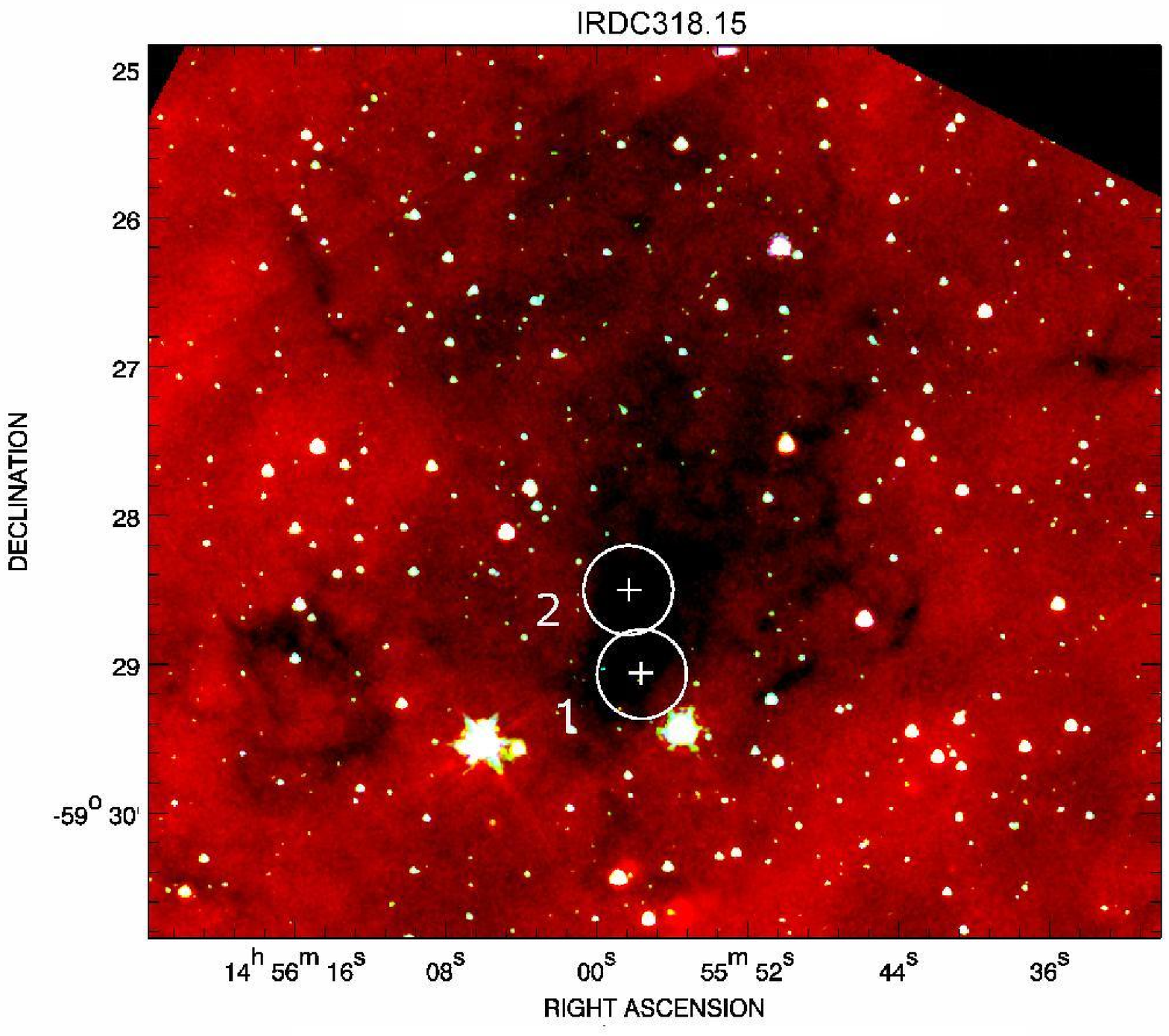}
     \includegraphics[width=9cm]{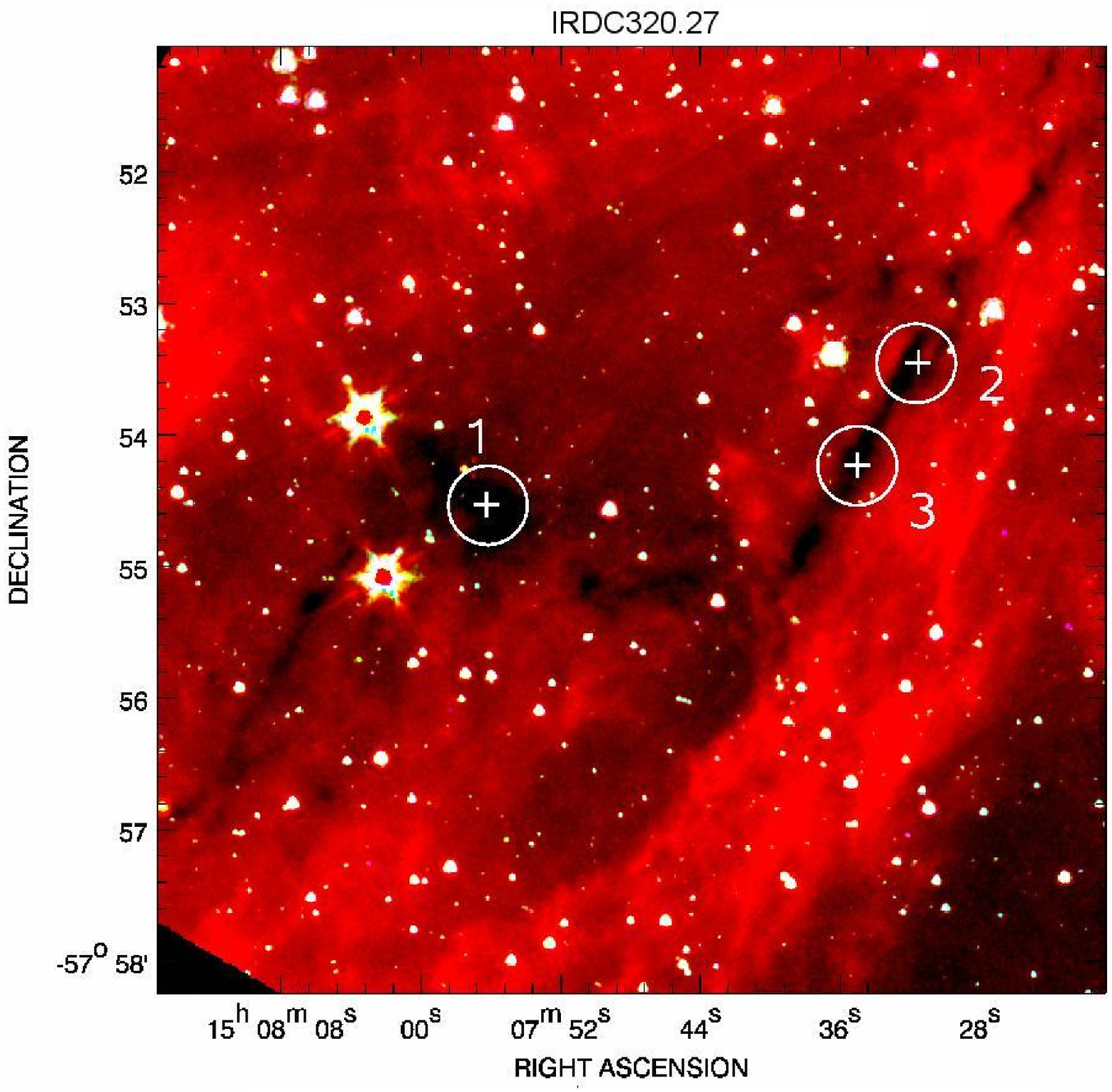}
     \includegraphics[width=9cm]{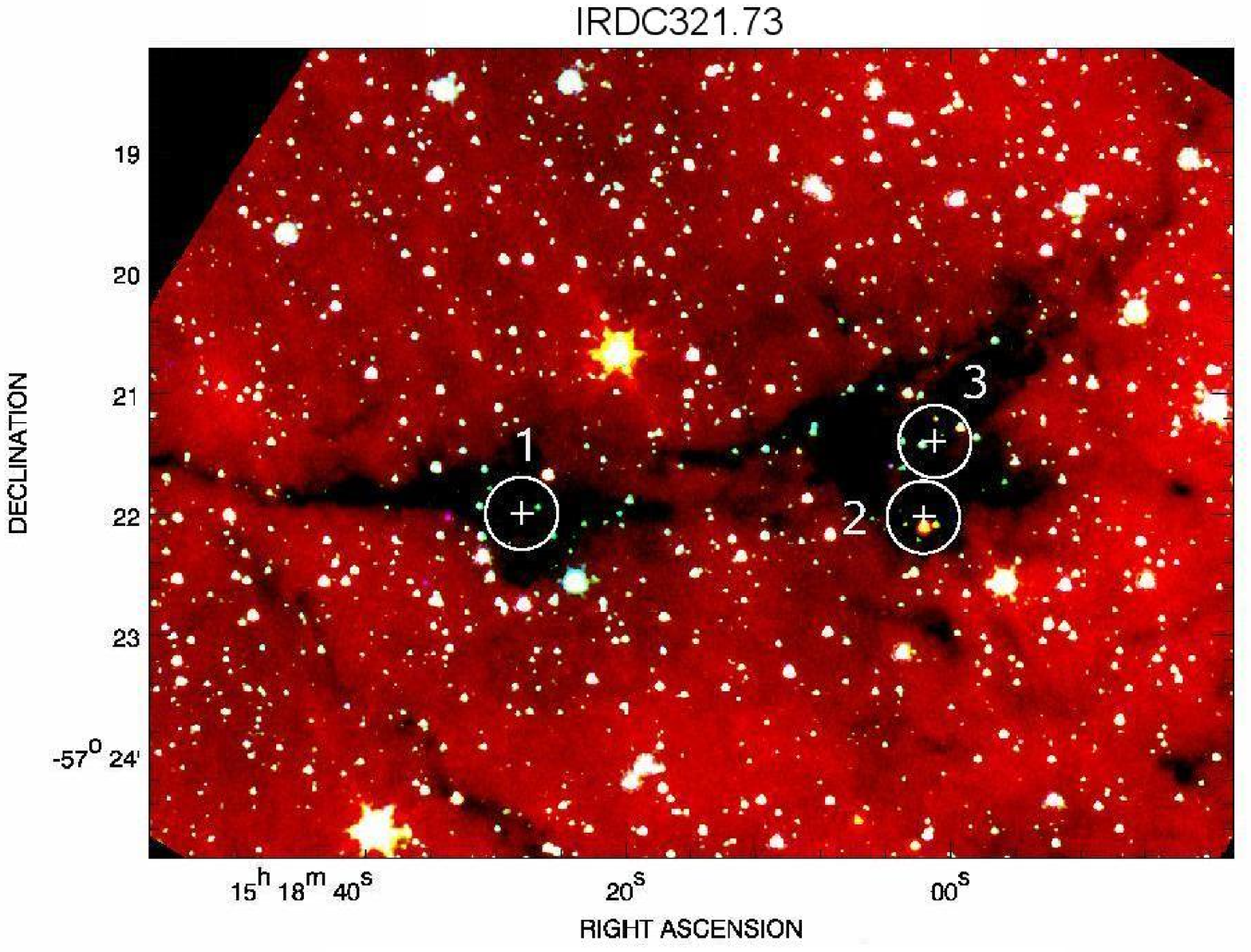}
     \caption{3-color Spitzer/GLIMPSE image of the Infrared Dark Cloud, where 3.6 $\mu$m is blue, 
              4.5$\mu$m is green and 8$\mu$m is red. Circles mark observed positions and 
	      show the beam size.}
     \label{3color3}
   \end{figure*}
\clearpage

\end{appendix}


\begin{appendix} 
\section{Spectra images.}
...
 
\begin{figure*}
\centering
\includegraphics[width=9cm]{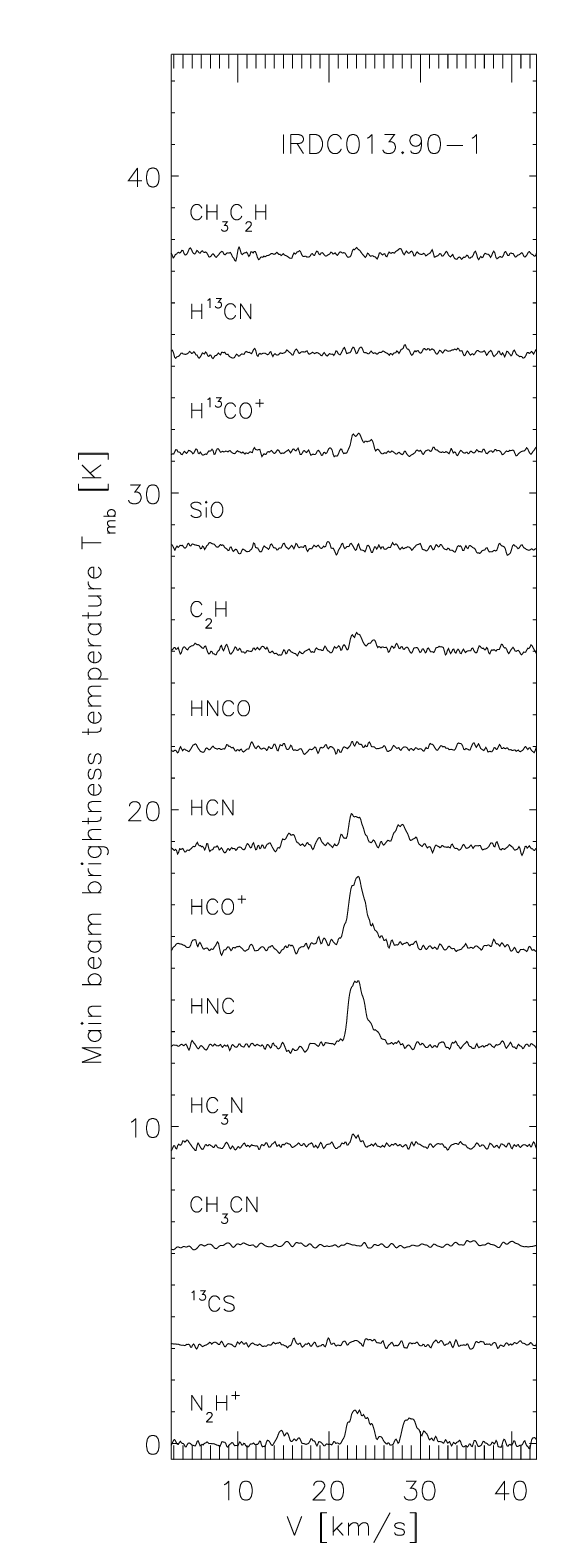}
\includegraphics[width=9cm]{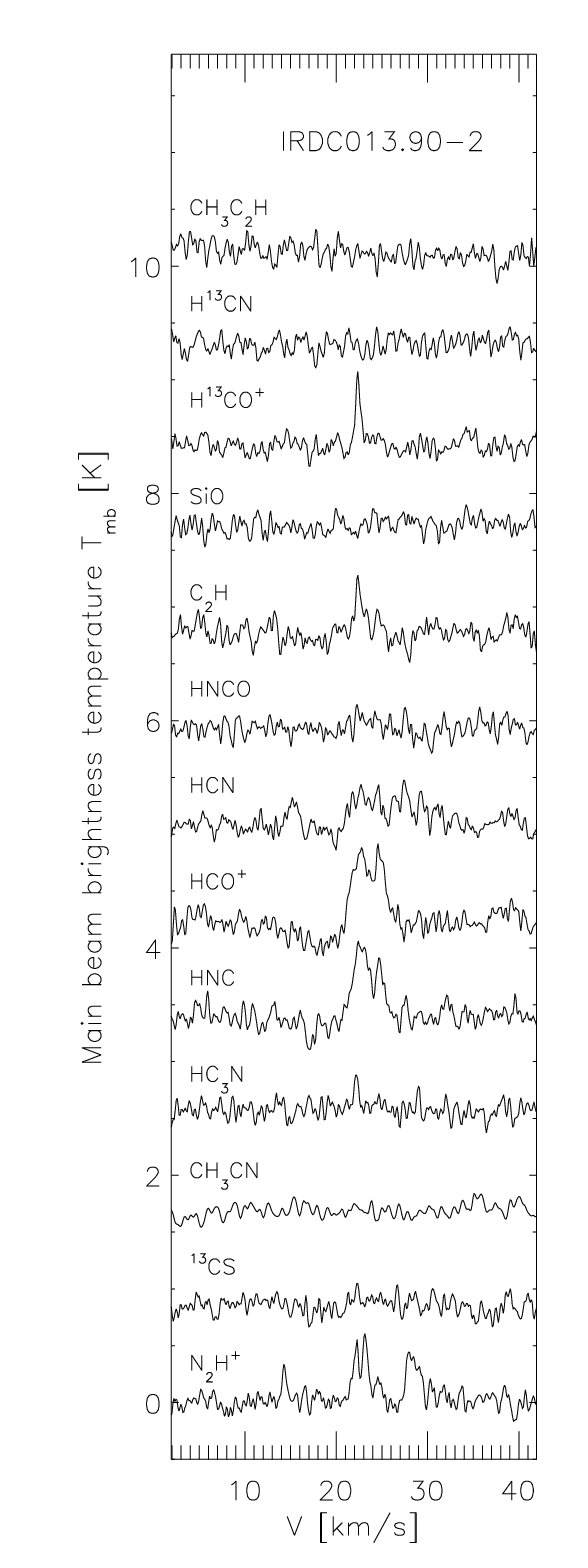}
\caption{Observed lines}
\label{appfig1}
\end{figure*} 
\clearpage

\begin{figure*}
\centering
\includegraphics[width=9cm]{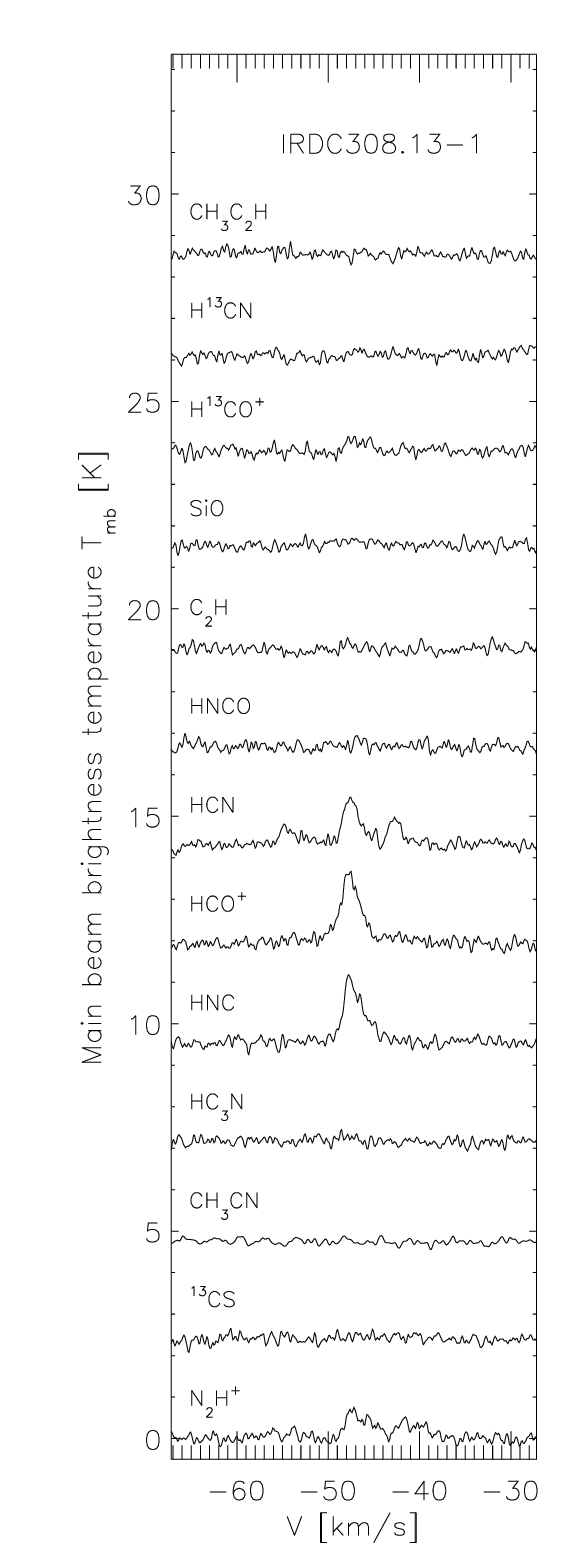}
\includegraphics[width=9cm]{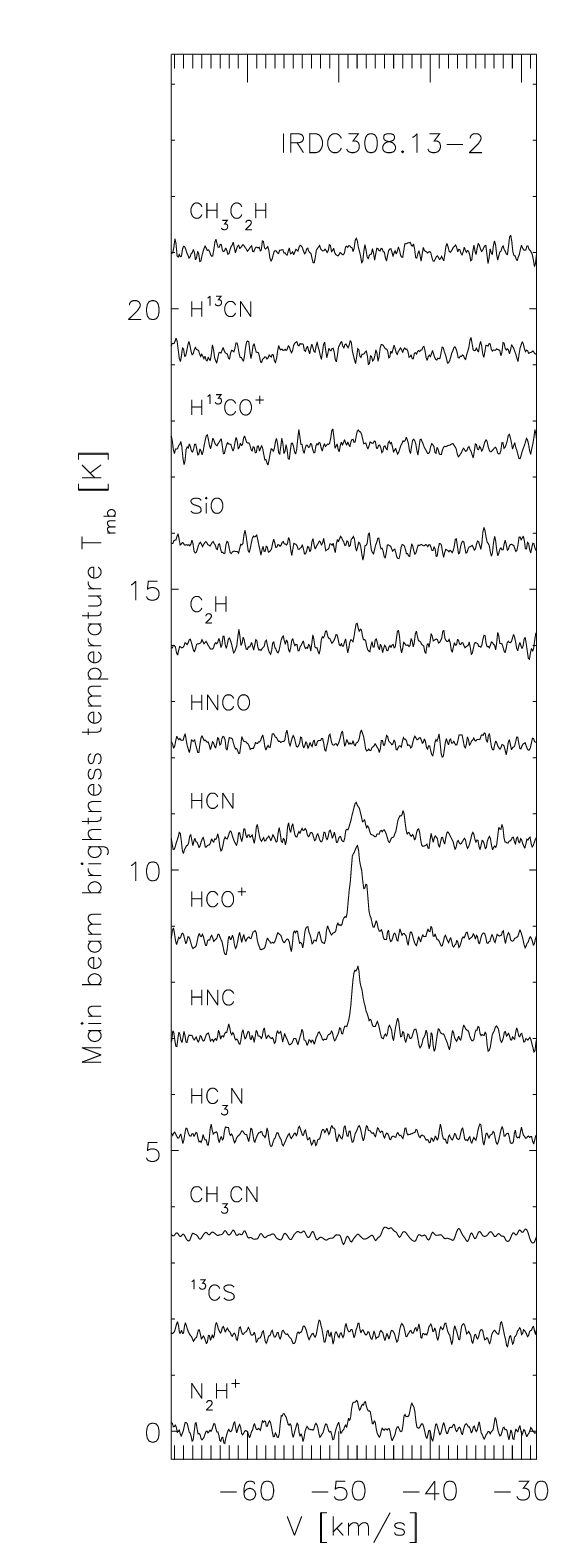}
\caption{Observed lines}
\label{appfig2}
\end{figure*}
\clearpage

\begin{figure*}
\centering
\includegraphics[width=9cm]{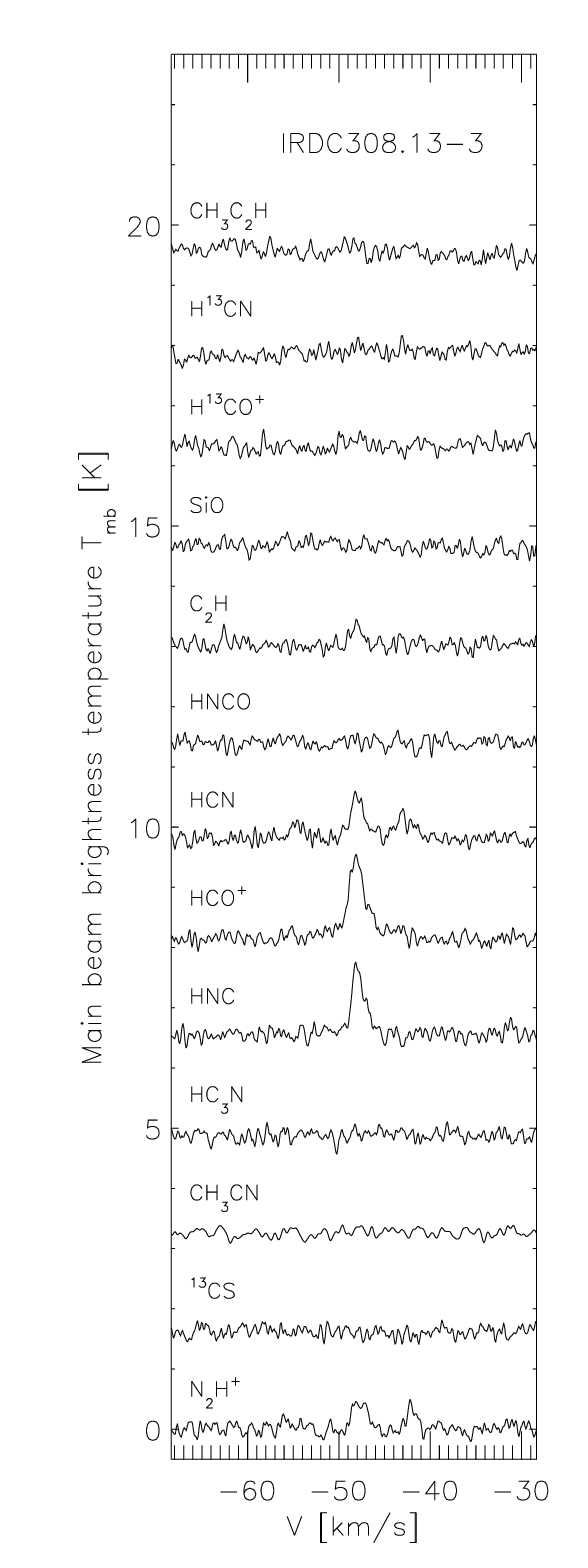}
\includegraphics[width=9cm]{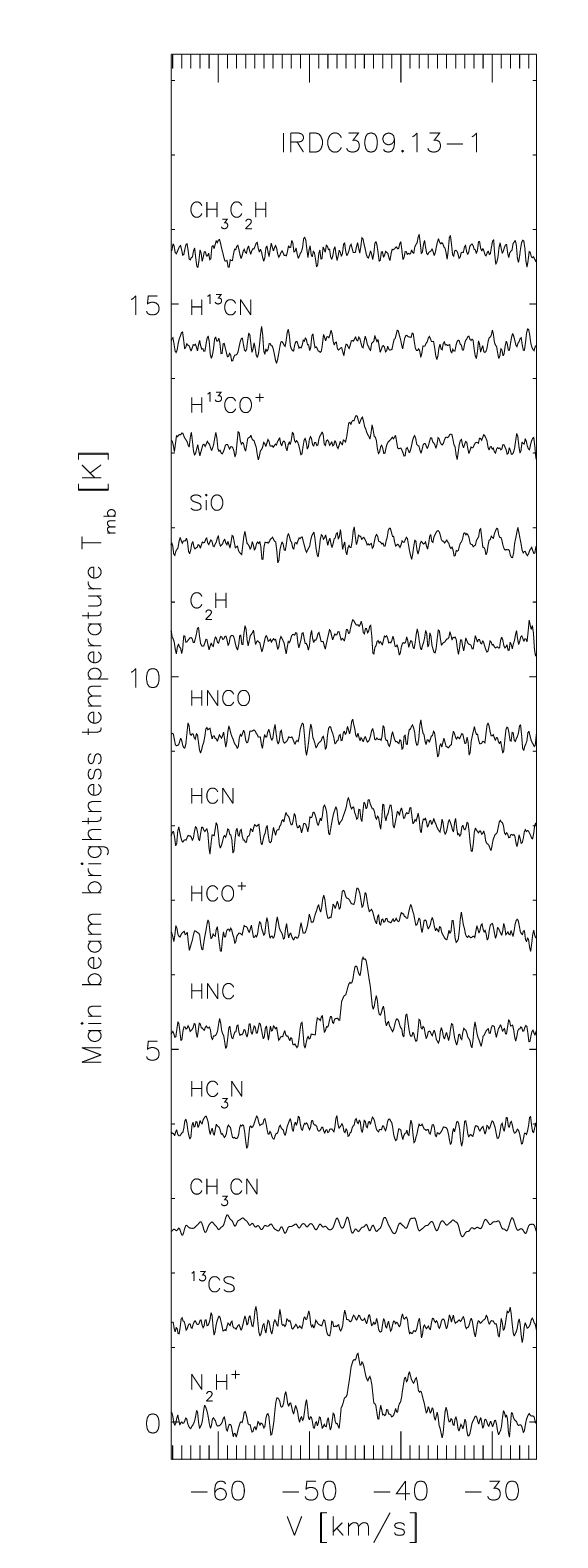}
\caption{Observed lines}
\label{appfig3}
\end{figure*}
\clearpage

\begin{figure*}
\centering
\includegraphics[width=9cm]{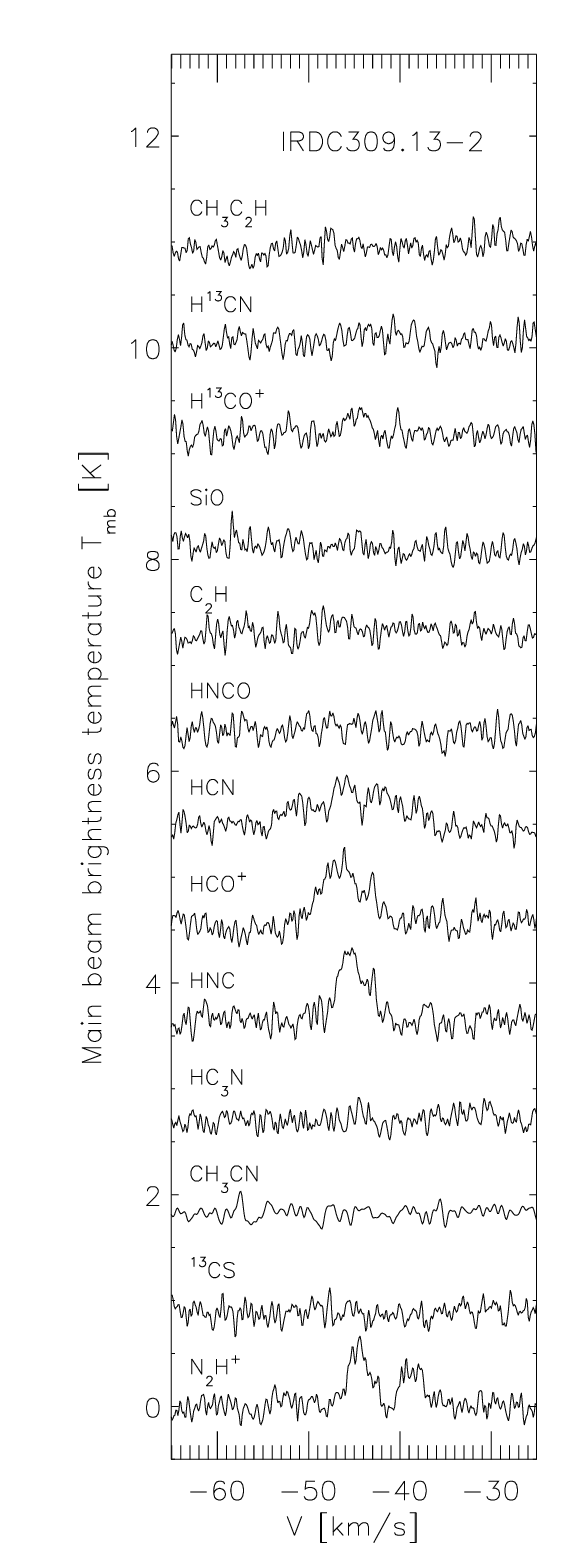}
\includegraphics[width=9cm]{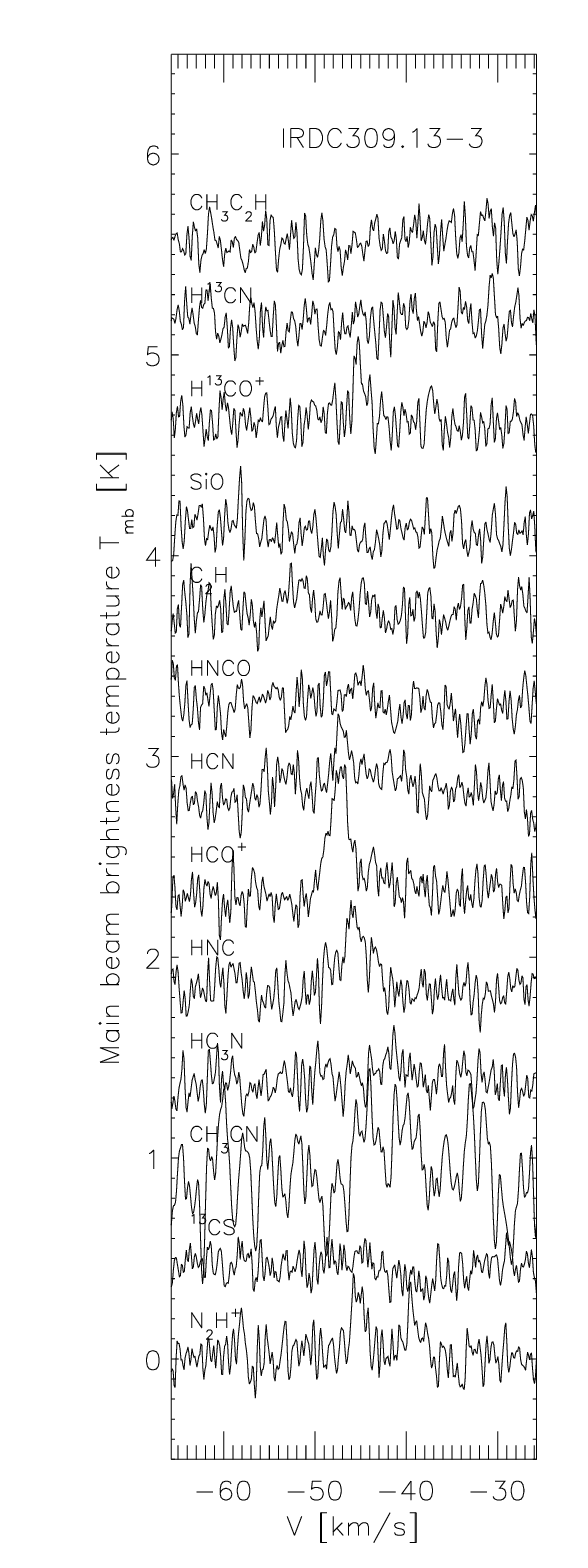}
\caption{Observed lines}
\label{appfig4}
\end{figure*}
\clearpage

\begin{figure*}
\centering
\includegraphics[width=9cm]{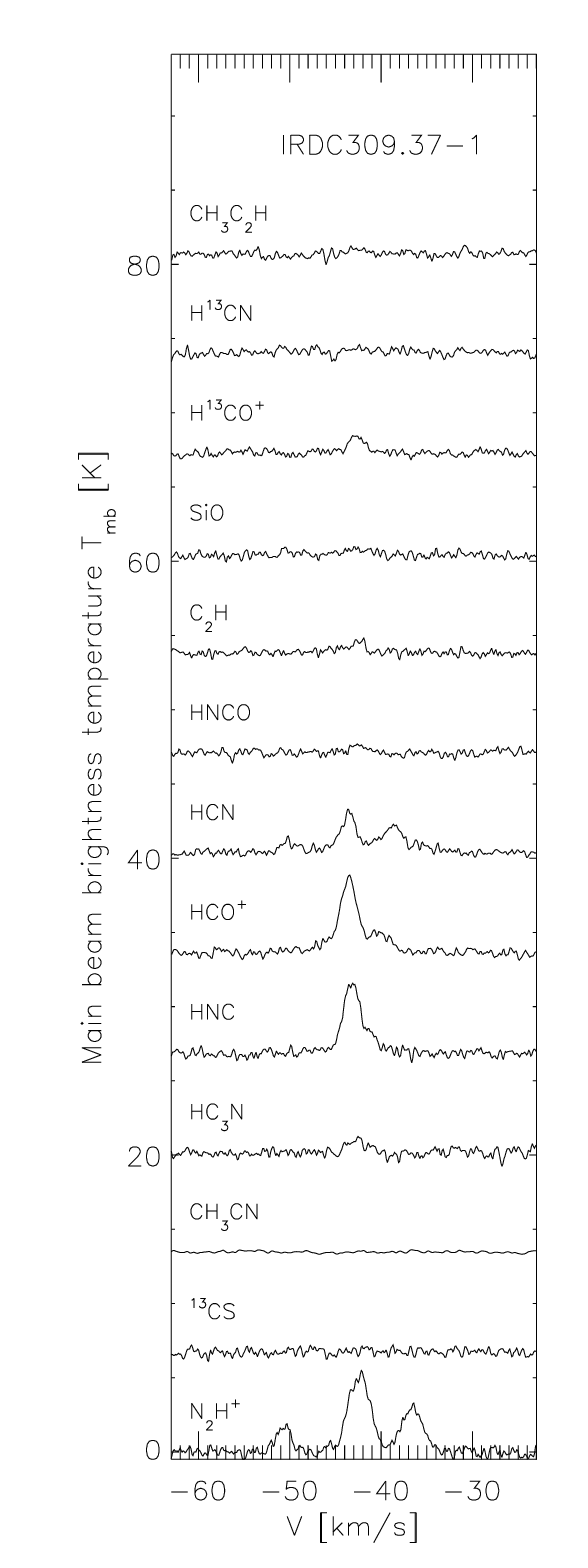}
\includegraphics[width=9cm]{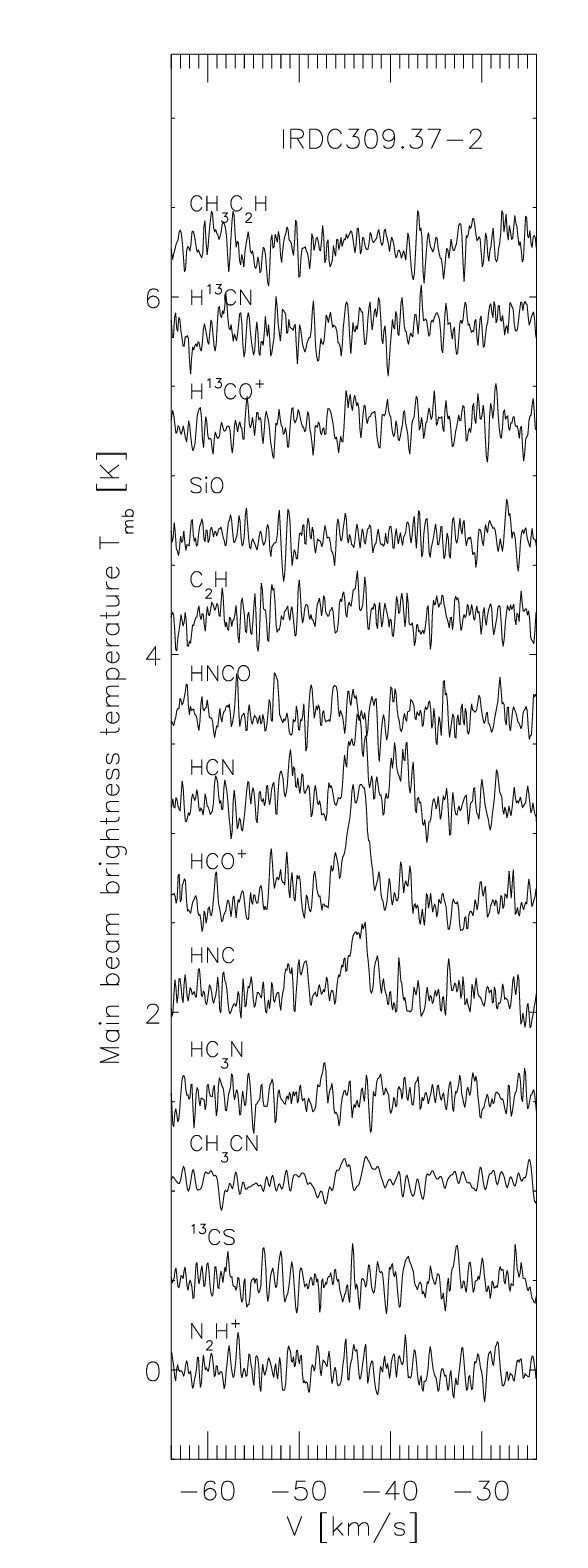}
\caption{Observed lines}
\label{appfig5}
\end{figure*}
\clearpage

\begin{figure*}
\centering
\includegraphics[width=9cm]{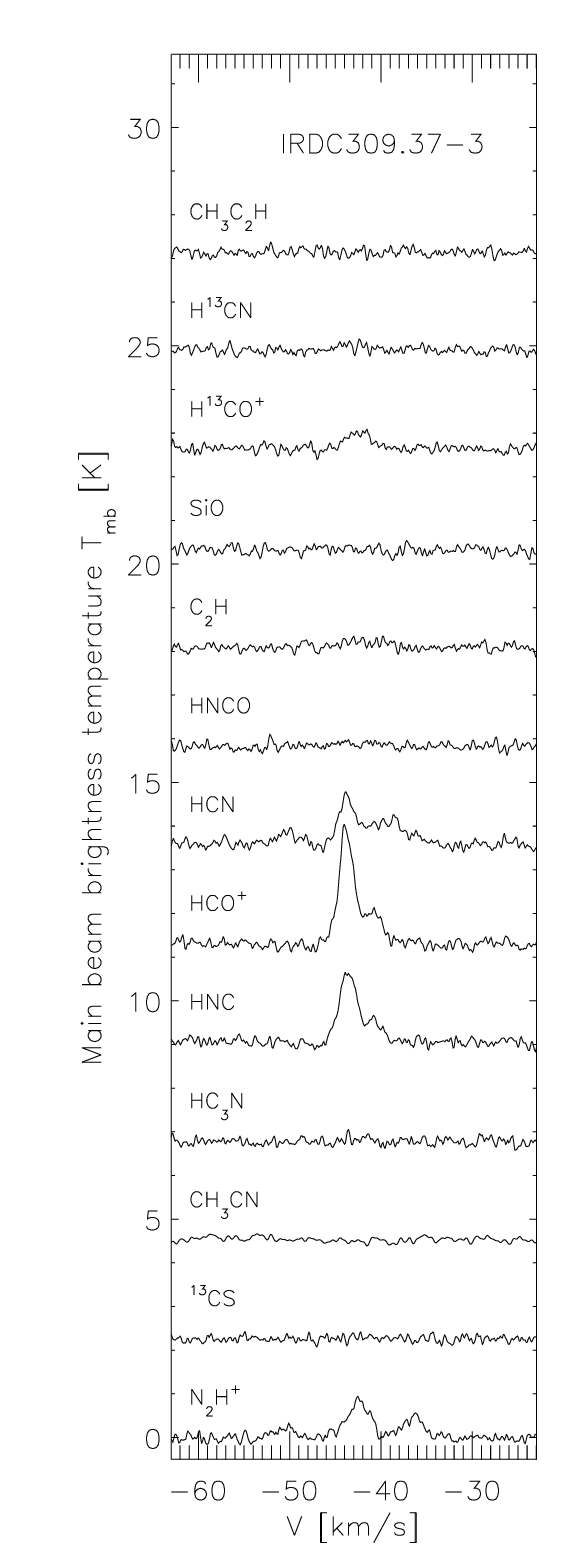}
\includegraphics[width=9cm]{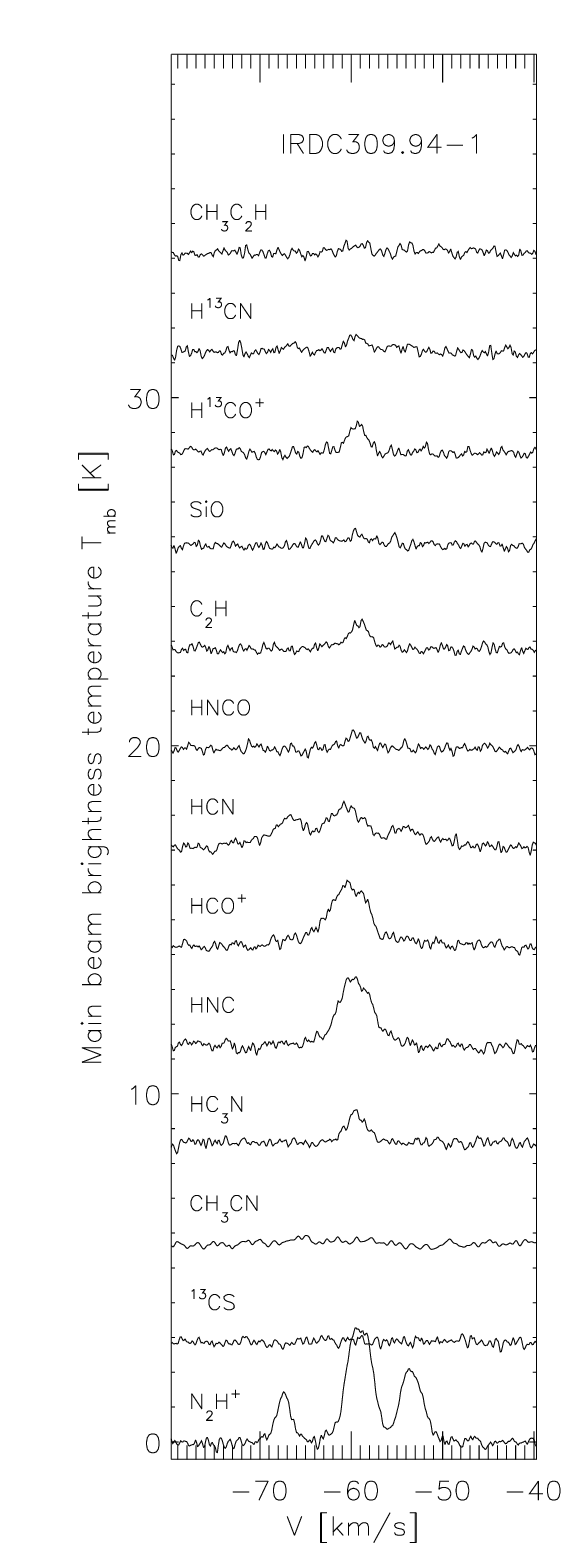}
\caption{Observed lines}
\label{appfig}
\end{figure*}
\clearpage

\begin{figure*}
\centering
\includegraphics[width=9cm]{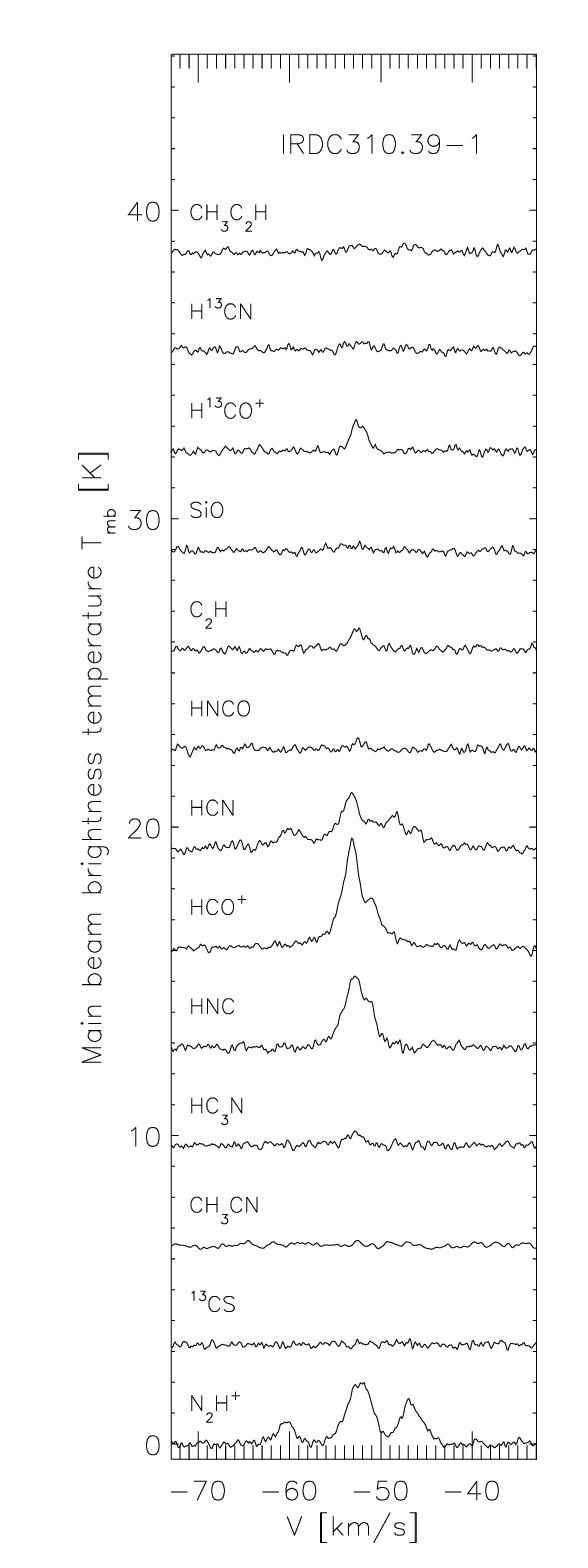}
\includegraphics[width=9cm]{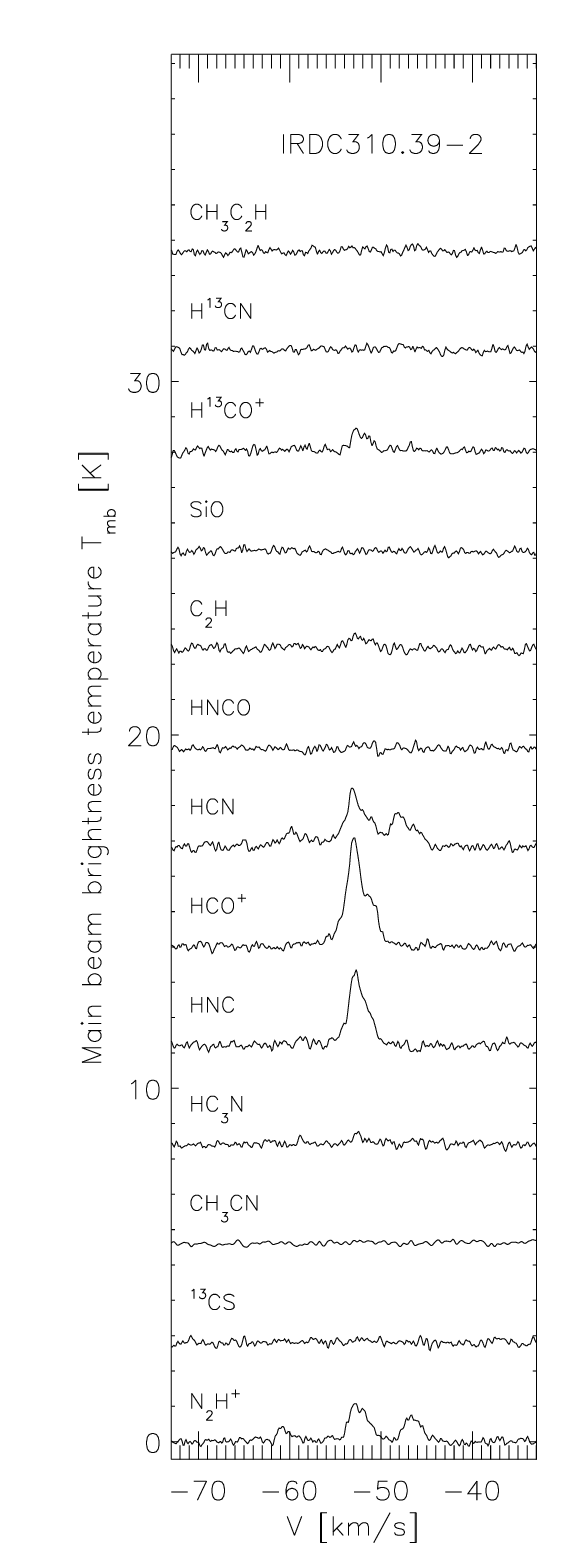}
\caption{Observed lines}
\label{appfig6}
\end{figure*}
\clearpage

\begin{figure*}
\centering
\includegraphics[width=9cm]{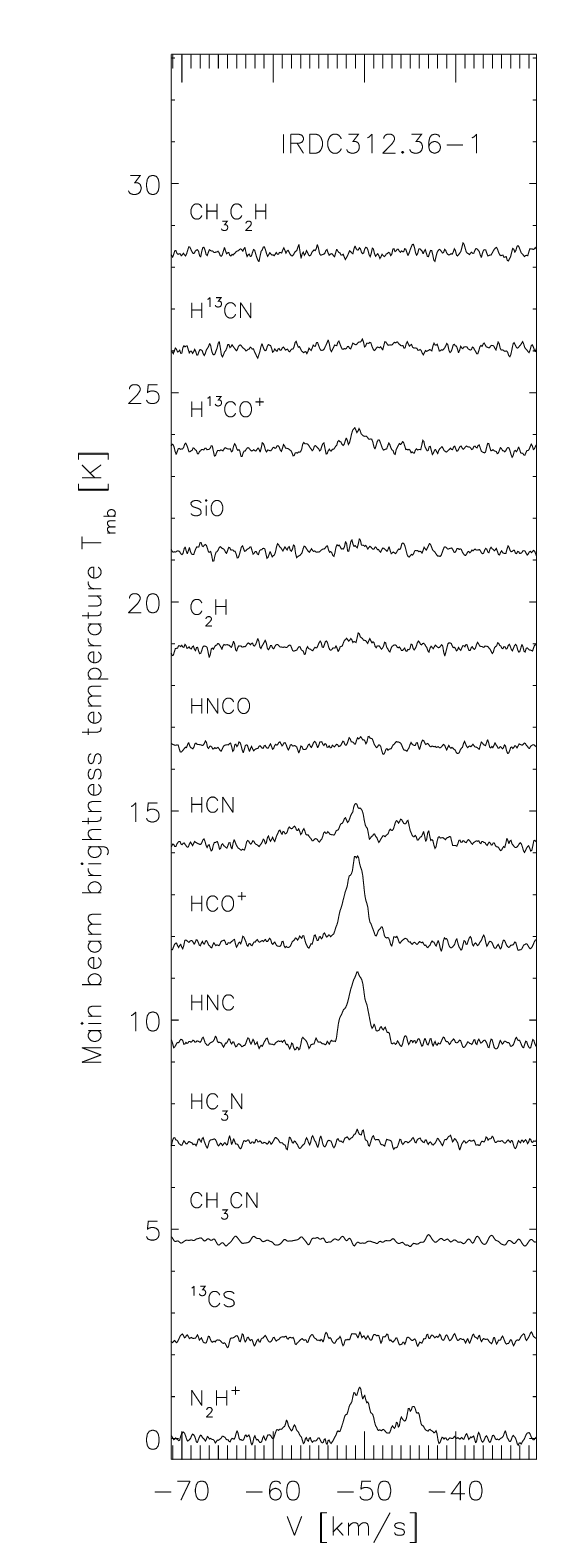}
\includegraphics[width=9cm]{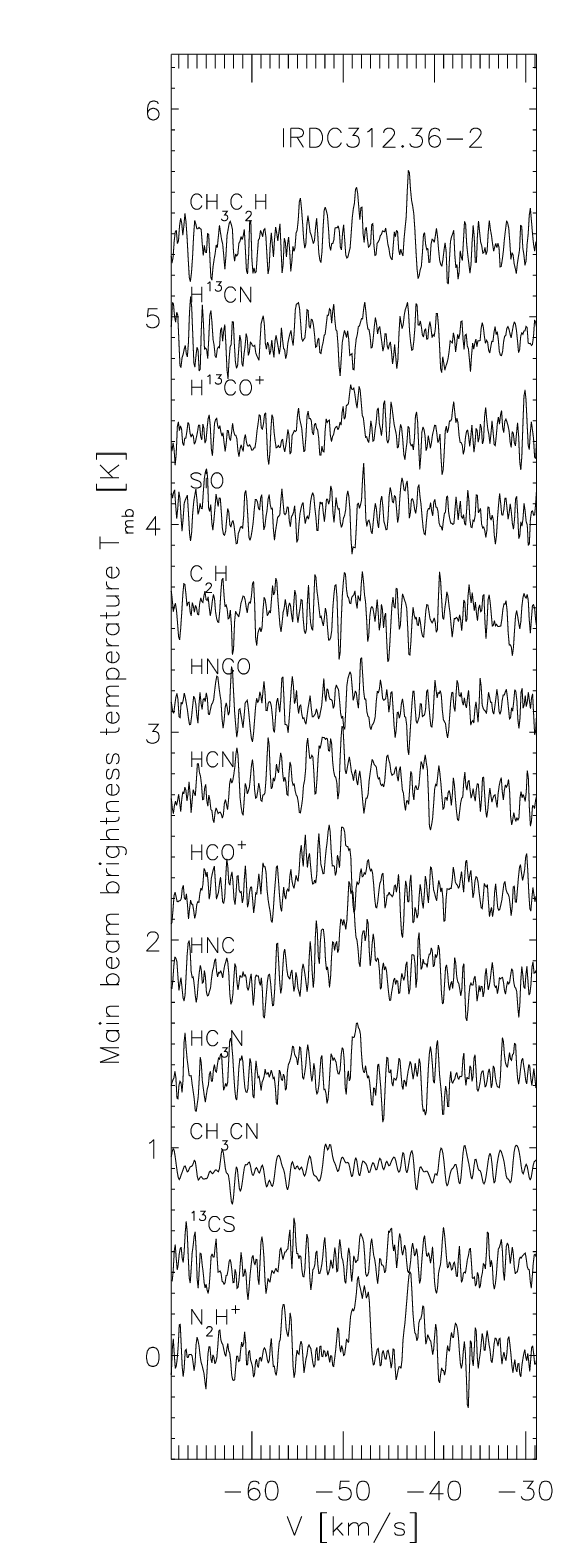}
\caption{Observed lines}
\label{appfig7}
\end{figure*}
\clearpage

\begin{figure*}
\centering
\includegraphics[width=9cm]{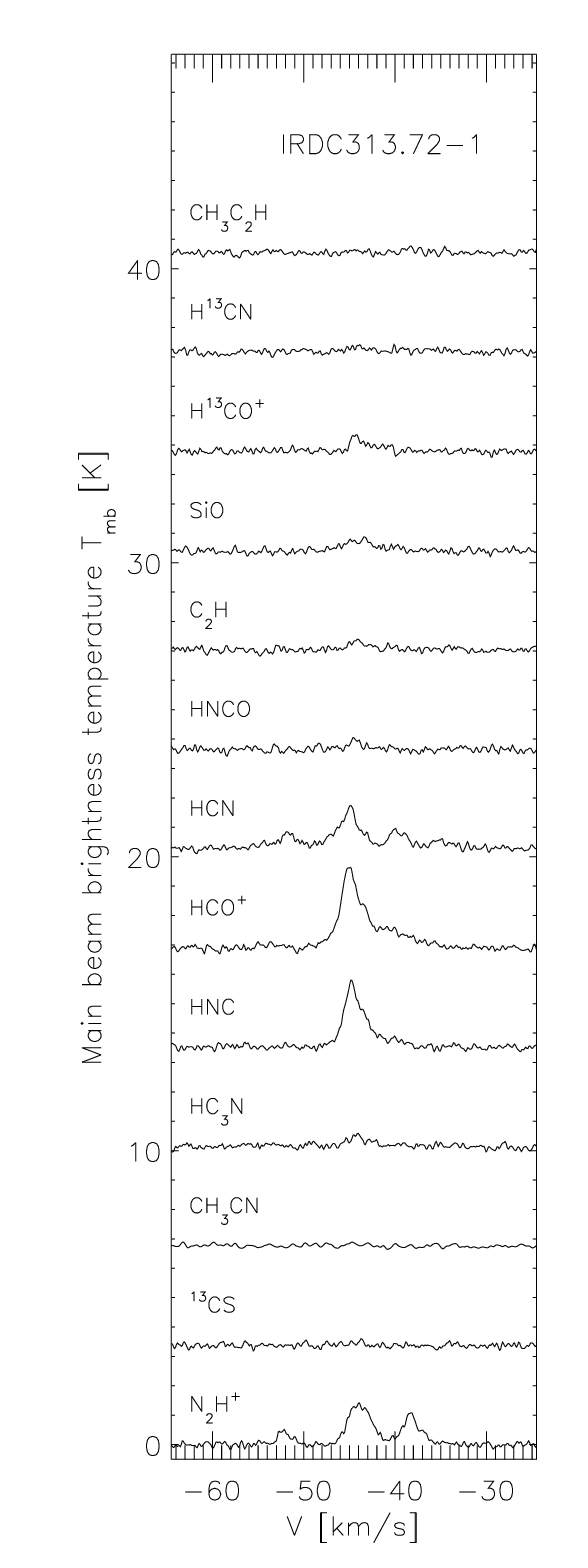}
\includegraphics[width=9cm]{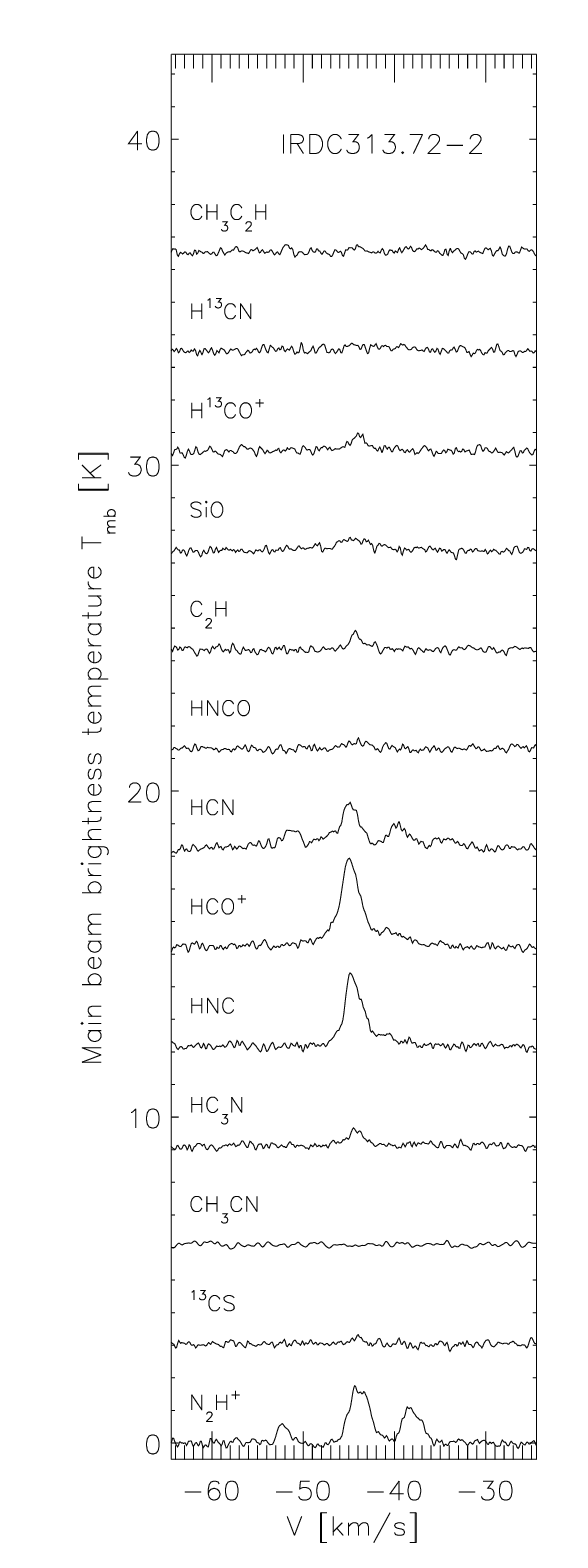}
\caption{Observed lines}
\label{appfig8}
\end{figure*} 
 \clearpage
          
\begin{figure*}
\centering
\includegraphics[width=9cm]{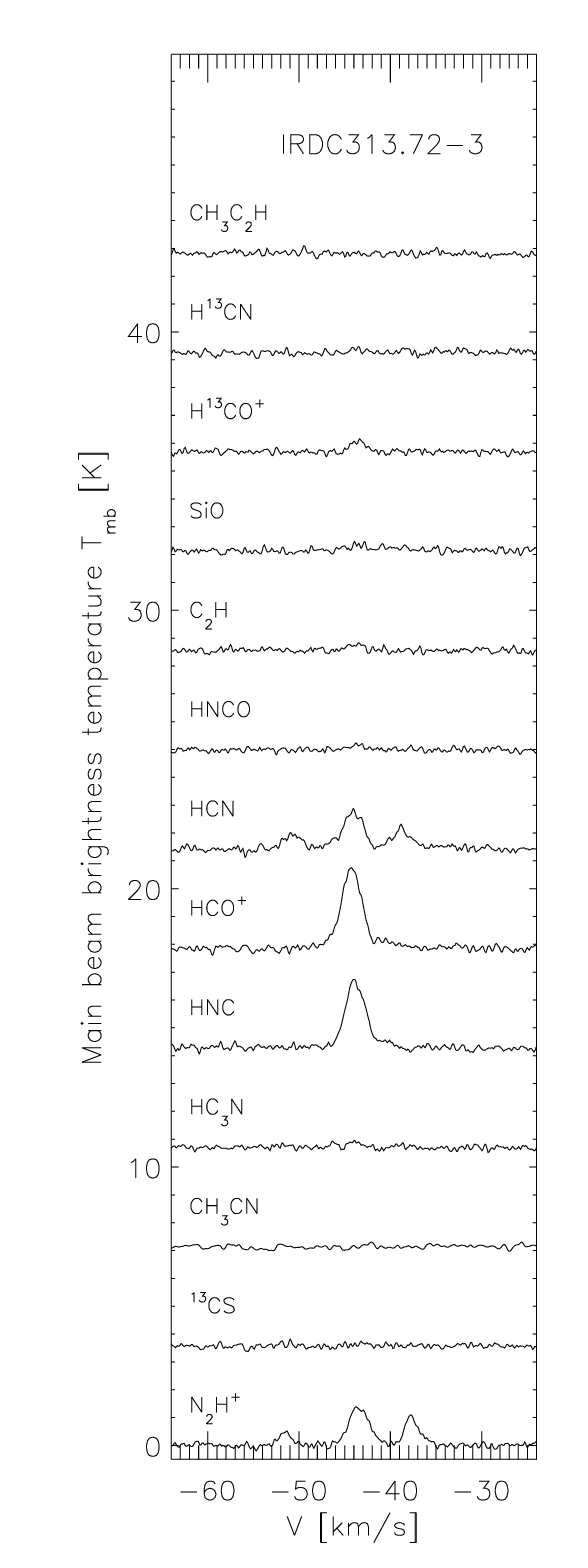}
\includegraphics[width=9cm]{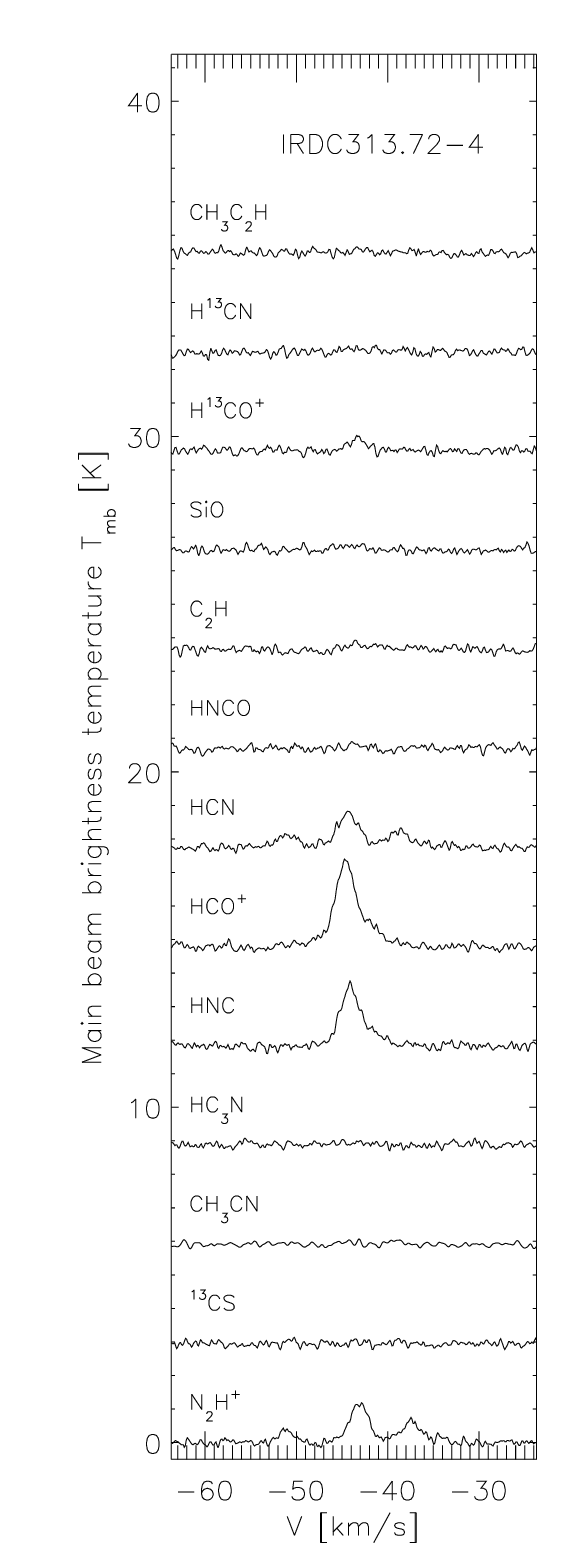}
\caption{Observed lines}
\label{appfig9}
\end{figure*}
\clearpage

\begin{figure*}
\centering
\includegraphics[width=9cm]{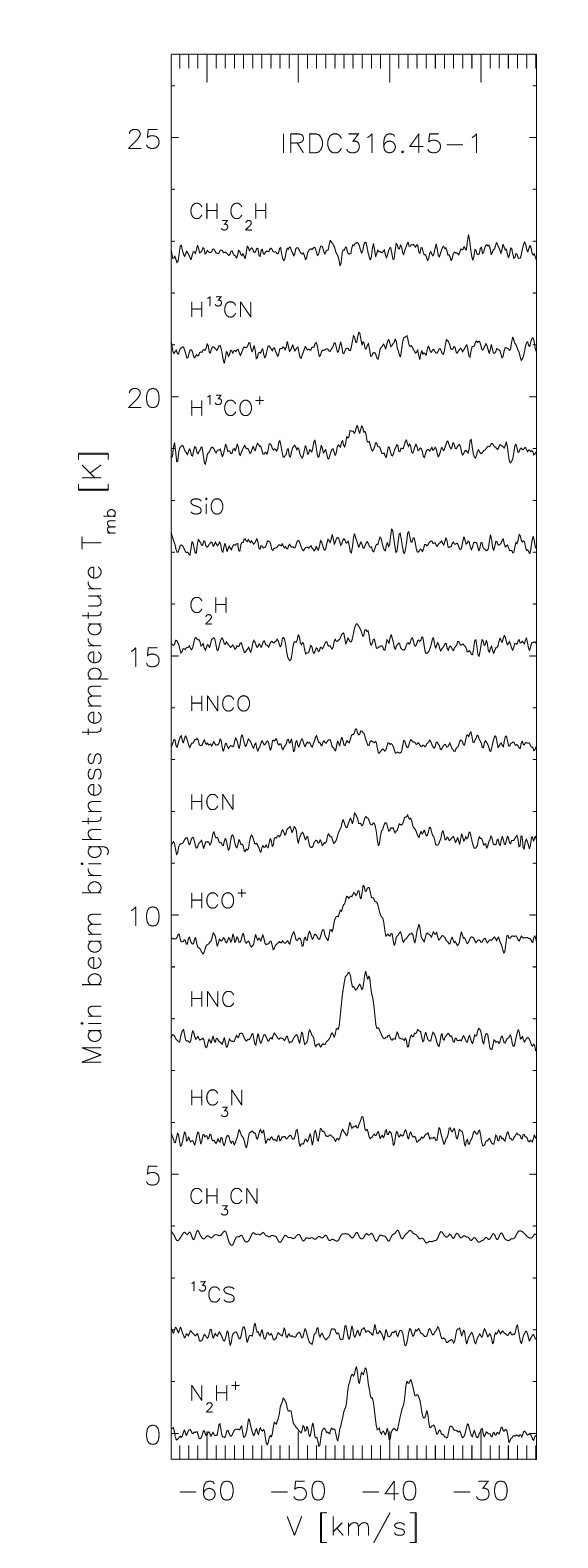}
\includegraphics[width=9cm]{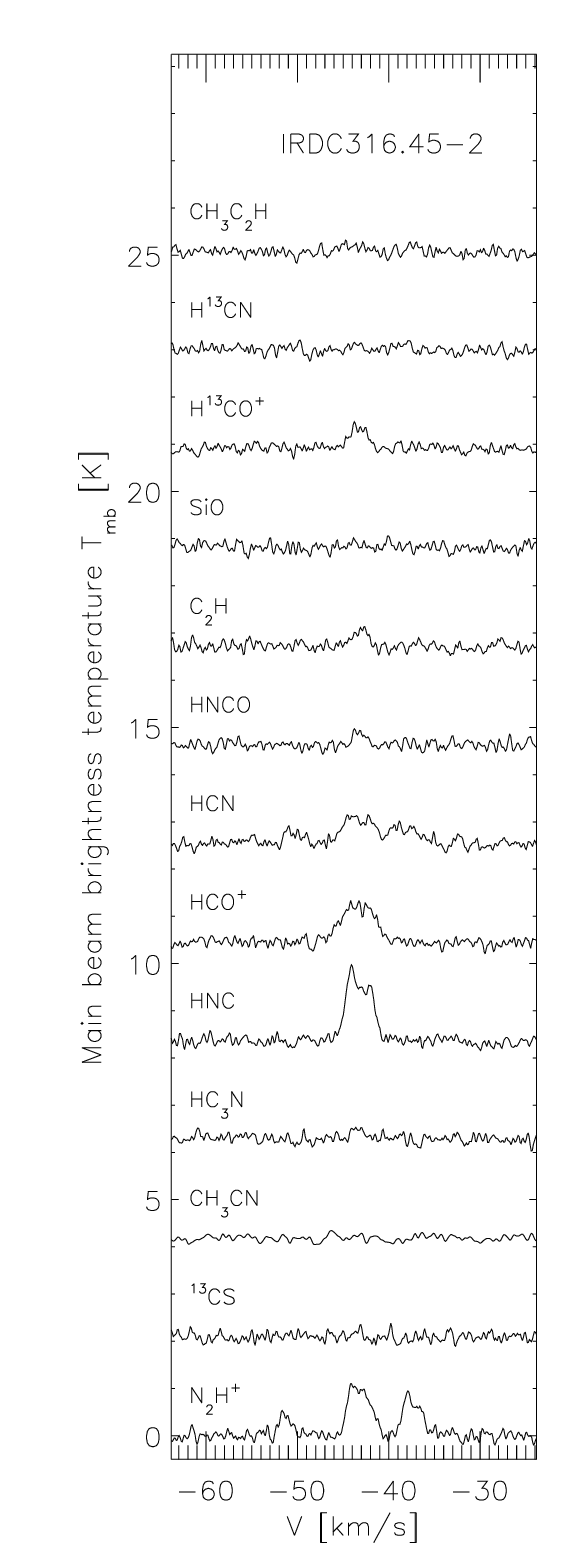}
\caption{Observed lines}
\label{appfig10}
\end{figure*}
\clearpage

\begin{figure*}
\centering
\includegraphics[width=9cm]{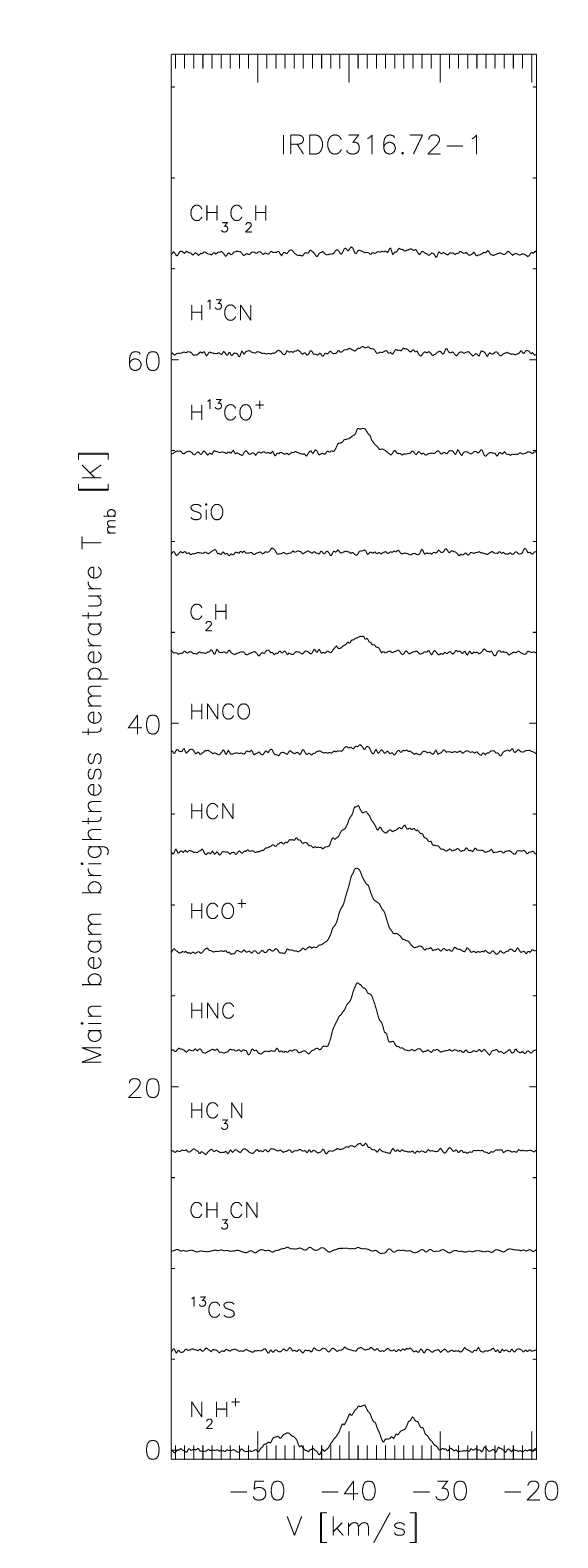}
\includegraphics[width=9cm]{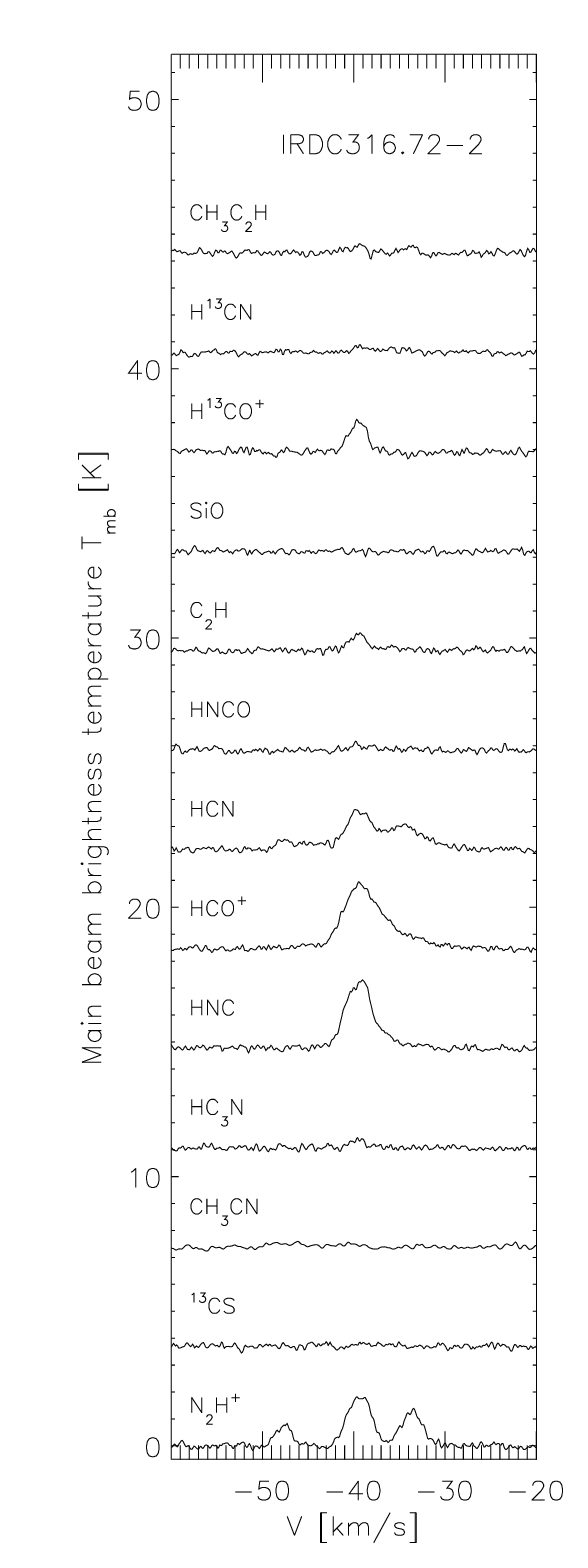}
\caption{Observed lines}
\label{appfig11}
\end{figure*}
\clearpage

\begin{figure*}
\centering
\includegraphics[width=9cm]{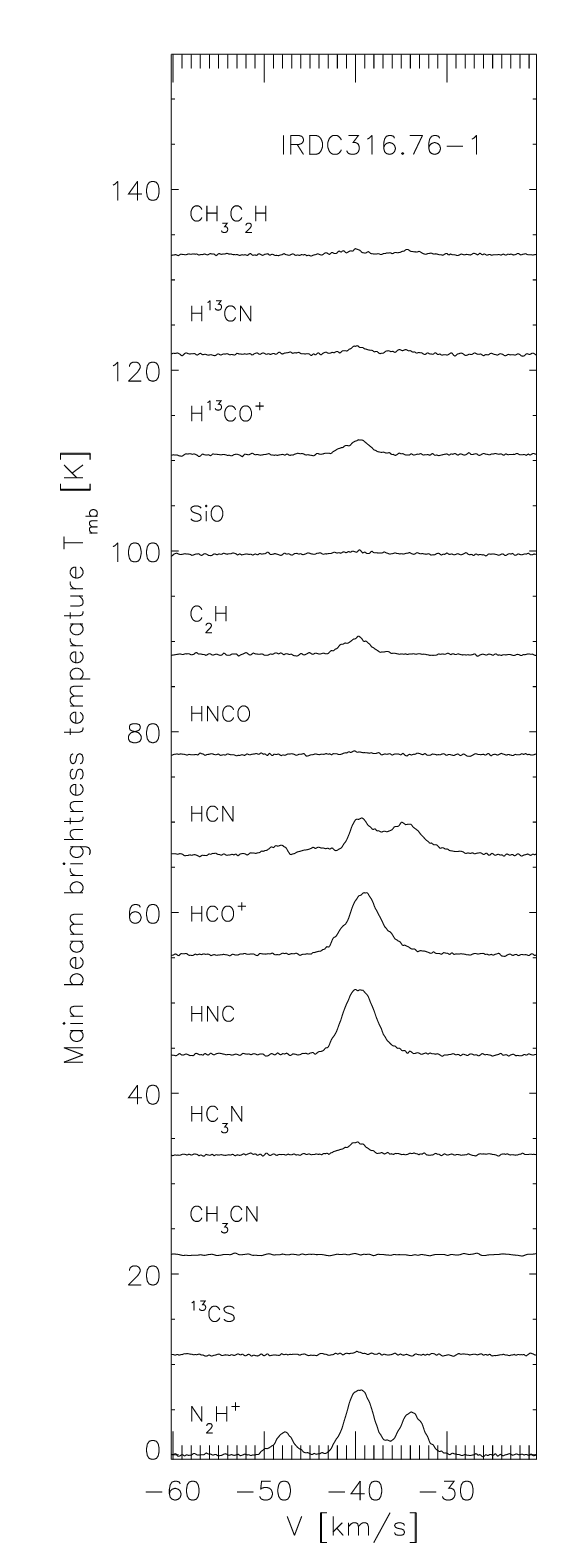}
\includegraphics[width=9cm]{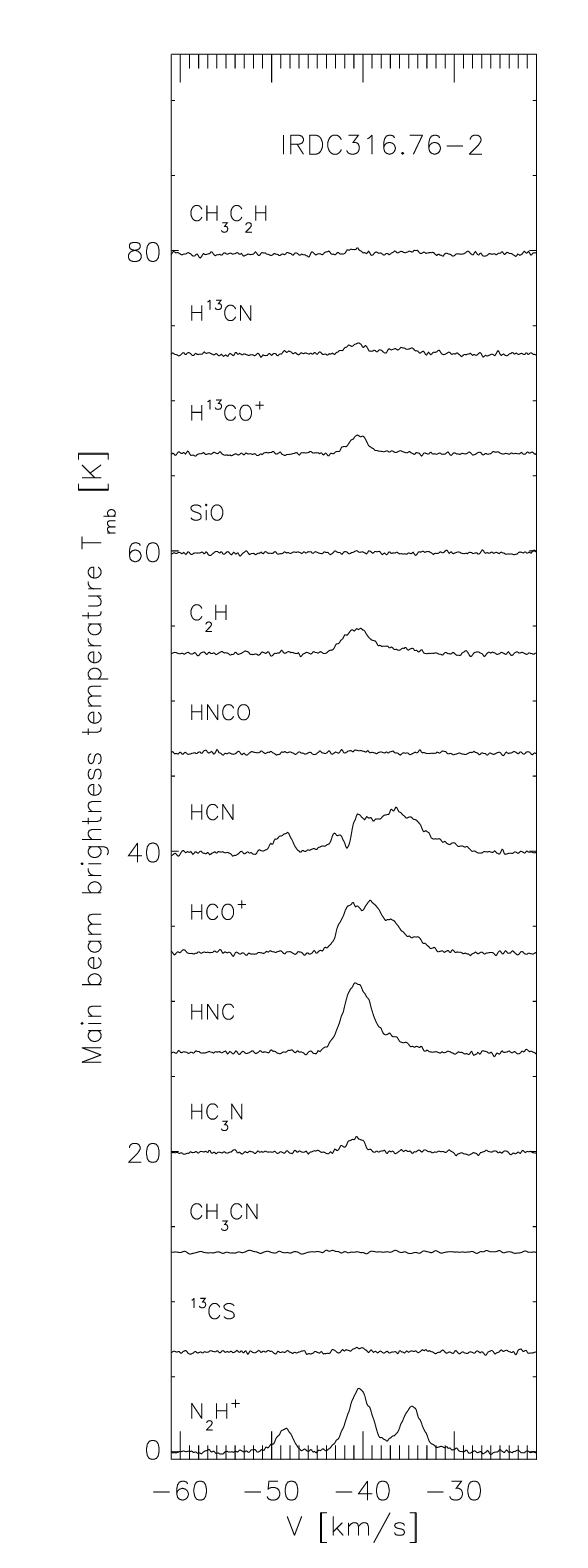}
\caption{Observed lines}
\label{appfig12}
\end{figure*}
\clearpage

\begin{figure*}
\centering
\includegraphics[width=9cm]{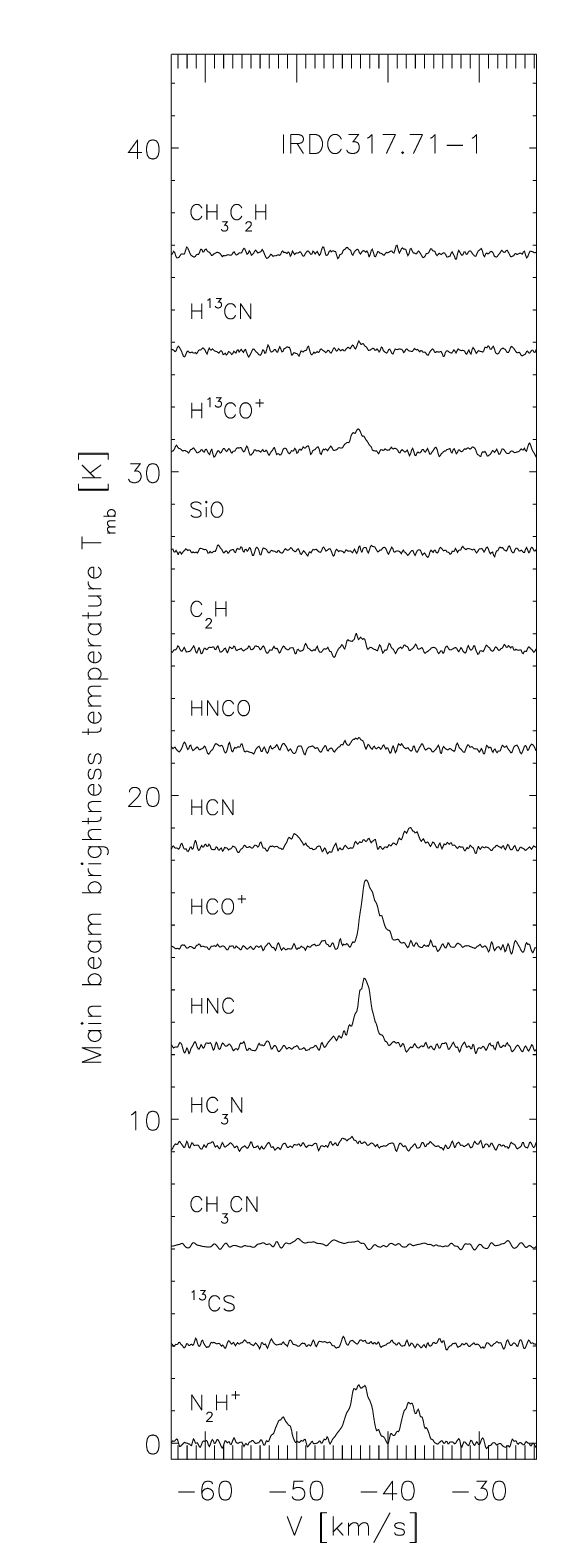}
\includegraphics[width=9cm]{IRDC317.71-2.ps}
\caption{Observed lines}
\label{appfig13}
\end{figure*}
\clearpage

\begin{figure*}
\centering
\includegraphics[width=9cm]{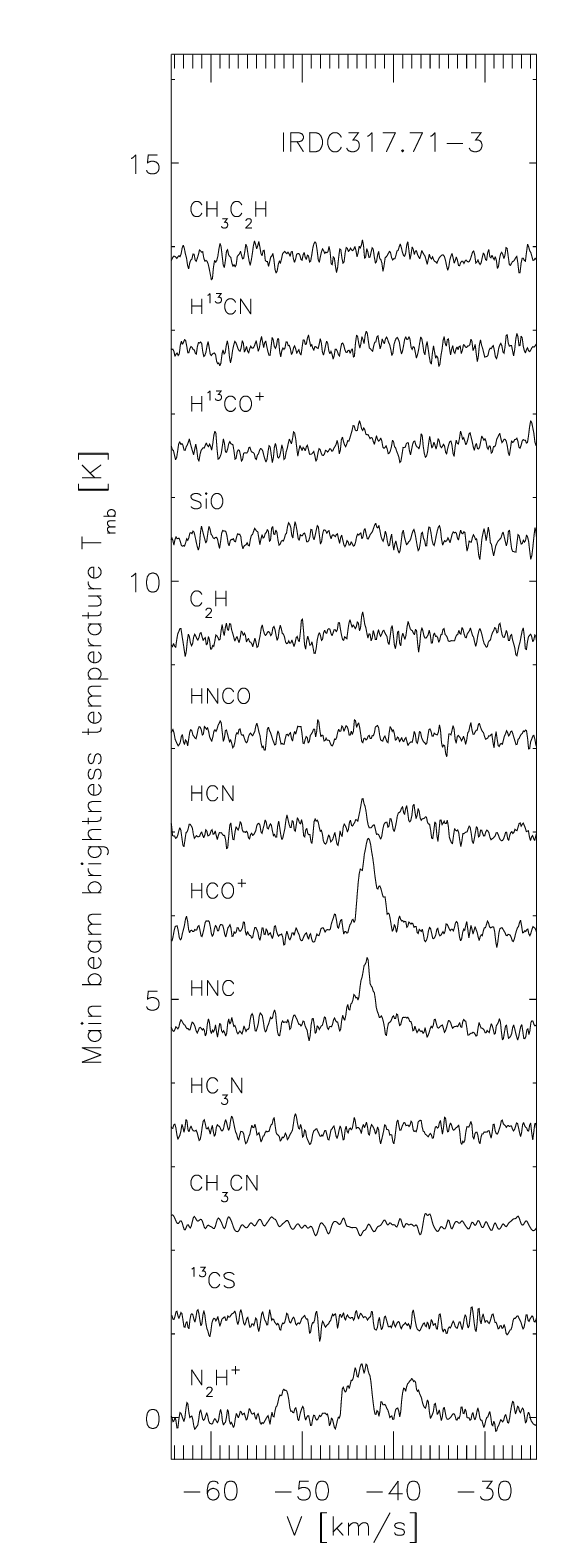}
\includegraphics[width=9cm]{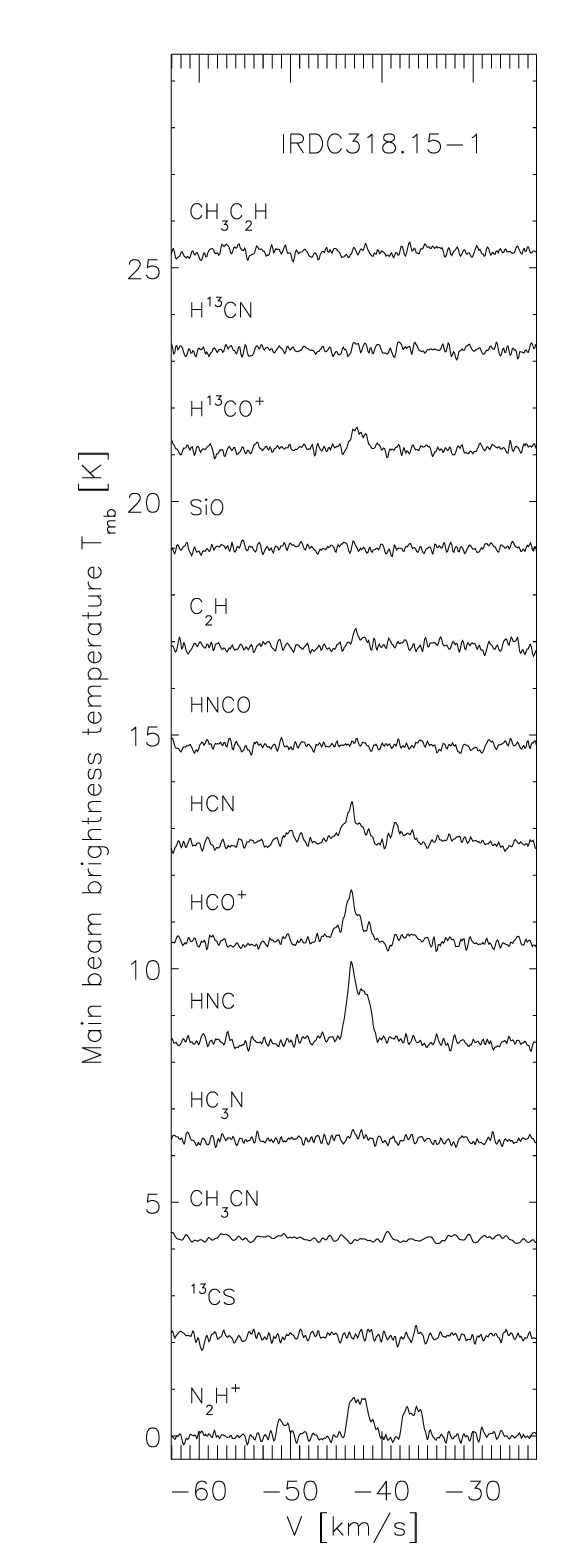}
\caption{Observed lines}
\label{appfig14}
\end{figure*}
\clearpage

\begin{figure*}
\centering
\includegraphics[width=9cm]{IRDC318.15-2.ps}
\includegraphics[width=9cm]{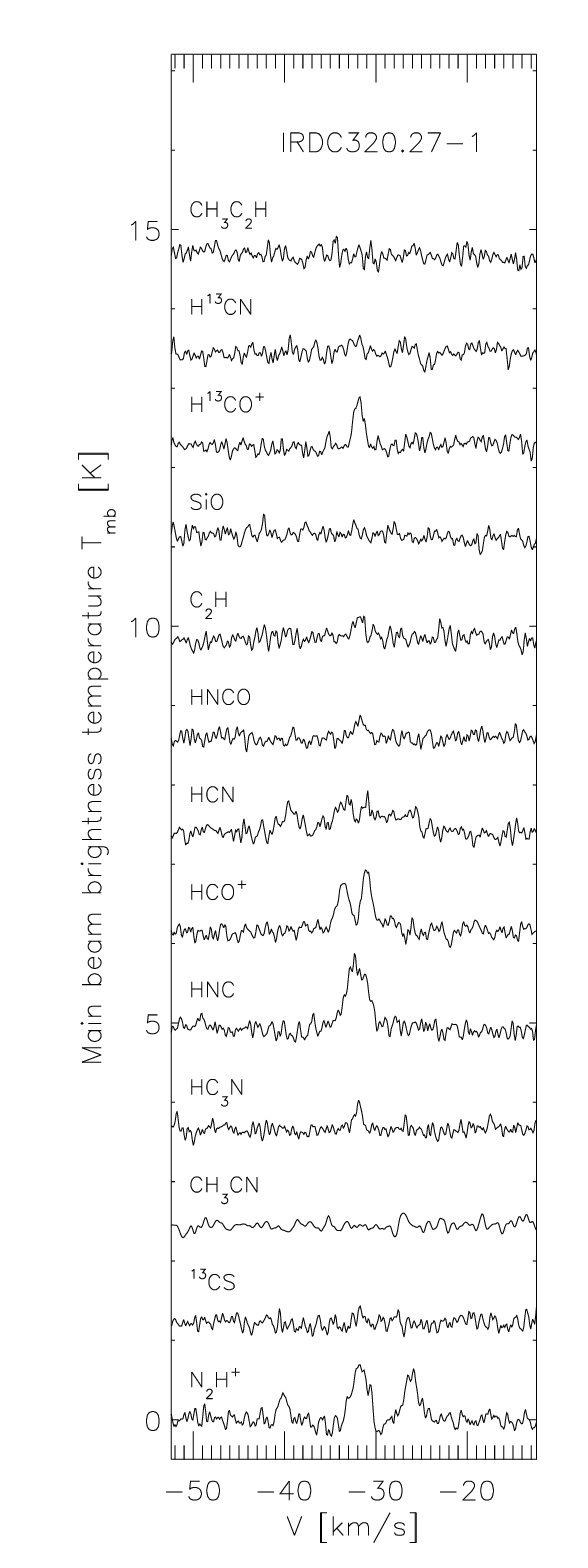}
\caption{Observed lines}
\label{appfig15}
\end{figure*}
\clearpage

\begin{figure*}
\centering
\includegraphics[width=9cm]{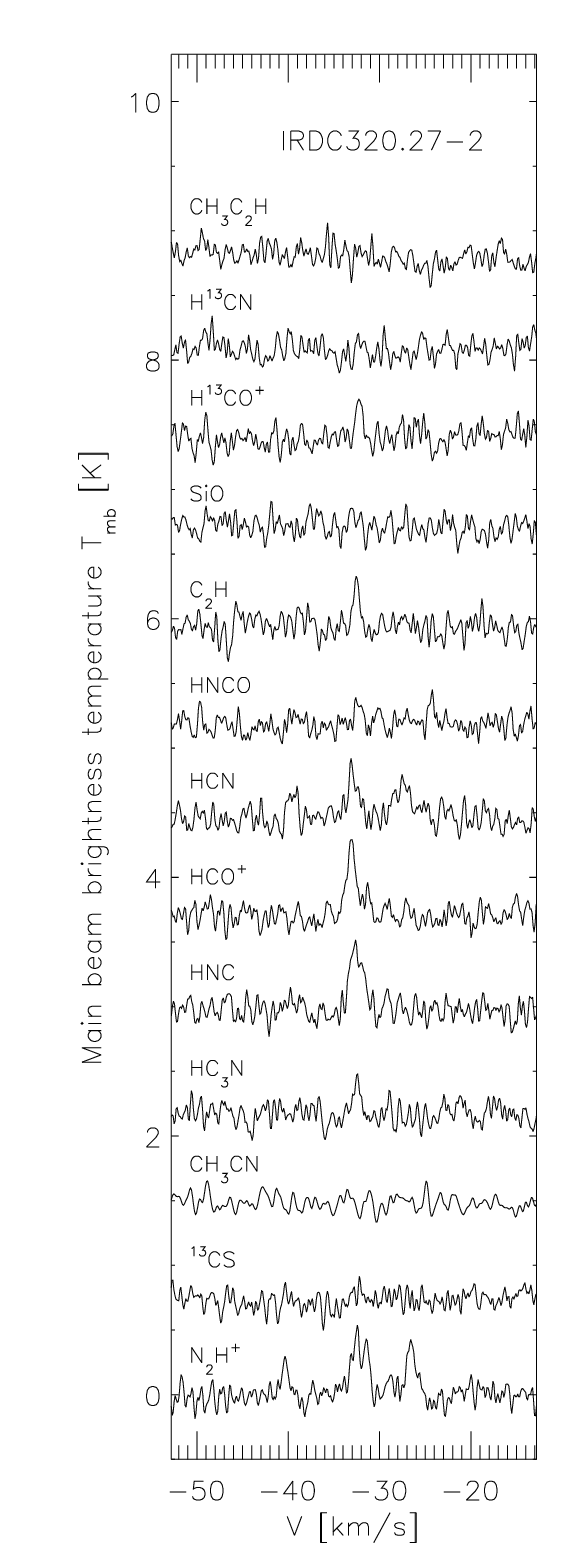}
\includegraphics[width=9cm]{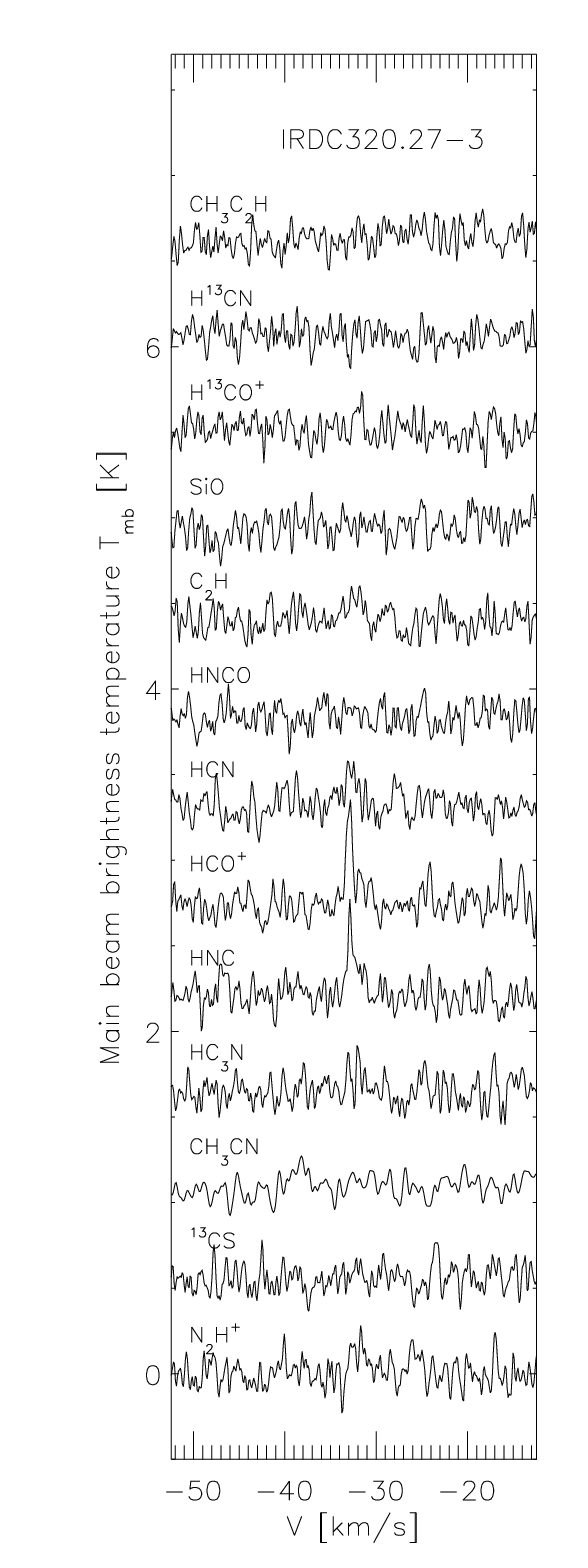}
\caption{Observed lines}
\label{appfig16}
\end{figure*}
\clearpage

\begin{figure*}
\centering
\includegraphics[width=9cm]{IRDC321.73-1.ps}
\includegraphics[width=9cm]{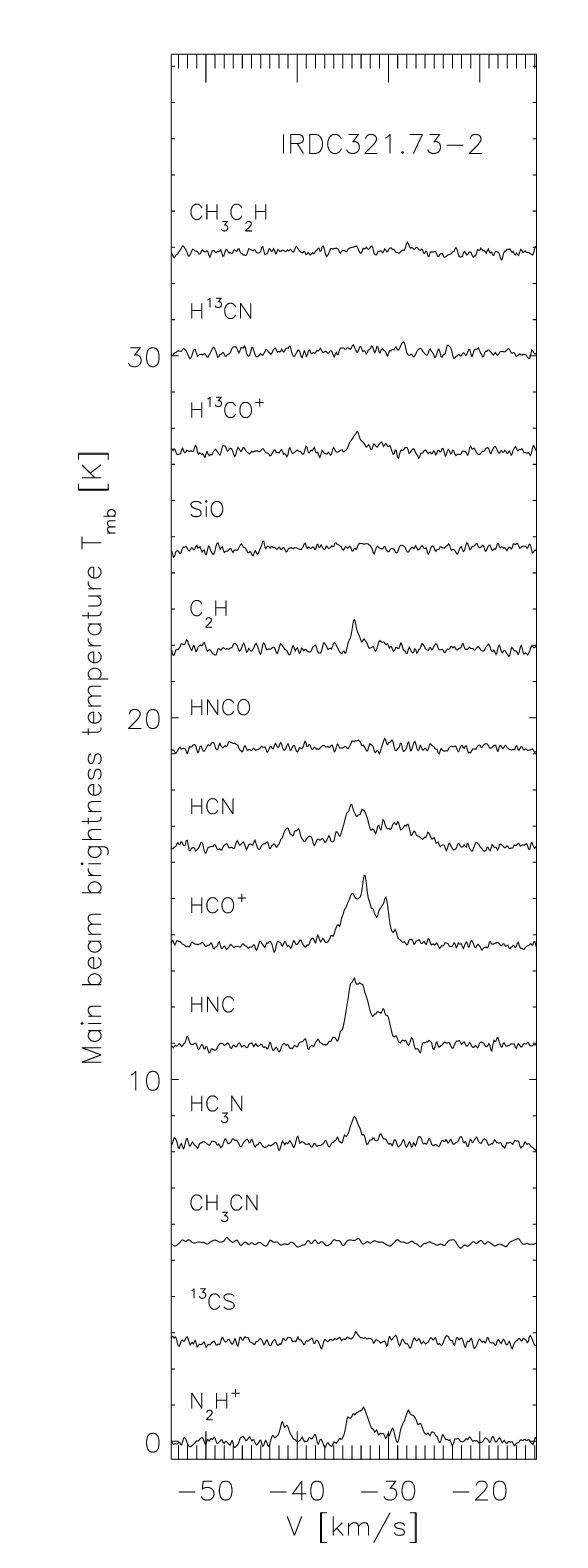}
\caption{Observed lines}
\label{appfig17}
\end{figure*}
\clearpage

\begin{figure*}
\centering
\includegraphics[width=9cm]{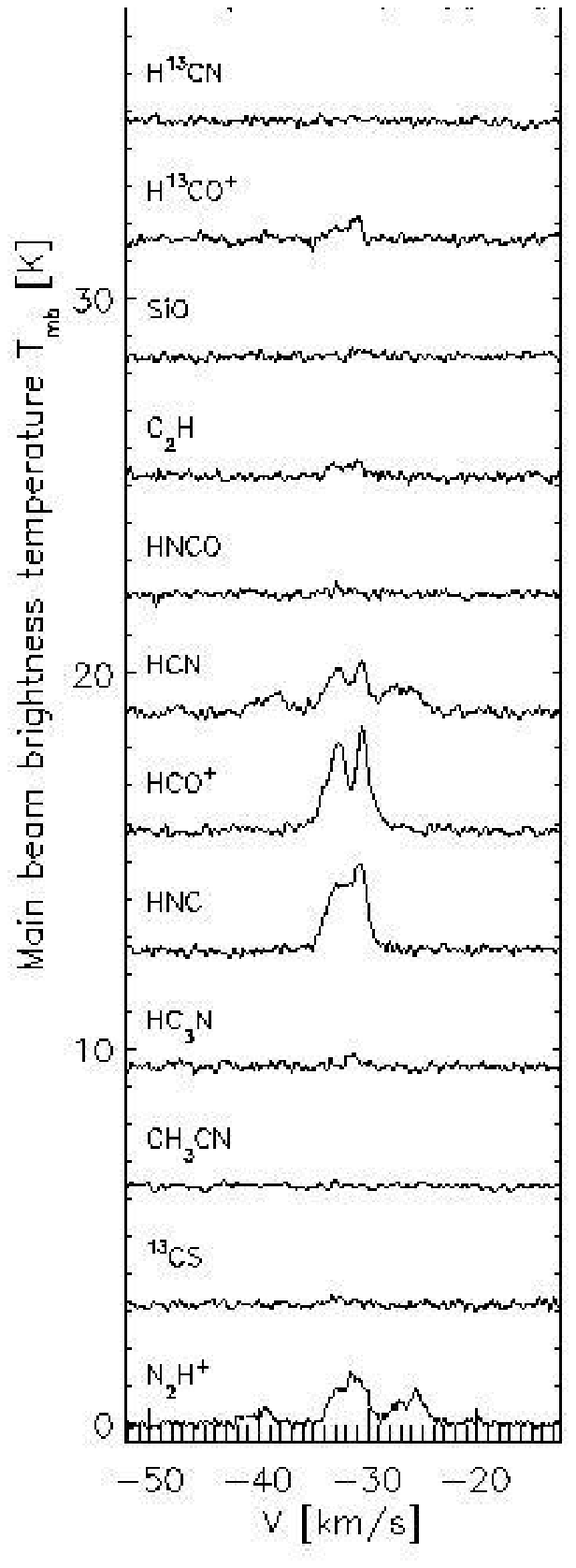}
\caption{Observed lines}
\label{appfig18}
\end{figure*}
\clearpage

\end{appendix}


\begin{appendix} 
\section{Line parameters.}

\begin{table*}

\caption{N$_2$H$^+$ line parameters.} 
\label{table:N2H}      
\begin{tabular}{r c c c c c  }        
\hline\hline             
\noalign{\smallskip}
Name &V$_{lsr}$  & $\Delta$ V  & $\tau_{main}$ & 1$\sigma$ rms \\
     & (km/s)	 &  (km/s)     &	       & (K)	       \\   

\noalign{\smallskip}   
\hline                        
\noalign{\smallskip}

IRDC308.13-1 &   -46.8  (0.09) &   3.0   (0.24) &   	         & 0.15 \\ 
IRDC308.13-2 &   -47.9  (0.04) &   0.7   (0.11) &   	         & 0.16 \\
IRDC308.13-3 &   -47.8  (0.03) &   0.9   (0.08) &   	         & 0.14 \\
IRDC309.13-1 &   -44.7  (0.04) &   1.8   (0.16) &   	         & 0.15 \\
IRDC309.13-2 &   -44.6  (0.06) &   2.3   (0.12) &   	         & 0.13 \\
IRDC309.13-3 &   -45.3  (0.04) &   0.9   (0.13) &   	         & 0.13 \\
IRDC309.37-1 &   -42.5  (0.02) &   2.4   (0.00) &   0.2  (0.007) & 0.42 \\
IRDC309.37-3 &   -42.5  (0.05) &   2.4   (0.13) &   	         & 0.13 \\
IRDC310.39-1 &   -52.5  (0.01) &   2.6   (0.06) &   0.9  (0.260) & 0.12  \\
IRDC310.39-2 &   -52.5  (0.03) &   1.8   (0.12) &   1.6  (0.744) & 0.13 \\
IRDC312.36-1 &   -50.7  (0.03) &   2.4   (0.08) &   	         & 0.12 \\
IRDC312.36-2 &   -51.1  (0.02) &   0.7   (0.04) &   	         & 0.11 \\
IRDC313.72-1 &   -44.1  (0.02) &   2.3   (0.09) &   	         & 0.12 \\
IRDC313.72-2 &   -44.0  (0.01) &   2.0   (0.07) &   	         & 0.12 \\
IRDC313.72-3 &   -43.5  (0.02) &   1.7   (0.09) &   	         & 0.12 \\
IRDC313.72-4 &   -41.2  (0.03) &   2.1   (0.10) &   	         & 0.12 \\
IRDC316.72-1 &   -39.0  (0.01) &   2.9   (0.04) &   1.9  (0.214) & 0.12  \\
IRDC316.76-1 &   -39.7  (0.00) &   2.8   (0.00) &   0.8  (0.015) & 0.12 \\
IRDC316.72-2 &   -39.5  (0.02) &   2.2   (0.06) &   2.3  (0.372) & 0.12  \\
IRDC316.76-2 &   -40.5  (0.01) &   2.5   (0.03) &   1.4  (0.166) & 0.12 \\
IRDC317.71-1 &   -43.3  (0.02) &   2.1   (0.06) &   1.8  (0.389) & 0.12  \\
IRDC317.71-2 &   -43.6  (0.00) &   2.4   (0.00) &   0.9  (0.129) & 0.12 \\
IRDC317.71-3 &   -43.9  (0.01) &   1.8   (0.14) &   1.4  (0.156) & 0.12 \\
IRDC320.27-1 &   -31.9  (0.04) &   1.0   (0.09) &   4.3  (1.918) & 0.13 \\
IRDC320.27-2 &   -32.4  (0.02) &   0.5   (0.06) & 	         & 0.13 \\
IRDC320.27-3 &   -32.0  (0.06) &   0.5   (0.06) & 	         & 0.07 \\
IRDC321.73-1 &   -32.3  (0.01) &   1.7   (0.06) &   1.4  (0.386) & 0.13  \\
IRDC321.73-2 &   -33.3  (0.03) &   1.5   (0.13) &   4.5  (1.172) & 0.12 \\
IRDC321.73-3 &   -31.8  (0.03) &   2.9   (0.07) &   	         & 0.12 \\
IRDC013.90-1 &    23.1  (0.03) &   1.6   (0.10) &   1.8  (0.635) & 0.12 \\
IRDC013.90-2 &    22.5  (0.03) &   1.3   (0.10) &   2.6  (0.967) & 0.11 \\
IRDC316.45-1 &   -43.5  (0.02) &   1.4   (0.08) &   5.4  (0.994) & 0.13 \\
IRDC316.45-2 &   -43.4  (0.03) &   1.6   (0.10) &   4.3  (1.063) & 0.14 \\
IRDC318.15-1 &   -42.6  (0.03) &   1.4   (0.11) &   3.4  (1.273) & 0.12 \\
IRDC318.15-2 &   -42.0  (0.02) &   1.0   (0.08) &   2.7  (1.093) & 0.12 \\
IRDC309.94-1 &   -59.2  (0.01) &   2.5   (0.05) &   1.1  (0.246) & 0.16 \\

\noalign{\smallskip}   
\hline                                   
\end{tabular}

Notes: Columns are name, LSR velocity, full linewidth at half
	maximum, main group optical depth, 1$\sigma$ rms values.

\end{table*}

\begin{table*}

\caption{$^{13}$CS line parameters.} 
\label{table:13CS}      
\begin{tabular}{r c c c c c }        
\hline\hline             
\noalign{\smallskip}
Name & $\int \! T_{MB} \, dv $  & V$_{lsr}$  & $\Delta$ V  & T $_{mb}$ & 1$\sigma$ rms \\
& (K km/s)  & (km/s) & (km/s) & (K ) & (K) \\     

\noalign{\smallskip}   
\hline                        
\noalign{\smallskip}

IRDC313.72-2 &0.3  (0.02)&   -44.1 (0.1)&    1.3 (0.2) &  0.2 & 0.06 \\
IRDC316.76-1 &0.4  (0.01)&   -39.8 (0.1)&    1.0 (0.0) &  0.4 & 0.06 \\
IRDC316.76-2 &0.5  (0.02)&   -40.6 (0.1)&    1.9 (0.2) &  0.3 & 0.06 \\

\noalign{\smallskip}   
\hline                                   
\end{tabular}

Notes: Columns are name, integrated area obtained from the gaussian fits, LSR velocity, full linewidth at half
	maximum, main beam brightness temperature, 1$\sigma$ rms values.

\end{table*}


\begin{table*}

\caption{HC$_3$N line parameters.} 
\label{table:HCCCN }      
\begin{tabular}{r c c c c c  }        
\hline\hline             
\noalign{\smallskip}
Name & $\int \! T_{MB} \, dv $  & V$_{lsr}$  & $\Delta$ V   & T $_{mb}$ & 1$\sigma$ rms  \\
& (K km/s)  & (km/s)  & (km/s)  & (K )  & (K)\\     

\noalign{\smallskip}   
\hline                        
\noalign{\smallskip}

IRDC309.37-1 &   2.1 (0.09)&  -42.6 (0.10)&   2.3   (0.2)&  0.8 & 0.42  \\
IRDC310.39-1 &   1.0 (0.03)&  -52.8 (0.06)&   2.4   (0.1)&  0.4 & 0.13  \\
IRDC310.39-2 &   0.4 (0.02)&  -52.5 (0.07)&   1.5   (0.2)&  0.2 & 0.13  \\
IRDC312.36-1 &   0.3 (0.02)&  -50.7 (0.09)&   1.4   (0.2)&  0.2 & 0.12  \\
IRDC313.72-1 &   1.0 (0.03)&  -44.0 (0.09)&   3.0   (0.2)&  0.3 & 0.13 \\
IRDC313.72-2 &   1.0 (0.02)&  -44.2 (0.05)&   2.1   (0.1)&  0.4 & 0.12  \\
IRDC316.72-1 &   1.0 (0.02)&  -39.0 (0.07)&   2.6   (0.1)&  0.3 & 0.12  \\
IRDC316.76-1 &   4.2 (0.03)&  -39.9 (0.02)&   3.0   (0.1)&  1.2 & 0.12  \\
IRDC316.72-2 &   0.5 (0.02)&  -39.5 (0.06)&   1.8   (0.1)&  0.3 & 0.12  \\
IRDC316.76-2 &   2.3 (0.02)&  -40.8 (0.02)&   2.2   (0.0)&  0.9 & 0.12  \\
IRDC317.71-1 &   0.5 (0.02)&  -44.2 (0.12)&   2.3   (0.2)&  0.2 & 0.13  \\
IRDC317.71-2 &   3.0 (0.02)&  -44.1 (0.01)&   2.0   (0.0)&  1.4 & 0.13  \\
IRDC320.27-1 &   0.2 (0.01)&  -31.9 (0.05)&   0.8   (0.1)&  0.3 & 0.13  \\
IRDC321.73-1 &   1.3 (0.02)&  -32.3 (0.03)&   1.8   (0.1)&  0.6 & 0.13  \\
IRDC321.73-2 &   1.0 (0.02)&  -33.7 (0.03)&   1.4   (0.1)&  0.6 & 0.13  \\
IRDC321.73-3 &   0.7 (0.04)&  -31.3 (0.10)&   2.4   (0.5)&  0.2 & 0.13  \\
IRDC013.90-1 &   0.4 (0.02)&   22.8 (0.05)&   1.2   (0.1)&  0.3 & 0.12  \\
IRDC316.45-1 &   0.6 (0.03)&  -43.4 (0.09)&   2.2   (0.2)&  0.2 & 0.15  \\
IRDC309.94-1 &   2.7 (0.04)&  -59.5 (0.04)&   2.9   (0.1)&  0.8 & 0.16 \\

\noalign{\smallskip}   
\hline                                   
\end{tabular}

Notes: Columns are name, integrated area obtained from the gaussian fits, LSR velocity, full linewidth at half
	maximum, main beam brightness temperature, 1$\sigma$ rms values.

\end{table*}


\begin{table*}

\caption{HNC line parameters.} 
\label{table:HNC}      
\begin{tabular}{r c c c c c }        
\hline\hline             
\noalign{\smallskip}
Name & $\int \! T_{MB} \, dv $  & V$_{lsr}$  & $\Delta$ V   & T $_{mb}$ & 1$\sigma$ rms\\
& (K km/s)  & (km/s)  & (km/s)  & (K )  & (K)\\   

\noalign{\smallskip}   
\hline                        
\noalign{\smallskip}

IRDC308.13-1 &   3.5  (0.02)& -47.3 (0.02) &  2.1 (0.00)&  1.5 & 0.16 \\ 
IRDC308.13-2 &   2.1  (0.03)& -47.9 (0.02) &  1.7 (0.07)&  1.1 & 0.17 \\
IRDC308.13-3 &   2.1  (0.03)& -47.8 (0.02) &  1.8 (0.08)&  1.0 & 0.16 \\
IRDC309.13-1 &   3.2  (0.04)& -44.4 (0.04) &  3.4 (0.10)&  0.8 & 0.15 \\
IRDC309.13-2 &   2.4  (0.04)& -45.2 (0.06) &  3.7 (0.15)&  0.6 & 0.14 \\
IRDC309.13-3 &   1.1  (0.05)& -45.6 (0.15) &  3.4 (0.40)&  0.3 & 0.16 \\
IRDC309.37-1 &  12.2  (0.10)& -43.1 (0.02) &  2.5 (0.05)&  4.4 & 0.41 \\
IRDC309.37-2 &   1.0  (0.03)& -43.5 (0.08) &  2.7 (0.20)&  0.3 & 0.13 \\
IRDC309.37-3 &   4.3  (0.03)& -43.4 (0.02) &  2.7 (0.06)&  1.5 & 0.13 \\
IRDC310.39-1 &   7.9  (0.03)& -52.6 (0.01) &  3.4 (0.03)&  2.1 & 0.13 \\
IRDC310.39-2 &   5.0  (0.02)& -52.5 (0.01) &  2.5 (0.03)&  1.8 & 0.13 \\
IRDC312.36-1 &   4.5  (0.03)& -50.8 (0.01) &  2.7 (0.04)&  1.5 & 0.13 \\
IRDC312.36-2 &   1.6  (0.04)& -49.4 (0.13) &  5.3 (0.36)&  0.2 & 0.12 \\
IRDC313.72-1 &   4.7  (0.02)& -44.5 (0.01) &  2.0 (0.00)&  2.2 & 0.12 \\
IRDC313.72-2 &   5.5  (0.03)& -44.4 (0.01) &  2.5 (0.03)&  2.0 & 0.12 \\
IRDC313.72-3 &   6.5  (0.02)& -43.8 (0.01) &  2.5 (0.02)&  2.3 & 0.12 \\
IRDC313.72-4 &   4.1  (0.02)& -44.0 (0.01) &  2.0 (0.00)&  1.9 & 0.12 \\
IRDC316.72-1 &  16.6  (0.03)& -38.8 (0.01) &  4.2 (0.02)&  3.6 & 0.12 \\
IRDC316.76-1 &  32.7  (0.03)& -39.6 (0.01) &  4.1 (0.01)&  7.3 & 0.12 \\
IRDC316.72-2 &   9.4  (0.03)& -39.3 (0.01) &  3.6 (0.03)&  2.4 & 0.12 \\
IRDC316.76-2 &  19.4  (0.04)& -40.4 (0.01) &  4.1 (0.02)&  4.4 & 0.13 \\
IRDC317.71-1 &   4.3  (0.02)& -42.5 (0.01) &  2.0 (0.00)&  2.0 & 0.13 \\
IRDC317.71-2 &   6.1  (0.02)& -42.7 (0.01) &  1.8 (0.00)&  3.1 & 0.12 \\
IRDC317.71-3 &   1.4  (0.02)& -43.0 (0.03) &  1.8 (0.00)&  0.7 & 0.13 \\
IRDC320.27-1 &   2.4  (0.03)& -32.1 (0.03) &  2.8 (0.09)&  0.8 & 0.13 \\
IRDC320.27-2 &   0.9  (0.02)& -32.5 (0.04) &  1.7 (0.10)&  0.4 & 0.13 \\
IRDC320.27-3 &   0.5  (0.03)& -32.6 (0.07) &  1.3 (0.24)&  0.3 & 0.13 \\
IRDC321.73-1 &   6.6  (0.03)& -32.4 (0.01) &  2.6 (0.02)&  2.3 & 0.13 \\
IRDC321.73-2 &   5.8  (0.02)& -33.1 (0.01) &  3.0 (0.00)&  1.8 & 0.13 \\
IRDC321.73-3 &   8.3  (0.03)& -31.7 (0.01) &  3.7 (0.03)&  2.1 & 0.13 \\
IRDC013.90-1 &   5.4  (0.04)&  23.1 (0.02) &  2.4 (0.05)&  2.0 & 0.12 \\
IRDC013.90-2 &   1.8  (0.03)&  23.1 (0.05) &  3.1 (0.11)&  0.5 & 0.12 \\
IRDC316.45-1 &   4.5  (0.03)& -43.5 (0.02) &  3.3 (0.05)&  1.2 & 0.15 \\
IRDC316.45-2 &   4.7  (0.03)& -43.3 (0.02) &  3.2 (0.05)&  1.3 & 0.14 \\
IRDC318.15-1 &   3.7  (0.02)& -42.7 (0.01) &  2.5 (0.04)&  1.4 & 0.13 \\
IRDC318.15-2 &   3.3  (0.02)& -42.0 (0.01) &  2.1 (0.04)&  1.4 & 0.13 \\
IRDC309.94-1 &   8.5  (0.05)& -59.5 (0.02) &  4.2 (0.05)&  1.8 & 0.16 \\

\noalign{\smallskip}   
\hline                                   
\end{tabular}

Notes: Columns are name, integrated area obtained from the gaussian fits, LSR velocity, full linewidth at half
	maximum, main beam brightness temperature, 1$\sigma$ rms values.

\end{table*}


\begin{table*}

\caption{HCO$^+$ line parameters.} 
\label{table:HCO}      
\begin{tabular}{r c c c c c }        
\hline\hline             
\noalign{\smallskip}
Name & $\int \! T_{MB} \, dv $  & V$_{lsr}$  & $\Delta$ V   & T $_{mb}$ & 1$\sigma$ rms \\
& (K km/s)  & (km/s)  & (km/s)  & (K )  & (K )\\   

\noalign{\smallskip}   
\hline                        
\noalign{\smallskip}

IRDC308.13-1 &  4.7 (0.04)&-47.6 (0.02)&  2.7 (0.07)&  1.5 & 0.17 \\ 
IRDC308.13-2 &  3.4 (0.03)&-47.9 (0.02)&  2.1 (0.05)&  1.5 & 0.16\\
IRDC308.13-3 &  3.0 (0.03)&-48.0 (0.02)&  2.3 (0.07)&  1.2 & 0.16\\
IRDC309.13-1 &  4.1 (0.07)&-45.0 (0.15)&  9.1 (0.41)&  0.4 & 0.16\\
IRDC309.13-2 &  3.2 (0.05)&-46.2 (0.09)&  5.6 (0.20)&  0.5 & 0.14\\
IRDC309.13-3 &  1.6 (0.03)&-47.3 (0.05)&  2.6 (0.14)&  0.5 & 0.15\\
IRDC309.37-1 & 13.9 (0.14)&-43.4 (0.02)&  2.8 (0.08)&  4.5 & 0.31\\
IRDC309.37-2 &  1.8 (0.03)&-43.6 (0.05)&  2.8 (0.13)&  0.6 & 0.14\\
IRDC309.37-3 &  5.9 (0.05)&-43.7 (0.02)&  2.2 (0.06)&  2.4 & 0.13\\
IRDC310.39-1 &  8.5 (0.05)&-53.0 (0.02)&  2.3 (0.00)&  3.4 & 0.12\\
IRDC310.39-2 &  6.4 (0.02)&-52.8 (0.01)&  2.0 (0.00)&  3.0 & 0.12\\
IRDC312.36-1 &  5.7 (0.03)&-51.0 (0.01)&  2.7 (0.04)&  1.9 & 0.12\\
IRDC312.36-2 &  1.1 (0.04)&-51.4 (0.21)&  5.7 (0.44)&  0.1 & 0.12\\
IRDC313.72-1 &  7.2 (0.06)&-44.8 (0.02)&  2.5 (0.00)&  2.7 & 0.11\\
IRDC313.72-2 &  7.1 (0.04)&-44.8 (0.02)&  2.5 (0.00)&  2.7 & 0.13\\
IRDC313.72-3 &  8.1 (0.04)&-44.2 (0.01)&  2.6 (0.03)&  2.8 & 0.12\\
IRDC313.72-4 &  7.0 (0.03)&-44.5 (0.01)&  2.5 (0.00)&  2.6 & 0.12\\
IRDC316.72-1 & 19.6 (0.04)&-38.7 (0.01)&  4.0 (0.00)&  4.6 & 0.12\\
IRDC316.76-1 & 30.0 (0.04)&-39.0 (0.01)&  4.0 (0.00)&  7.0 & 0.12\\
IRDC316.72-2 & 12.9 (0.04)&-38.9 (0.01)&  5.4 (0.05)&  2.2 & 0.12\\
IRDC316.76-2 & 23.2 (0.04)&-39.3 (0.01)&  6.5 (0.03)&  3.3 & 0.12\\
IRDC317.71-1 &  4.5 (0.02)&-41.9 (0.01)&  2.1 (0.00)&  2.0 & 0.13\\
IRDC317.71-2 &  4.5 (0.02)&-42.1 (0.01)&  2.1 (0.00)&  2.0 & 0.12\\
IRDC317.71-3 &  2.3 (0.02)&-42.7 (0.02)&  2.1 (0.00)&  1.0 & 0.12\\
IRDC320.27-1 &  1.1 (0.03)&-31.0 (0.04)&  1.5 (0.00)&  0.7 & 0.12\\
IRDC320.27-2 &  0.7 (0.02)&-33.0 (0.03)&  1.3 (0.10)&  0.5 & 0.12\\
IRDC320.27-3 &  0.4 (0.01)&-32.9 (0.02)&  0.7 (0.06)&  0.5 & 0.14\\
IRDC321.73-1 &  9.2 (0.03)&-32.7 (0.01)&  2.4 (0.02)&  3.4 & 0.13\\
IRDC321.73-2 &  8.0 (0.04)&-32.7 (0.02)&  5.0 (0.06)&  1.4 & 0.12\\
IRDC321.73-3 &  4.3 (0.10)&-30.6 (0.05)&  1.5 (0.00)&  2.7 & 0.14\\
IRDC013.90-1 &  5.4 (0.03)& 23.1 (0.01) & 2.4 (0.04) & 2.1 & 0.45\\
IRDC013.90-2 &  2.1 (0.04)& 23.5 (0.07) & 3.4 (0.12) & 0.5 & 0.20\\
IRDC316.45-1 &  4.7 (0.04)&-43.2 (0.03)&  4.3 (0.09)&  1.0 & 0.12\\
IRDC316.45-2 &  3.6 (0.03)&-43.3 (0.04)&  4.1 (0.09)&  0.8 & 0.13\\
IRDC318.15-1 &  1.9 (0.02)&-43.3 (0.02)&  1.8 (0.00)&  0.9 & 0.12\\
IRDC318.15-2 &  0.9 (0.02)&-43.2 (0.06)&  1.8 (0.00)&  0.5 & 0.12\\
IRDC309.94-1 &  9.9 (0.06)&-60.3 (0.03)&  5.4 (0.08)&  1.7 & 0.15\\

\noalign{\smallskip}   
\hline                                   
\end{tabular}

Notes: Columns are name, integrated area obtained from the gaussian fits, LSR velocity, full linewidth at half
	maximum, main beam brightness temperature, 1$\sigma$ rms values.

\end{table*}


\begin{table*}

\caption{HNCO line parameters.} 
\label{table:HNCO}      
\begin{tabular}{r c c c c c }        
\hline\hline             
\noalign{\smallskip}
Name & $\int \! T_{MB} \, dv $  & V$_{lsr}$  & $\Delta$ V   & T $_{mb}$ & 1$\sigma$ rm \\
& (K km/s)  & (km/s)  & (km/s)  & (K ) & (K )\\   

\noalign{\smallskip}   
\hline                        
\noalign{\smallskip}

IRDC309.37-1 &  1.3   (0.07) & -42.4   (0.12) &  2.3   (0.36) &  0.5  & 0.15 \\
IRDC310.39-1 &  0.3   (0.02) & -53.8   (0.08) &  1.3   (0.20) &  0.2  & 0.07 \\
IRDC312.36-1 &  0.4   (0.02) & -50.4   (0.14) &  2.5   (0.26) &  0.1  & 0.06 \\
IRDC313.72-1 &  0.5   (0.02) & -44.3   (0.06) &  1.4   (0.19) &  0.3  & 0.06 \\
IRDC313.72-2 &  0.6   (0.02) & -44.1   (0.13) &  2.9   (0.30) &  0.1  & 0.05 \\
IRDC313.72-3 &  0.3   (0.02) & -43.6   (0.11) &  2.0   (0.36) &  0.1  & 0.06 \\
IRDC316.72-1 &  0.9   (0.03) & -39.1   (0.08) &  2.6   (0.18) &  0.3  & 0.06 \\
IRDC316.76-1 &  1.0   (0.03) & -39.7   (0.10) &  3.3   (0.29) &  0.2  & 0.06 \\
IRDC317.71-1 &  0.6   (0.03) & -43.5   (0.09) &  2.0   (0.23) &  0.3  & 0.07 \\
IRDC317.71-2 &  1.1   (0.04) & -43.7   (0.10) &  3.4   (0.31) &  0.3  & 0.06 \\
IRDC320.27-1 &  0.3   (0.01) & -31.7   (0.07) &  1.3   (0.17) &  0.2  & 0.06 \\
IRDC316.45-1 &  0.3   (0.02) & -43.4   (0.08) &  1.4   (0.19) &  0.2  & 0.07 \\
IRDC316.45-2 &  0.4   (0.02) & -43.2   (0.07) &  1.5   (0.16) &  0.3  & 0.07 \\
IRDC309.94-1 &  1.1   (0.03) & -59.4   (0.08) &  2.6   (0.22) &  0.3  & 0.07 \\

\noalign{\smallskip}   
\hline                                   
\end{tabular}

Notes: Columns are name, integrated area obtained from the gaussian fits, LSR velocity, full linewidth at half
	maximum, main beam brightness temperature, 1$\sigma$ rms values.

\end{table*}


\begin{table*}

\caption{SiO line parameters.} 
\label{table:SiO }      
\begin{tabular}{r c c c c c  }        
\hline\hline             
\noalign{\smallskip}
Name & $\int \! T_{MB} \, dv $  & V$_{lsr}$  & $\Delta$ V   & T $_{mb}$ & 1$\sigma$ rms  \\
& (K km/s) & (km/s)  & (km/s)  & (K ) & (K )\\     

\noalign{\smallskip}   
\hline                        
\noalign{\smallskip}

IRDC309.37-1 &  0.9  (0.08)&   -42.6  (0.2)&   2.8    (0.5)	&  0.3   & 0.18 \\
IRDC310.39-1 &  0.6  (0.03)&   -53.3  (0.2)&   3.9    (0.4)	&  0.1   & 0.08 \\
IRDC312.36-1 &  0.2  (0.03)&   -50.8  (0.2) &  2.0    (0.5)	&  0.1   & 0.08 \\
IRDC313.72-1 &  1.6  (0.04)&   -43.9  (0.1)&   4.5    (0.3)	&  0.3   & 0.08 \\
IRDC313.72-2 &  1.7  (0.05)&   -44.4  (0.1)&   5.1    (0.4)	&  0.3   & 0.08 \\
IRDC313.72-3 &  1.3  (0.06)&   -41.9  (0.3)&   8.2    (0.8)	&  0.1   & 0.08 \\
IRDC313.72-4 &  0.5  (0.03)&   -44.1  (0.2)&   3.4    (0.4)	&  0.1   & 0.08 \\
IRDC316.76-1 &  2.4  (0.07)&   -39.5  (0.2) &  8.9    (0.7)	&  0.2   & 0.08 \\
IRDC317.71-3 &  0.3  (0.03)&   -42.1  (0.2)&   1.9    (0.5)	&  0.1   & 0.08 \\
IRDC321.73-1 &  2.1  (0.05)&   -32.1  (0.1) &  4.8    (0.3)	&  0.4   & 0.08 \\
IRDC309.94-1 &  2.4  (0.05)&   -59.6  (0.2) &  5.0    (0.0)	&  0.4   & 0.10 \\

\noalign{\smallskip}   
\hline                                   
\end{tabular}

Notes: Columns are name, integrated area obtained from the gaussian fits, LSR velocity, full linewidth at half
	maximum, main beam brightness temperature, 1$\sigma$ rms values.

\end{table*}

\begin{table*}

\caption{H$^{13}$CO$^+$ line parameters.} 
\label{table:H13CO}      
\begin{tabular}{r c c c c c c c  }        
\hline\hline             
\noalign{\smallskip}
Name & $\int \! T_{MB} \, dv $  & V$_{lsr}$  & $\Delta$ V   & T $_{mb}$ & $\tau$ &1$\sigma$ rms & $\delta$V \\
& (K km/s)  & (km/s)  & (km/s)  & (K )&  & (K ) & \\     

\noalign{\smallskip}   
\hline                        
\noalign{\smallskip}
									    
IRDC308.13-1 & 0.6  (0.06) & -46.7  (0.28)&  2.5   (0.41) &  0.2 & 0.07  &    0.08  &	-0.8  (0.12)\\ 
IRDC309.13-1 & 0.8  (0.05) & -44.7  (0.14)&  2.1   (0.31) &  0.3 & 0.20  &    0.07  &	-0.2  (0.09) \\
IRDC309.13-3 & 0.5  (0.05) & -45.2  (0.12)&  1.5   (0.39) &  0.3 & 0.17  &    0.08  &	-2.1  (0.35)\\
IRDC309.37-1 & 2.6  (0.13) & -42.6  (0.09)&  2.1   (0.26) &  1.1 & 0.20  &    0.16  &	-0.8  (0.06) \\
IRDC309.37-3 & 1.3  (0.06) & -42.4  (0.15)&  3.2   (0.34) &  0.3 & 0.10  &    0.08  &	-1.2  (0.06) \\
IRDC310.39-1 & 1.9  (0.04) & -52.4  (0.04)&  1.9   (0.11) &  0.9 & 0.19  &    0.08  &	-0.6  (0.03) \\
IRDC310.39-2 & 1.2  (0.05) & -52.3  (0.09)&  2.2   (0.20) &  0.5 & 0.12  &    0.08  &	-0.4  (0.04) \\
IRDC312.36-1 & 1.3  (0.06) & -50.6  (0.13)&  3.0   (0.35) &  0.4 & 0.12  &    0.08  &	-0.3  (0.04) \\
IRDC312.36-2 & 0.2  (0.03) & -48.8  (0.20)&  1.4   (0.38) &  0.1 & 0.11  &    0.08  &	-2.5  (0.49)\\
IRDC313.72-1 & 1.1  (0.08) & -43.9  (0.19)&  2.5   (0.63) &  0.4 & 0.10  &    0.08  &	-0.8  (0.11) \\
IRDC313.72-2 & 1.1  (0.05) & -43.9  (0.10)&  2.2   (0.31) &  0.4 & 0.11  &    0.08  &	-0.8  (0.06) \\
IRDC313.72-3 & 0.7  (0.04) & -43.5  (0.10)&  1.8   (0.21) &  0.4 & 0.09  &    0.08  &	-0.7  (0.07) \\
IRDC313.72-4 & 0.8  (0.05) & -43.3  (0.12)&  2.2   (0.30) &  0.3 & 0.09  &    0.08  &	-1.2  (0.09) \\
IRDC316.72-1 & 4.4  (0.06) & -38.8  (0.04)&  3.1   (0.10) &  1.2 & 0.22  &    0.08  &	 0.1  (0.01) \\
IRDC316.76-1 & 5.2  (0.06) & -39.7  (0.03)&  3.2   (0.09) &  1.5 & 0.18  &    0.08  &	 0.7  (0.01) \\
IRDC316.72-2 & 2.7  (0.05) & -39.5  (0.04)&  2.2   (0.09) &  1.1 & 0.33  &    0.08  &	 0.6  (0.02) \\
IRDC316.76-2 & 3.1  (0.05) & -40.5  (0.04)&  2.4   (0.10) &  1.1 & 0.26  &    0.08  &	 1.1  (0.02) \\
IRDC317.71-1 & 1.3  (0.04) & -43.3  (0.07)&  1.9   (0.17) &  0.6 & 0.18  &    0.08  &	 1.3  (0.07) \\
IRDC317.71-2 & 1.7  (0.05) & -43.8  (0.05)&  1.9   (0.16) &  0.8 & 0.23  &    0.08  &	 1.7  (0.07) \\
IRDC317.71-3 & 0.4  (0.04) & -43.7  (0.20)&  1.7   (0.47) &  0.2 & 0.08  &    0.09  &	 1.0  (0.19)\\
IRDC320.27-1 & 0.7  (0.03) & -31.9  (0.05)&  1.1   (0.12) &  0.5 & 0.28  &    0.09  &	 0.8  (0.10)\\
IRDC321.73-1 & 1.6  (0.04) & -32.1  (0.04)&  1.5   (0.09) &  0.9 & 0.20  &    0.08  &	-0.5  (0.03) \\
IRDC321.73-2 & 0.6  (0.04) & -33.4  (0.08)&  1.3   (0.22) &  0.4 & 0.16  &    0.08  &	 0.7  (0.10)\\
IRDC321.73-3 & 1.2  (0.06) & -31.6  (0.14)&  2.5   (0.31) &  0.4 & 0.10  &    0.09  &	 1.0  (0.07) \\
IRDC013.90-1 & 1.0  (0.04) &  23.3  (0.09)&  1.9   (0.18) &  0.5 & 0.14  &    0.08  &	-0.2  (0.04) \\
IRDC013.90-2 & 0.3  (0.02) &  22.3  (0.03)&  0.4   (0.10) &  0.6 & 0.35  &    0.08  &	 1.2  (0.68)\\
IRDC316.45-1 & 0.8  (0.05) & -43.5  (0.12)&  1.9   (0.25) &  0.4 & 0.16  &    0.08  &	 0.2  (0.06) \\
IRDC316.45-2 & 1.1  (0.05) & -43.3  (0.10)&  2.2   (0.23) &  0.4 & 0.21  &    0.08  &	-0.0  (0.05)\\
IRDC318.15-1 & 0.7  (0.04) & -42.5  (0.09)&  1.6   (0.21) &  0.4 & 0.17  &    0.08  &	-0.7  (0.08)\\
IRDC318.15-2 & 0.6  (0.04) & -42.0  (0.10)&  1.5   (0.20) &  0.3 & 0.19  &    0.08  &	-1.1  (0.13)\\
IRDC309.94-1 & 1.7  (0.06) & -59.3  (0.07)&  2.1   (0.17) &  0.7 & 0.25  &    0.11  &	-0.9  (0.05)\\

\noalign{\smallskip}   
\hline                                   
\end{tabular}

Notes: Columns are name, integrated area obtained from the gaussian fits, LSR velocity, full linewidth at half
	maximum, main beam brightness temperature, optical depth, 1$\sigma$ rms values, 
	$\delta V =  (V_{HCO^+}-V_{H^{13}CO^+}) / \Delta V_{H^{13}CO^+} $.

\end{table*}


\begin{table*}

\caption{Integrated areas,  measured by summing the 
	channels between suitable velocity limits under the line.} 
\label{table:area}      
\begin{tabular}{r c c c c c c c c c c c   }        
\hline\hline             
\noalign{\smallskip}
Name & N$_2$H$^+$ &$^{13}$CS &HC$_3$N& HNC& HCO$^+$& HCN &HNCO& C$_2$H &SiO  &H$^{13}$CN &CH$_3$C$_2$H \\   
& (K km/s)& (K km/s)& (K km/s)& (K km/s)& (K km/s)& (K km/s)& (K km/s)& (K km/s)& (K km/s)& (K km/s)& (K km/s) \\
\noalign{\smallskip}   
\hline                        
\noalign{\smallskip}
                   
IRDC308.13-1 & 4.2     &       &        &   4.8  &  6.2  &    5.3 &  	      &       &      &       &         \\ 
IRDC308.13-2 & 2.1     &       &        &   2.6  &  4.2  &    2.9 & 	      & 0.6   &      &       &        \\ 
IRDC308.13-3 & 1.6     &       &        &   2.9  &  3.7  &    3.3 & 	      &       &      &       &      	\\ 
IRDC309.13-1 & 5.0     &       &        &   3.1  &  4.1  &    4.5 & 	      & 0.4   &      &       &      	\\
IRDC309.13-2 & 2.8     &       &        &   2.5  &  3.2  &    3.9 & 	      &       &      &       &      	\\ 
IRDC309.13-3 & 1.5     &       &        &   0.8  &  1.9  &    2.3 & 	      &       &      &       &         \\ 
IRDC309.37-1 & 29.1    &       &  1.6   &  13.6  & 17.9  &   15.4 &   1.4     & 3.2   &      &       &       \\ 
IRDC309.37-2 &         &       &        &   1.2  &  2.4  &    2.4 &  	      &       &      &         &    	\\ 
IRDC309.37-3 &  4.6    &       &        &   4.8  &  6.8  &    6.0 & 	      &       &      &         &      \\ 
IRDC310.39-1 &  13.4   &       &  1.9   &   8.1  & 13.3  &   12.5 &   0.2     & 1.7   &  0.4 &         &    	 \\ 
IRDC310.39-2 &  5.4    &       &  1.2   &   5.5  &  8.7  &    9.0 &  	      & 1.4   &      &         &    	 \\ 
IRDC312.36-1 &  6.2    &       &  0.2   &   4.9  &  6.6  &    6.6 &   0.6     & 0.7   &      &         &    	   \\
IRDC312.36-2 &  1.3    &       &        &   1.9  &  1.0  &    2.1 &  	      &       &      &         &       \\ 
IRDC313.72-1 &  8.0    &       &  1.7   &   6.9  & 11.3  &    7.8 &   0.6     & 1.1   &  1.6 &         &     \\ 
IRDC313.72-2 &  8.6    &  0.3  &  1.2   &   6.7  & 10.6  &    8.0 &   0.5     & 0.9   &  1.9 &         &     \\ 
IRDC313.72-3 &  7.1    &       &  0.6   &   7.3  &  9.1  &    7.3 &   0.4     & 0.6   &  1.3 &         &     \\ 
IRDC313.72-4 &  5.9    &       &        &   5.7  &  9.3  &    5.9 &   	      &       &  0.6 &         &       \\ 
IRDC316.72-1 &  18.4   &       &  1.1   &  17.1  & 23.2  &   18.6 &   0.8     & 2.8   &      &  1.9    & 0.3	 \\ 
IRDC316.76-1 &  50.7   &  0.9  &  4.9   &  34.1  & 34.6  &   33.8 &   1.1     & 7.1   &  2.3 &  4.8    & 0.7   \\ 
IRDC316.72-2 &  11.8   &       &  0.8   &  10.2  & 14.1  &   10.9 &   	      & 1.9   &      &         & 0.3   \\ 
IRDC316.76-2 &  27.4   &  0.5  &  2.9   &  21.4  & 23.3  &   27.4 &  	      & 7.1   &      &  3.3    & 0.3   \\ 
IRDC317.71-1 &  10.5   &       &  0.4   &   5.0  &  5.6  &    2.7 &   0.7     & 0.8   &      &         &     \\ 
IRDC317.71-2 &  21.9   &       &  2.9   &   8.1  &  5.6  &    4.9 &   1.3     & 1.5   &      &  1.5    & 0.57	  \\ 
IRDC317.71-3 &  3.5    &       &        &   2.3  &  2.5  &    1.8 &  	      &       &  1.0 &         &    	    \\
IRDC320.27-1 &  2.5    &       &        &   2.6  &  2.6  &    3.8 & 	      &       &      &         &    	   \\ 
IRDC320.27-2 &  1.7    &       &        &   1.1  &  0.7  &    1.7 & 	      &       &      &         &    	   \\
IRDC320.27-3 &  0.6    &       &        &   0.8  &  0.3  &    0.8 & 	      &       &      &         &    	   \\ 
IRDC321.73-1 &  11.0   &       &  1.3   &   7.0  & 10.5  &   11.1 & 	      & 1.4   &  2.5 &         &    	  \\
IRDC321.73-2 &  5.5    &       &  1.7   &   6.9  &  8.4  &    7.9 & 	      & 0.9   &      &         &    	   \\
IRDC321.73-3 &  7.5    &       &  1.3   &   8.4  & 10.4  &    8.9 & 	      & 0.8   &      &         &    	   \\
IRDC013.90-1 &  5.6    &       &  0.7   &   5.6  &  6.3  &    5.4 & 	      & 0.9   &      &         &    	 \\
IRDC013.90-2 &  2.4    &       &        &   1.4  &  0.8  &    1.9 & 	      &       &      &         &    	 \\ 
IRDC316.45-1 &  7.0    &       &  1.0   &   4.7  &  5.1  &    4.0 &   0.1     & 0.7   &      &         &       \\ 
IRDC316.45-2 &  6.2    &       &        &   4.9  &  3.4  &    4.0 &   0.4     & 0.7   &      &         &    	 \\ 
IRDC318.15-1 &  3.7    &       &        &   3.9  &  3.0  &    3.3 &   	      & 0.7   &      &         &     \\ 
IRDC318.15-2 &  4.6    &       &        &   3.9  &  1.8  &    2.9 &  	      & 0.5   &      &         &     \\ 
IRDC309.94-1 &  22.0   &       &  3.7   &   9.4  & 11.0  &   11.8 &   0.9     & 2.4   &      &  2.1   &     \\

\noalign{\smallskip}   
\hline                        
\noalign{\smallskip}
                              
\end{tabular}

Notes: Columns are name, integrated area for every species.

\end{table*}


\end{appendix}

\end{document}